\documentclass[aps,prl,twocolumn,superscriptaddress,nofootinbib,notitlepage,longbibliography,reprint]{revtex4-2}
\usepackage{graphicx}% Include figure files
\usepackage{dcolumn}% Align table columns on decimal point
\usepackage{color} % Required for inserting images
\usepackage{amsmath}
\usepackage{bm}
\usepackage{extpfeil}
\usepackage{indentfirst}
\usepackage{nicematrix}
\usepackage{bbm} % to draw empty letter/number
\usepackage{centernot} % to negate a symbol
\usepackage[thicklines]{cancel} % to cancel terms in equation
\usepackage{mathrsfs} % to use \mathscr
\usepackage{comment} % to comment paragraph
\usepackage[normalem]{ulem}
\usepackage{mathtools} % For $:=$
\usepackage[rightcaption]{sidecap} % allows side captions
\usepackage{xcolor}
\usepackage{tocbasic}
\usepackage{hyperref}
\usepackage{bookmark}

\DeclarePairedDelimiter\ket{\lvert}{\rangle}
\DeclarePairedDelimiter\bra{\langle}{\lvert}
\DeclarePairedDelimiter\braket{\langle}{\rangle}
\newcommand*\diff{\mathop{}\!\mathrm{d}} % integral d
\begin{document}
	
	% --- Main text (two columns) ---
	%\documentclass[aps,prl,twocolumn,superscriptaddress,nofootinbib,notitlepage,longbibliography,reprint]{revtex4-2}
%\usepackage{graphicx}% Include figure files
%\usepackage{dcolumn}% Align table columns on decimal point
%\usepackage{graphicx,color} % Required for inserting images
%\usepackage{amsmath}
%\usepackage{bm}
%\usepackage{extpfeil}
%\usepackage{indentfirst}
%\usepackage{nicematrix}
%\usepackage{bbm} % to draw empty letter/number
%\usepackage{centernot} % to negate a symbol
%\usepackage[thicklines]{cancel} % to cancel terms in euqation
%\usepackage{mathrsfs} % to use \mathscr
%\usepackage{comment} % to comment paragraph
%\usepackage[normalem]{ulem}
%\DeclarePairedDelimiter\ket{\lvert}{\rangle}
%\DeclarePairedDelimiter\bra{\langle}{\lvert}
%\DeclarePairedDelimiter\braket{\langle}{\rangle}
%\newcommand*\diff{\mathop{}\!\mathrm{d}} % integral d
%\newcommand*\Diff[1]{\mathop{}\!\mathrm{d^#1}} % integral d^2
%\usepackage{mathtools} % For $:=$
%\newcommand{\JX}[1]{\textcolor{red}{#1}}
%\newcommand{\XL}[1]{\textcolor{blue}{#1}}
%\usepackage[rightcaption]{sidecap} % allows side captions
%\usepackage{xcolor}

%\begin{document}
	
	\title{Quantum Geometric Friedel Oscillations}
	
	\author{Xing-Lei MA}
	\affiliation{Department of Physics, Hong Kong University of Science and Technology, Clear Water Bay, Hong Kong, China}
	\author{Jinchao Zhao}
	\affiliation{Department of Physics, Hong Kong University of Science and Technology, Clear Water Bay, Hong Kong, China}
	\author{Bo-Qing Wu}
	\affiliation{Department of Physics, Hong Kong University of Science and Technology, Clear Water Bay, Hong Kong, China}
	\author{Patrick A. Lee} \thanks{palee@mit.edu}
	\affiliation{Department of Physics, Massachusetts Institute of Technology, Cambridge, MA 02139, USA}
	
	\author{K. T. Law} \thanks{phlaw@ust.hk}
	\affiliation{Department of Physics, Hong Kong University of Science and Technology, Clear Water Bay, Hong Kong, China}

	%\date{\today}
	\begin{abstract}
		In conventional Friedel oscillations, the real-space charge density oscillations induced by an impurity are characterized by an oscillation period set by the Fermi momentum. In this work, we demonstrate that in metals with an isolated (nearly) flat band at the Fermi energy, quantum geometry induces a distinct type of oscillation, which we call \emph{quantum geometric Friedel oscillations} (QGFOs). The period of the QGFOs is set by the momentum-space separation of the quantum metric hot spots of the isolated band. The conventional and quantum geometry-induced oscillations coexist at low temperatures. At higher temperatures, the conventional Friedel oscillation amplitudes away from the impurity site are set by the thermal length such that the oscillations can be easily washed out by temperature effects. Remarkably, the decay length of the QGFOs is set by the quantum metric length, which is defined by the integration of the quantum metric of the isolated band. As a result, the QGFOs can persist even at temperatures much larger than the bandwidth of the isolated flat band. Moreover, the decay length is invariant for a wide range of temperatures, which is a striking result. Our work suggests that the measurement of the QGFOs is a powerful way to detect the quantum metric length (which is associated with the integral of the quantum metric) and the quantum metric hot spot separations (which are associated with the distribution of the quantum metric in the momentum space). 
		
	\end{abstract}
	
	\maketitle
	
%	\emph{\textcolor{blue}{Introduction.---}}
	\section{Introduction}
	Friedel oscillations are intriguing phenomena caused by the interference of electronic wavefunctions induced by charge impurities~\cite{friedel1958metallic}, and their measurement is a powerful tool for probing the Fermi surface properties of metals~\cite{bena2016friedel}. These charge density oscillations share the same root as other important phenomena such as the Ruderman-Kittel-Kasuya-Yosida (RKKY) interaction~\cite{ruderman1954indirect,kasuya1956theory, yosida1957magnetic}, Thomas-Fermi screening~\cite{ashcroft1976solid}, and the Kohn-Luttinger mechanism of superconductivity~\cite{kohn1965new}. Conventionally, these real-space oscillations are characterized by a spatial period determined by the Fermi momentum, $2k_F$~\cite{PhysRevB.72.045127}. The oscillation amplitudes are susceptible to effects such as finite temperature and electron-electron interactions~\cite{affleck2008friedel,grassme1993friedel,cavaliere2014thermally,egger1996friedel}, which smear the Fermi surface. At finite temperatures, for example, the Friedel oscillation amplitude is suppressed by an exponential decay factor where the decay length is controlled by the thermal length $\xi_T = \hbar v_F/{2\pi k_BT}$ ~\cite{grassme1993friedel}.  This fragility is particularly pronounced in flat-band metals~\cite{geim2013van,cao2018unconventional,cai2023signatures,leykam2018artificial,neves2024crystal,ye2024hopping} with a quenched Fermi velocity $v_F$ as the thermal energy $k_BT$ can easily wash out the Friedel oscillations.
	
	%The questions are: Are there Friedel oscillations in flat band materials with small $v_F$? If the answer is confirmatory, what are the governing length scales of the charge oscillations? This work is devoted to answering the above questions.
	The questions are: Are there Friedel oscillations in flat band materials with quenched $v_F$? If the answer is confirmatory, what are the governing length scales of the charge oscillations? In this work, we answer these questions and show that the conventional Friedel oscillation theory, which ignores the quantum geometric effects~\cite{provost1980riemannian,resta2011insulating,torma2023essay,yu2025quantum,liu2025quantum,verma2026quantum,peotta2015superfluidity,PhysRevLett117_045303,PhysRevLett.123.237002,xie2020topology,herzog2022superfluid, torma2022superconductivity,PhysRevLett.132.026002,PhysRevLett133_106701,du2021quantum,jiang2025revealing,souza2000polarization,PhysRevResearch.7.023158,ahn2022riemannian,verma2025instantaneous,PhysRevX.14.041004,zhang2026identifying,rossi2021quantum,Sukhachov_2025,5fks-rrvg,PhysRevLett.131.016002,sun2025flat,PhysRevLett.132.036001,shavit2025quantum,kitamura2024spin,hu2026ferromagnetism}, is incomplete. We point out that the nontrivial Bloch wavefunctions of the flat bands, manifested through the quantum metric and the quantum metric hot spot separations, induce a new type of charge density oscillation, which we call \emph{quantum geometric Friedel oscillations} (QGFOs). At low temperatures, QGFOs coexist with the conventional Friedel oscillations and give rise to extra oscillating components. At high temperatures, however, the conventional Friedel oscillations are greatly suppressed, whereas the QGFOs persist even when the thermal energy ($k_B T$) far exceeds the flat-band bandwidth. Moreover, the decay length of the oscillation amplitude as well as the spatial spread of the QGFOs [see Eq.~\eqref{Eq_QGFOs_extent_QML}] is determined by the quantum metric length $\ell_{\mathrm{QM}}$~\cite{PhysRevLett.132.026002,hu2025anomalous,guo2025majorana,PhysRevResearch.7.023273,chau2026quantummetriclengthfundamental,ma2025universalboundarymodeslocalizationquantum,dai2026quantummetriclocalizationquantum,zhao2026quantummetricboundstate,chen2026quantumgeometricquadrupolecooper,lee2026embedding}, which is the integral of the quantum metric of the flat band, as defined in Eq.~\eqref{QML}.
	
	%In recent years, Hilbert-space geometry~\cite{provost1980riemannian,resta2011insulating} has emerged as a cetral theme in understanding flat-band physics~\cite{torma2023essay,yu2025quantum,liu2025quantum,verma2026quantum}. While the imaginary part of the quantum geometric tensor—the Berry curvature—is renowned for its topological consequences~\cite{xiao2010berry}, the physical implications of its real part—the quantum metric—have been fully appreciated only relatively recently. It reshapes subjects ranging from the superfluid stiffness~\cite{peotta2015superfluidity,PhysRevLett117_045303,PhysRevLett.123.237002,xie2020topology,herzog2022superfluid, torma2022superconductivity,PhysRevLett.132.026002}, nonlinear transports~\cite{PhysRevLett133_106701,du2021quantum,jiang2025revealing} and optical responses~\cite{souza2000polarization,PhysRevResearch.7.023158,ahn2022riemannian,verma2025instantaneous}, to various interaction-driven phases~\cite{PhysRevX.14.041004,zhang2026identifying,rossi2021quantum,Sukhachov_2025,5fks-rrvg,PhysRevLett.131.016002,sun2025flat,PhysRevLett.132.036001}. However, a fundamental question remains: in nearly flat bands where the Fermi momentum is smeared, does quantum geometry permit a mechanism that produces real-space oscillations?
	
\begin{figure}[tbp]
		\centering
		\includegraphics[width=1\linewidth]{ 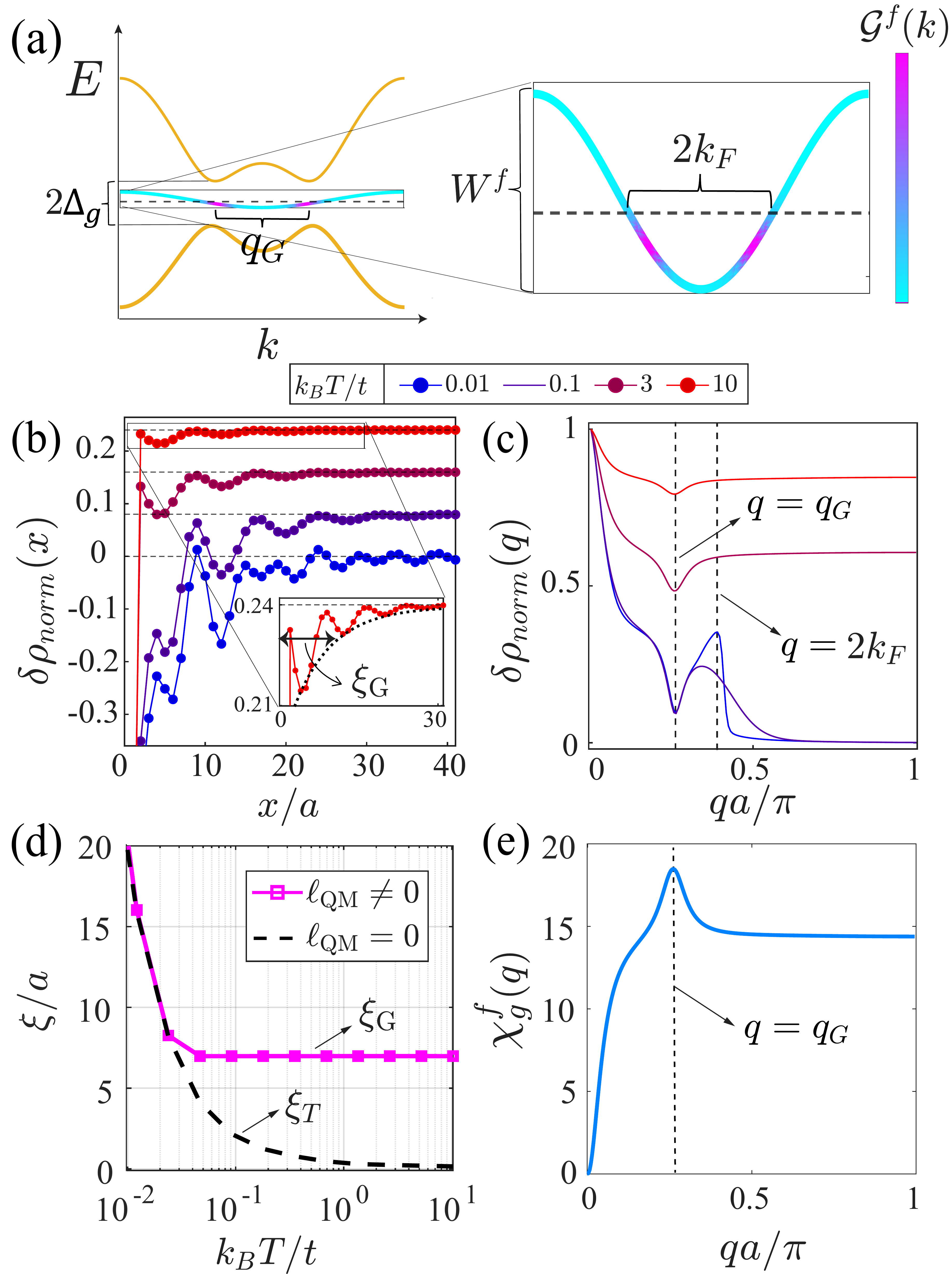}
		\caption{Quantum geometric Friedel oscillations (QGFOs) in a three-band model.
			(a) Schematic band structure of the three-band model as described by Eq.~\eqref{Eq_Model_Hamilt} in the Materials and Methods. The colors of the central band encode the quantum metric distribution $\mathcal{G}^f(k)$ of the middle band. The momentum-space separation between quantum metric hot spots defines a characteristic geometric wavevector $q_G$. The Fermi level (dashed line) cuts through the flat band, with the two Fermi momenta separated by the wavevector $2k_F$. $W^f$ denotes the flat-band bandwidth and $\Delta_g$ denotes the band gap between the flat band and other bands.
			(b) Normalized real-space density modulations $\delta\rho_{norm}(x) \equiv \delta \rho(x)/\rho_{0}$ induced by a local impurity. At low temperatures (blue curve), conventional $2k_F$ oscillations and QGFOs coexist. At high temperatures ($k_B T > W^f$, red curve), only the QGFOs survive. Curves are vertically offset for clarity. The envelope decay function at high temperatures exhibits an exponential decay with decay length $\xi_G$ governed by the quantum metric length $\ell_{\mathrm{QM}}$.
			(c) Normalized Fourier spectra of the density modulations at various temperatures, $\delta\rho_{\text{norm}}(q) \equiv \delta\rho(q)/\delta\rho(0)$. Dashed lines mark the momenta at which prominent features appear.
			(d) The long-range spatial decay length $\xi$ extracted as a function of temperature for finite (solid magenta curve) and zero quantum metric length (dashed black curve), respectively. A crossover from the thermal length $\xi_T$ to a plateaued $\xi_G$ is identified for $\ell_{QM}\neq 0$ by increasing the temperature. (e) Quantum geometric susceptibility in the flat band at temperature $k_B T = 10t$, which exhibits a peak at wavevector $q_G$. Parameters for (b-d) are $J=1$, $t=10^{-4}J$, $\mu = -1.6t$, $\lambda=0.02$, $\delta = 0.03$, $\delta' = 5\delta$, $U_0 = 0.01J$.}
		\label{Fig_1}
	\end{figure}

	To be specific, we calculate the charge density of a three-band tight-binding model with an impurity site (at $x=0$). The details of the model and the calculation methods are given in the Materials and Methods. The band structure of the model and the quantum metric of the flat band $\mathcal{G}^f(k)$ are illustrated in Fig.~\ref{Fig_1}(a). The chemical potential cuts the middle flat band, which is isolated from dispersive bands by a band gap $\Delta_g$ and has a bandwidth $W^{f}$ satisfying $W^{f} \ll \Delta_g$. Normalized charge density oscillations $\delta\rho_{norm}(x)$ near the impurity site and their Fourier spectra $\delta\rho_{norm}(q)$ are depicted in Fig.~\ref{Fig_1}(b) and (c), respectively. The following are the key observations concerning the charge oscillations. First, there are additional oscillating periods at low temperatures ($k_BT \ll W^{f}$), as shown in Fig.~\ref{Fig_1}(b). The Fourier spectra exhibit sharp features at $2k_F$ as well as $q_G$, where $q_G$ is the quantum metric hot spots separation defined in Fig.~\ref{Fig_1}(a). In other words, $q_G$ induces new components of Friedel oscillations.
	
	%First, we show that the QGFO period is not determined by the Fermi surface, but is instead governed by a characteristic wavevector $q_G$. As depicted in Fig.~\ref{Fig_1} (a), this wavevector corresponds to the momentum-space separation between quantum metric hot spots. %—points where the band-averaged quantum distance exhibits pronounced extrema. This purely geometric wavevector effectively replaces the role of the conventional Fermi momentum, becoming the dominant real-space oscillation frequency evident in the Fourier spectrum, Fig.~\ref{Fig_1}(c).
	
	Second, because the QGFOs are rooted in the intrinsic quantum geometry of the Bloch states rather than the sharpness of the Fermi surface, they exhibit remarkable thermodynamic stability. As shown in Fig.~\ref{Fig_1}(b) and (c), while the conventional oscillations are completely washed out by temperature, the QGFOs persist robustly even when the temperature far exceeds the flat-band bandwidth. Importantly, as shown in the inset of Fig.~\ref{Fig_1}(b), in the high temperature regime when $W^{f} \ll k_BT \ll \Delta_g$, the oscillation amplitude decay $\xi_G$ of the QGFOs is set by the quantum metric length $\ell_{\mathrm{QM}}$. Incredibly, as shown in Fig.~\ref{Fig_1}(d), the extracted decay length $\xi$ exhibits a crossover from $\xi_T$ (which is proportional to $1/T$) to $\xi_G$ (which is temperature independent) as temperature increases. The decay length plateau (as a function of temperature), depicted in Fig.~\ref{Fig_1}(d), is a striking experimental signature of the quantum metric.  In sharp contrast, the conventional Friedel oscillations with zero quantum metric length quickly approach zero at finite temperatures [dashed black curve in Fig.~\ref{Fig_1}(d)]. The key connection between $q_G$ and the QGFOs is through the quantum geometric component of the charge susceptibility $\chi^f_g(q)$ in Eq.~\eqref{Eq_chi_c_chi_g}, which peaks at $q=q_G$, as shown in Fig.~\ref{Fig_1}(e). In the following sections, the theory of QGFOs, explaining the numerical results, is unfolded. 
	
	\section{Results}
	\subsection{Origin of QGFOs}
	%\section{Origin of the QGFOs}
	When a local impurity $U_0 \delta(\bm{r})$ is introduced in a flat-band metal, the induced density modulations $\delta \rho(\bm{r})$ can be related to the static susceptibility function $\chi(\bm q)$ as
	\begin{equation}
		\begin{aligned}
			\delta \rho(\bm{r}) =  U_0 \sum_{\bm{q}} \chi(\bm{q})  e^{i\bm{q} \cdot \bm{r}}.
		\end{aligned}
		\label{Eq_delta_rho_k}
	\end{equation}
	%Here, $\chi(\bm{q})$ is the susceptibility function in response to a localized defect potential $U_0 \delta(\bm{r})$ (diagonal in orbital basis).
	The susceptibility takes a generic form~\cite{dutreix2016friedel,singh2003ferromagnetism,chen2024impurity} as
	\begin{equation}
		\begin{aligned}
			\chi(\bm{q}) 
			= & -\frac{1}{\pi} \Im \left[ \frac{1}{N} \sum_{\bm{k}}  \int_\omega f(\omega)  \operatorname*{Tr} \left[G_{\omega}(\bm{k}+\bm{q}) \mathcal{T}_{\omega} G_{\omega}(\bm{k}) \right] \right] ,
		\end{aligned}
		\label{Eq_chi_q_FullBand}
	\end{equation}
	where $\mathcal{T}_{\omega} = \frac{1}{1-U_0 G_0(\omega)}$ is the $\mathcal{T}$-matrix, $G_0(\omega) = \frac{1}{N} \sum_{\bm{k}} G_{\omega}(\bm{k})$ the local Green's function, $G_{\omega}(\bm{k}) = (\omega^+-H(\bm{k}))^{-1}$ with $\omega^+ = \omega+i0^+$ and $f(\omega)=\frac{1}{1+\mathrm{e}^{\omega/k_BT}}$ is the Fermi-Dirac distribution at temperature $T$. 
	For an isolated nearly flat band with $U_0, k_BT \ll \Delta_g$, $\chi(\bm{q})$ is dominated by the flat band contribution (see SI Appendix, Section III), which, after band projection, gives the flat-band static susceptibility~\cite{sablikov2025friedel}:
	\begin{equation}
		\begin{aligned}
			\chi^f(\bm{q})&=  -\frac{1}{\pi}
			\Im \left[ \frac{1}{N} \sum_{\bm{k}} \int_\omega  \frac{f(\omega) \Lambda^f_{\bm{k}+\bm{q}, \bm{k}} \tilde{\Lambda}^f_{\bm{k}, \bm{k}+\bm{q}}}{(\omega^+-E^f_{\bm{k}+\bm{q}}) (\omega^+-E^f_{\bm{k}}) } \right].
		\end{aligned}
		\label{Eq_chi_q_projection}
	\end{equation}
	Here, the projected form factors $\Lambda^f_{\bm{k}, \bm{k}'} = \braket{u_f(\bm{k})|u_f(\bm{k'})}$ and $\tilde{\Lambda}^f_{\bm{k}, \bm{k}'} =  \braket{u_f(\bm{k})|\mathcal{T}_{\omega}|u_f(\bm{k}')}$ in the numerator encode the quantum geometry of the flat band. For weak impurities with $U_0 \ll \Delta_g$, $\mathcal{T}_{\omega} \approx \mathcal{I}$ is a reasonable approximation. Physically, the presence of the form factor indicates that during each scattering event, the quantum distance between the final and initial states $d_{\bm{k}, \bm{k}'} \equiv 1-|\Lambda_{\bm{k}, \bm{k}'}|^2$ determines the scattering probability. As a result, the quantum distance fundamentally reshapes the Friedel oscillations. To see this, we rearrange Eq.~\eqref{Eq_chi_q_projection} as
	\begin{equation}
		\chi^f(\bm{q}) =  \chi^f_c(\bm{q}) + \chi^f_g(\bm{q}),
	\end{equation}
	with
	\begin{equation}
		\chi^f_c(\bm{q}) 
		=  \frac{1}{N} \sum_{\bm{k}} C^f_{\bm{k}, \bm{q}}, \quad
		\chi^f_g(\bm{q}) 
		= -\frac{1}{N} \sum_{\bm{k}} C^f_{\bm{k}, \bm{q}} d_{\bm{k},\bm{k}+\bm{q}}.
		\label{Eq_chi_c_chi_g}
	\end{equation}
	We introduce $\chi^f_c$ and $\chi^f_g$ to separate the conventional contribution from the quantum geometric contribution. Here, $C^f_{\bm{k}, \bm{q}} \equiv [f(E^f_{\bm{k}+\bm{q}}) - f( E^f_{\bm{k}})]/(E^f_{\bm{k}+\bm{q}} - E^f_{\bm{k}})$ is the finite-temperature Lindhard weight. 
	The conventional part $\chi^f_c$ in Eq.~\eqref{Eq_chi_c_chi_g} gives rise to real-space oscillations characterized by a wavevector $2k_F$ and thermal length $\xi_T = \hbar v_F/2\pi k_B T$, with a well-known asymptotic form $\sim - \cos(2k_Fr) e^{-r/\xi_T}/r^{d-1}$ in $d$-dimensions~\cite{grassme1993friedel}. 
	
\begin{figure}[tbp]
		\centering
		\includegraphics[width=1\linewidth]{ 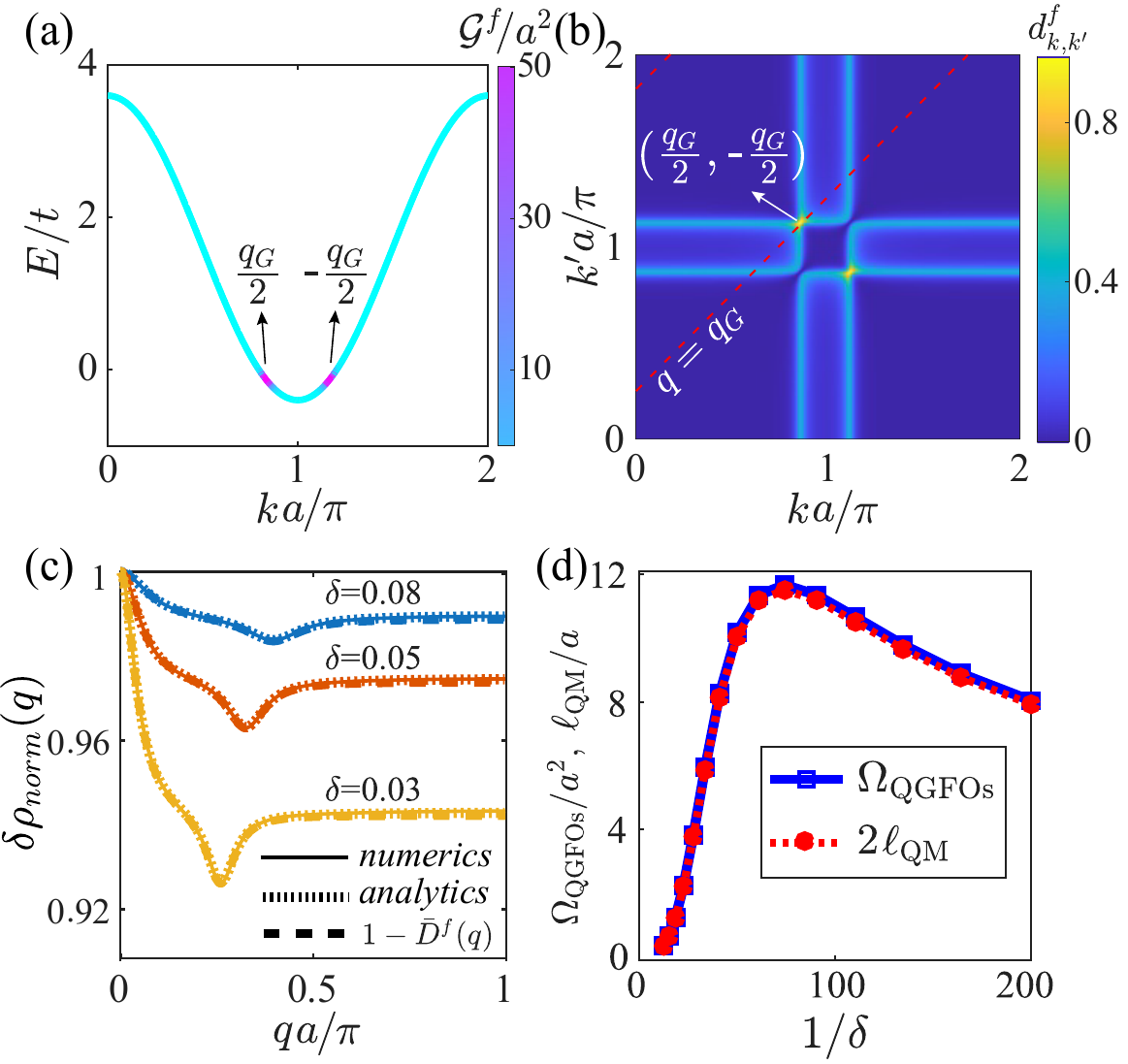}
		\caption{(a) The calculated quantum metric distribution of the middle flat band in the 1D three-band model, as described by Eq.~\eqref{Eq_Model_Hamilt} in the Materials and Methods. The color scale on the band depicts the quantum metric strength. There are two prominent quantum metric hot spots at $\pm \frac{q_G}{2}$. The parameters are the same as those in Fig.~\ref{Fig_1}, except the value of $\delta$ which is set at $\delta = 0.05$ for (a) and (b). (b) Quantum distance distribution $d_{k,k'}$ in the $k$-$k'$ plane. The red dashed lines mark the cut $q=q_G$ corresponding to the peak of $\bar{D}(q)$. (c) Fourier transform of the LDOS variations $\delta\rho_{norm}(q)$ for different model parameter $\delta$. Solid lines: obtained from real-space exact diagonalization (numerics); dotted lines: obtained from full-band formalism in Eq.~\eqref{Eq_chi_q_FullBand} with the Born approximation $\mathcal{T} = 1$. Both curves match very well with $1-\bar{D}^f(q)$ in the high temperature regime when $k_B T=10t$.
		A single impurity with strength $U_0=t$ is placed at the first site. (d) The spatial extent $\Omega_{\mathrm{QGFOs}}$ and the  quantum metric length are plotted as a function of $\delta$. The relation of $\Omega_{\mathrm{QGFOs}} = 2a \ell_{\mathrm{QM}}$ is clearly established. }
		\label{Fig_2}
	\end{figure}

	On the other hand, the quantum-geometric contribution $\chi^f_g$, which is related to the quantum distance $d_{\bm{k},\bm{k}'}$,  is greatly affected by the quantum geometric properties of the band.
	%Specifically, if $d_{\bm{k},\bm{k}'}$ peaks (dips) prominently in certain area, the points where they intersect with Fermi surface will contribute significantly to $\chi_g^f$. For example, when $d_{{k}, {k}'} = d_0 \sum_{s=\pm}[\delta( {k} + s {k}_g ) + \delta( {k}' + s {k}_g )] $ ($0 < k_g < k_F$), it leads to an oscillatory behavior with $\chi^f_g \sim \sum_{s=\pm} \cos[2(k_F + s k_g) r]/r$.
	Specifically, if $d_{\bm{k},\bm{k}'}$ peaks prominently at certain points in the $\bm{k}$-$\bm{k}'$ plane, e.g., $(\bm{k}_g, \bm{k}'_g)$, the corresponding wavevector $\bm{q}_G = \bm{k}'_g - \bm{k}_g$ will cause sharp features in $\chi^f_g(\bm{q})$. An example is given in Fig.~\ref{Fig_1}(e). Therefore, an oscillatory mode related to $\bm{q}_G$ is manifested in the local density of states (LDOS), namely, the QGFOs.
	In Fig.~\ref{Fig_2}(b), the quantum distance distribution in the flat band of the 1D three-band model is illustrated. The details of the model are given in the Materials and Methods. The quantum distance $d_{\bm{k},\bm{k}'}$ features two sets of prominent ridges along $k,k' = \pm q_G/2$ as well as two bright foci at $\pm(q_G/2,-q_G/2)$. These intense features originate from the quantum metric hot spots located at $\pm q_G/2$, as shown in Fig.~\ref{Fig_2}(a). Physically, a quantum metric hot spot indicates that the wavefunction at the hot spot momentum is very different from nearby Bloch states. In general, the quantum distance $d_{\bm{k},\bm{k}'}$ between states belonging to the two hot spots exhibits sharp features at $\bm{q}_G = \bm{k}'_g - \bm{k}_g$. Scatterings induced by impurities between the states of the hot spots are suppressed or enhanced. This special property of the hot spot states causes density oscillations with a period set by $\bm{q}_G$. It can be shown that the $\bm{q}_G$ component gets strengthened at stronger impurity beyond the Born approximation (see SI Appendix, Section I).

	%%%%%%%%%%%%%%%%%%%%High Temperature regime and dominant QGFOs
	While Eq.~\eqref{Eq_chi_c_chi_g} shows that both the conventional and quantum geometric components contribute to the Friedel oscillations on an equal footing, one contribution can dominate over the other depending on temperature and spatial range.
	At low temperatures when $k_B T \ll W^f$, both the $2k_F$ and $q_G$ components appear in the short-range modulations [Fig.~\ref{Fig_1}(b,c)], whereas at long distances only the $2k_F$ component persists, and the exponential decay of the oscillation amplitudes is governed by the thermal length $\xi_T \simeq a \frac{W^f}{k_BT} \gg a$ ($a$ is the lattice constant), as shown in the low temperature regime of Fig.~\ref{Fig_1}(d). 
	% In this case, the thermal length $\xi_T \simeq a \frac{W^f}{k_BT} \gg a$ ($a$ is the lattice constant) governs the long distance exponential decay of the oscillation amplitudes as shown in the low temperature regime of Fig.~\ref{Fig_1}(d). %$\xi_T \gg \hbar v_F/(2\pi W^f) \sim a$. 
	%The reason is that QGFOs decay exponentially [$\sim -(A+B\sin(q_G r + \phi) e^{-r/\xi_G}$ in 1D; see SM~\cite{supp}], whereas the conventional one follows a power law. 
	As temperature increases, the thermal length is reduced such that the conventional Friedel oscillations gradually fade away.
	
	Surprisingly, QGFOs remain robust and dominate even when the thermal energy exceeds the bandwidth, i.e., $ W^f \ll k_BT \ll \Delta_g$, as illustrated by the high-temperature plots in Fig.~\ref{Fig_1}(b). In this high temperature regime, we have
	%$f(\omega) \simeq \frac{1}{2}(1-\frac{\omega}{2T})$, 
	$C^f_{\bm{k},\bm{q}} \simeq  %\frac{f(E_{k+q}) - f(E_k)}{E_{k+q} - E_k} =
	-\frac{1}{4k_BT}$
	%\begin{equation}
	%	\begin{aligned}
		%%		f(\omega) &= \frac{1}{1+e^{\omega/T}} \approx \frac{1}{2}(1-\frac{\omega}{2T}) \\
		%		C_{\bm{q}} &= \frac{f(E_{k+q}) - f(E_k)}{E_{k+q} - E_k} = -\frac{1}{4T}.
		%	\end{aligned}
	%	\label{Eq_chi_c_chi_g}
	%\end{equation}
	such that the quantum geometric susceptibility $\chi^f_g(\bm{q})$ is to the averaged quantum distance $	\bar{D}^{f}(\bm{q}) \equiv \frac{1}{N} \sum_{\bm{k}} d^f_{\bm{k}, \bm{k}+\bm{q}}$, and the total flat-band susceptibility can be written as %$\chi^f_g(\bm{q}) = \frac{\bar{D}^f(\bm{q})}{4k_BT}$.
	%\begin{equation}
	%	\begin{aligned}
		%		%\chi^f_c( \bm{q}) &= -\frac{1}{4T} \\
		%		\chi^f_g(\bm{q}) &= \frac{\bar{D}^f(\bm{q})}{4T},
		%%\\  \Longrightarrow \chi^f(\bm{q}) &= - \frac{1}{4T}(1-\bar{D}(\bm{q})).
		%	\end{aligned}
	%	\label{Eq_chi_c_chi_g_High_Temperature}
	%\end{equation}
	\begin{equation}
		\begin{aligned}
			\chi^f(\bm{q}) &= - \frac{1}{4k_BT}\left[1-\bar{D}^f(\bm{q})\right].
		\end{aligned}
		\label{Eq_chi_tot_High_temperature}
	\end{equation}
	It is clear from Eq.~\eqref{Eq_chi_tot_High_temperature} that the $\bm{q}$-dependence of the high temperature flat-band susceptibility is determined by the averaged quantum distance $\bar{D}^{f}(\bm{q})$. Consequently, peaks and dips in $\bar{D}^f(\bm{q})$ yield distinctive features in the susceptibility, e.g., the peak at $q_G$ in $\chi^f_g(\bm{q})$ shown in Fig.~\ref{Fig_1}(e). These extrema in $\chi^f_g(\bm{q})$ result in charge density oscillations at high temperatures with spatial periods controlled by the quantum metric hot spot separations. 
	
	To show the validity of the analytical approach which results in Eq.~\eqref{Eq_chi_tot_High_temperature}, we calculate the normalized Fourier spectra of the charge density modulations $\delta\rho_{\text{norm}}(\bm{q}) \equiv \delta\rho(\bm{q})/\delta\rho(0)$ in the high temperature regime using three different methods.
	%we calculate the charge density modulations $\delta \rho(\bm{r})$ defined in Eq.~\eqref{Eq_delta_rho_k} and their Fourier spectra normalized as $\delta\rho_{\text{norm}}(q) \equiv \delta\rho(q)/\delta\rho(0)$. The Fourier spectra $\delta\rho_{\text{norm}}(q)$ is calculated using three different methods. 
	First, the charge density modulations and their Fourier components are calculated by numerical exact diagonalization [solid lines in Fig.~\ref{Fig_2}(c)] using the three-band model with an impurity site, as described in the Materials and Methods. Second, we employ Eq.~\eqref{Eq_chi_q_FullBand} with the three-band Green's function $G_{\omega}$ and the Born approximation (by setting $\mathcal{T}_{\omega}= 1$) to calculate $\delta\rho_{\text{norm}}(q)$ [dotted lines in Fig.~\ref{Fig_2}(c)]. Finally, we calculate the averaged quantum distance and plot $1-\bar{D}^f(\bm{q})$ in the projected single-band expression Eq.~\eqref{Eq_chi_tot_High_temperature} [dashed lines in Fig.~\ref{Fig_2}(c)].
	%to calculate $\delta\rho_{\text{norm}}(q)$ in the high temperature regime. The results are depicted in Fig.~\ref{Fig_2}(c). 
	It is remarkable that the analytical understanding from Eq.~\eqref{Eq_chi_tot_High_temperature} matches the exact diagonalization as well as the three-band calculations extremely well. Fig.~\ref{Fig_2}(c) clearly demonstrates the fact that the averaged quantum distance $\bar{D}^f(\bm{q})$ governs the high temperature QGFOs. 
	The parameter $\delta$, defined in Eq.~\eqref{Eq_Model_Hamilt}, tunes the quantum metric and its hot spots separation, thereby shifting the oscillation period $2\pi/q_{G}$ of the QGFOs in Fig.~\ref{Fig_2}(c).

	%%--------------------------------------------------------------
%	\emph{\color{blue}{Quantum metric length in QGFOs. ---}}
	\subsection{Quantum metric length in QGFOs}
	%\section{Quantum metric length in QGFOs}
	As shown in Fig.~\ref{Fig_1}(b), QGFOs exhibit remarkable robustness against temperature variations, as their origin lies in the quantum geometry of the isolated band rather than in the Fermi-surface sharpness responsible for the conventional Friedel oscillations. In the SI Appendix, Section I, we show that in the high temperature regime, the QGFOs in general take an exponentially decaying form, i.e., $\chi^f(r) \sim -[A+B\sin(q_G r + \phi)] e^{-r/\xi_G}$ in 1D, where $A$, $B$ and $\phi$ are model-dependent parameters. The decay length $\xi_G$, denoted in the inset of Fig.~\ref{Fig_1}(b), is governed by the quantum metric length $\ell_{\mathrm{QM}}$ defined as:
	%. Remarkably, $\xi_G \sim \ell_{\mathrm{QM}}$ where $\ell_{\mathrm{QM}}$ is the quantum metric length defined as:
	\begin{equation}
		\begin{aligned}
			\ell_{\mathrm{QM}} \equiv \frac{V_{cell}}{(2\pi)^d a} \int  \operatorname{Tr} [\mathcal{G}^f (\bm{k})] \diff \bm{k},
			%\frac{1}{Na} \sum_{\bm{k}} \operatorname{Tr} [\mathcal{G}^f (\bm{k})].
		\end{aligned}
		\label{QML}
	\end{equation}
	such that $\xi_G \approx \eta \ell_{\mathrm{QM}}$, where $\eta$ is an order 1 constant. Here, $V_{cell}$ is the unit-cell volume, $d$ is the spatial dimension, and $\mathcal{G}^f(\bm{k})$ is the quantum metric in the flat band defined at momentum $\bm{k}$. 
	Fig.~\ref{Fig_1}(d) shows that at low temperatures, the decay length of the Friedel oscillations $\xi$ is determined by the conventional thermal length $\xi_{T}$.
	As the temperature increases, the QGFOs component becomes dominant and the crossover occurs at a temperature satisfying $\xi_{T} = \xi_G$, after which the decay length enters a temperature-independent plateau set by $\xi_G$.
	%However, when the thermal energy of is larger than the bandwidth, the conventional Friedel oscillations are washed out and the QGFOs dominate. In this high temperature regime, the decay length is determined by $\xi_G$ which is temperature independent. 
	This decoupling between the decay length of the QGFOs and the thermal energy is one of the striking results of this work and is testable by scanning tunneling microscopy (STM) experiments.
	Notably, the quantum metric length~\cite{marzari1997maximally,marzari2012maximally,souza2001maximally} has been established as a fundamental length scale in flat-band materials, governing the spatial decay of various types of bound states~\cite{PhysRevResearch.7.023273,guo2025majorana,ma2025universalboundarymodeslocalizationquantum,lee2026embedding,zhao2026quantummetricboundstate}. Here, we demonstrate again that this same length scale also governs the spatial decay of the high-temperature charge density response, which involves not a single bound state, but all occupied states within an isolated band.

	Importantly, we show in the Materials and Methods that the spatial spread of the density modulations at high temperatures is also determined by the quantum metric length $\ell_{\mathrm{QM}}$:
	\begin{equation}
		\begin{aligned}
			\Omega_{\mathrm{QGFOs}}
			&= 2a \ell_{\mathrm{QM}}.
		\end{aligned}
		\label{Eq_QGFOs_extent_QML}
	\end{equation}
	Here, $\Omega_{\mathrm{QGFOs}} \equiv  \langle \bm{r}^2\rangle_{\delta\rho} - \langle \bm{r}\rangle_{\delta\rho}^2$, where $\langle \cdots \rangle_{\delta\rho}$ denotes the normalized average weighted by $\delta\rho(\bm{r})$. In Fig.~\ref{Fig_2}(d), we show that the spatial spread of QGFOs obtained from exact diagonalization of the three-band model agrees with $2a\ell_{\mathrm{QM}}$ throughout the parameter range of $\delta$, confirming the relation in Eq.~\eqref{Eq_QGFOs_extent_QML}.
	%is determined by the quantum metric length as suggested by Eq.~\eqref{Eq_QGFOs_extent_QML}.  
	Notably, Eq.~\eqref{Eq_QGFOs_extent_QML} holds generically in higher dimensions as well, and for a band with a finite Chern number $C$, it implies a topological lower bound $\Omega_{\mathrm{QGFOs}} \ge a^2 |C|/\pi $.
	
	%As a trivial example, in the atomic limit $\bar{D}^f(\bm{q})$ uniformly vanishes, and consequently, only the conventional term remains, producing Friedel oscillations with periods set by the nesting momentum on the Fermi surface, $2\bm{k}_F$. %and it will further influence the weight of the corresponding wavevector component $\bm{q} = \bm{k}'-\bm{k}$ in the density of states after averaging over different momentum $\bm{k}$, which leads to the averaged quantum distance $\bar{D}(\bm{q})$, as defined previously.
	%For more generic cases, at zero temperature, Eq.~\eqref{Eq_chi_c_chi_g} suggests that both the band dispersion and the quantum geometry should contribute to the flat band Friedel oscillations through the Fermi momentum $2\bm{k}_F$ and the averaged quantum distance $\bar{D}^f$, respectively.
	
	%\textcolor{red}{The third regime: $  E_F \lesssim T \ll W^f$. In this regime, Fermi surface is smeared out already so the QGFOs are visible. Interestingly, now it is the averaged quantum distance of the states around the Fermi surface in an energy range $(E_F-  T, E_F+T)$ that determines the QGFOs.}

	%%%%%%%%%%%%%%%%%%%%%%%%%%%%%%%%%%%%%%%%%%%%%%%	
%	\emph{\textcolor{blue}{Conclusions and Discussions.---}}
	\section{Conclusion and Discussions}
	%\section{Conclusions and Discussions}
	In this work, we show that quantum metric hot spots induce a new type of charge density oscillation, which we term quantum geometric Friedel oscillations (QGFOs). The oscillation periods are set by the momentum-space separations of quantum metric hot spots. A more generic proof linking the QGFO periods to quantum metric hot spot separations is provided in the SI Appendix, Section II.
	%Due to their quantum-geometric nature, QGFOs are robust against thermal smearing of the Fermi surface. 
	Importantly, when the conventional Friedel oscillations are washed out by temperature effects, the QGFOs can persist even when the thermal energy exceeds the bandwidth of the isolated band. In the high temperature regime, both the decay length of the oscillation amplitudes and the spatial spread of the QGFOs are governed by the quantum metric length, which remains temperature independent for a wide range of temperatures.
	Notably, owing to the quantum-geometric origin, the QGFOs not only remain robust against thermal smearing of the Fermi surface, but persist even in exactly flat bands where the Fermi surface is not well-defined. We further verify the robustness of QGFOs against variations in the strength and orbital dependence of the impurity, and the chemical potential (SI Appendix, Section I).
	Furthermore, we also apply the QGFOs to magnetic impurities and revealed thermally robust RKKY interactions, as detailed in the SI Appendix, Section VI.
	
	Interestingly, with a low filling fraction in the isolated band, the QGFOs can be predominantly observable at temperatures much lower than the bandwidth, provided that the quantum metric hot spots are located within the thermally accessible window. In the SI Appendix, Section IV, we demonstrate this regime using both 1D and 2D examples.
	%If the quantum metric hot spots are located above the Fermi energy, the QGFOs can only be observed when the temperature is high enough that the electrons can be excited to the states at the quantum metric hot spots.
	In the case that there is a single quantum metric hot spot, the charge density modulations will exhibit a purely exponential decay without oscillations, and the decay length is governed by the quantum metric length. 
	
	Our work uncovers the novel phenomenon of QGFOs, which can be tested by STM experiments. 
	Recently, quantum-geometric hot spots have been observed in flat band materials, including a moire flat band superconductor~\cite{liu2025electric} and the layered electride YCl~\cite{geng2026experimental,zhong2026yclelectridemultiorbitalcorrelated}. These systems offer promising platforms for detecting QGFOs.
	The results given here also open avenues for exploring QGFOs-mediated giant magnetoresistance effect, flat band Kondo physics~\cite{checkelsky2024flat}, unconventional superconductivity~\cite{shavit2025quantum}, and other correlation-driven phenomena.
	%Our work establishes QGFOs as the dominant real-space response when kinetic energy is overwhelmed by temperature. Extending this to interacting systems, where correlations may further reshape the spatial screening pattern, is an interesting future direction. The results here also open avenues for exploring thermally robust RKKY interactions, flat band Kondo physics~\cite{checkelsky2024flat}, unconventional superconductivity~\cite{shavit2025quantum}, and other correlation-driven phenomena.

	\section{Materials and Methods}

%	\twocolumngrid              % ← back to normal two-column text
%	\appendix
%	\setcounter{equation}{0}                     % reset equation counter
%	\renewcommand{\theequation}{A\arabic{equation}} % produce A1, A2, ...
	
	%%%%%%%%%%%%%%%%%%%%%%%%%%%%%%%%%%%%
	\begin{figure}[bbp]
		\centering
		\includegraphics[width=1\linewidth]{ 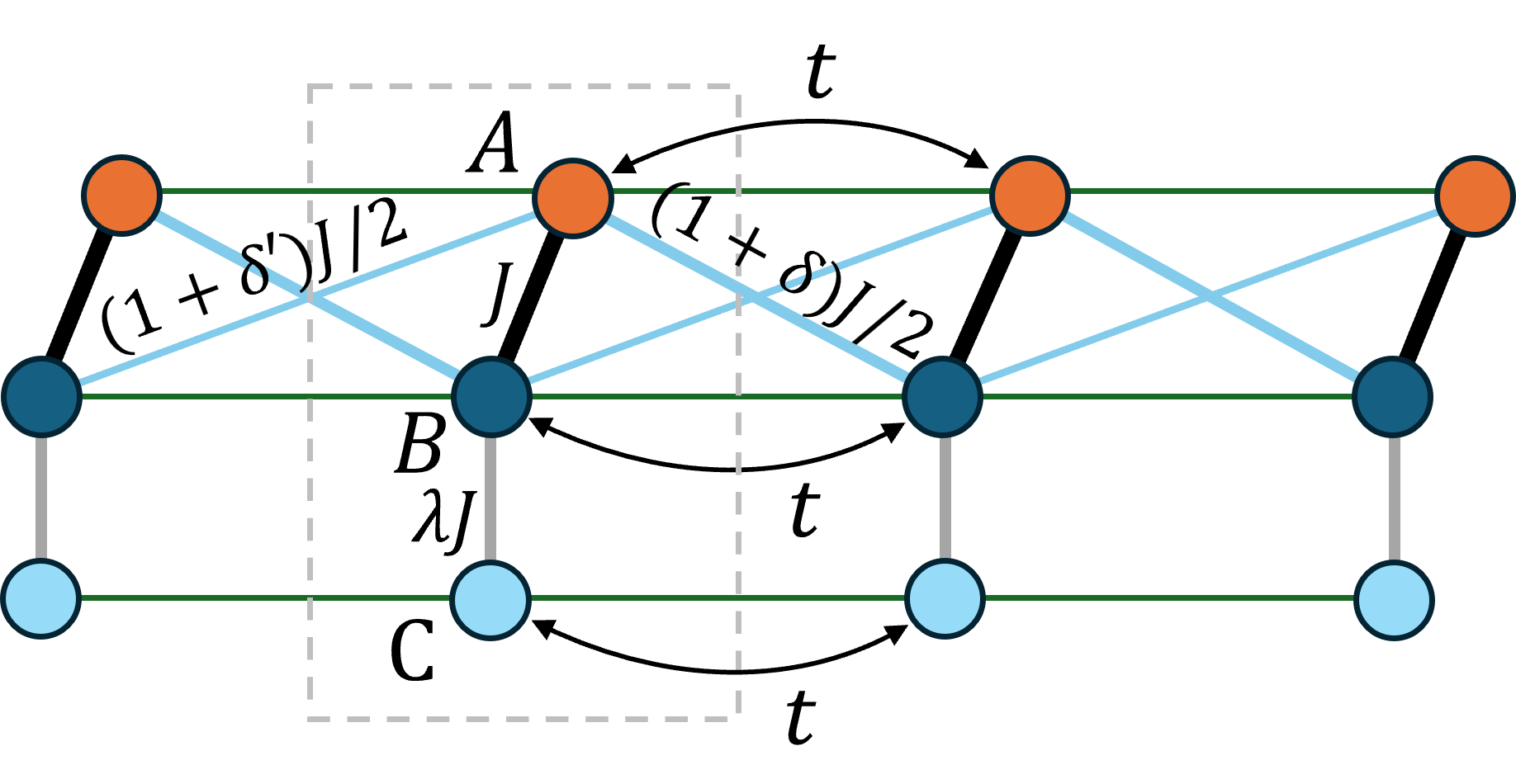}
		\caption{Real-space lattice and hopping structure of the 1D model. The gray dashed square denotes a unit cell.
		}
		\label{Fig_1D_Model_lattice}
	\end{figure}
%	\emph{1D Model and Methods.---}
	\subsection*{1D Model}
	We consider a 1D three-sublattice model with spin-diagonal Hamiltonian $H^{\text{1D}} = \sum_{k,\sigma=\pm} \Psi_{k,\sigma}^{\dagger} \mathcal{H}_{\sigma}(k) \Psi_{k,\sigma}$ where $\Psi_{k,\sigma} = (c_{k,\sigma,A}, c_{k,\sigma,B}, c_{k,\sigma,C})^{\operatorname{T}}$ and
	\begin{equation}
		\begin{aligned}
			& \mathcal{H}_{\sigma}(k) = \begin{pmatrix}
				E^f_k	 	& a_{k} & b_{k} \\
				a_{k}^{\dagger}  & E^f_k  & 0\\
				b_{k}^{\dagger}  &0   & E^f_k
			\end{pmatrix}, \\% + (E_f(k) -\mu) \mathcal{I}_3 \\
			\text{with}  
			\quad a_{k} &  = \left[ 1+ \frac{1+\delta}{2} e^{-ik} + \frac{1+\delta'}{2} e^{ik}  \right] J, \\
			\quad  b_{k} & = \lambda J, 
			\quad \text{and} \quad  
			E^f_k  = 2t \cos(k) -\mu .
		\end{aligned}
		\label{Eq_Model_Hamilt}
	\end{equation}
	Here, we set the lattice constant $a=1$ for simplicity, and $\mu$ is the chemical potential. Fig.~\ref{Fig_1D_Model_lattice} illustrates the real-space lattice structure. The model features three bands with one nearly flat band in the middle, as shown in Fig.~\ref{Fig_1}(a). The dispersive bands nearly touch the flat band at two momenta $\pm \frac{q_G}{2}$, resulting in quantum metric hot spots at these points. A zoomed-in view of the weakly dispersed flat band is provided in Fig.~\ref{Fig_2}(a), and we take $0 < t,|\mu| \ll J$ to ensure flatness. The parameters $\lambda, \delta, \delta' \ll 1$ control the band gap $\Delta_g \simeq \sqrt{(\delta'-\delta)^2(\delta+\delta')/4 + \lambda^2}$, and $\delta, \delta'$ determine $\frac{q_G}{2}$ as well as the quantum-geometric texture in the flat band. Throughout, time-reversal symmetry is preserved and any Kramers pair at $\pm k$ satisfies $u_{k} = u^*_{-k}$. When $\delta \neq \delta'$, the model lacks inversion symmetry, and the Bloch vectors for a Kramers pair are different, i.e., $u_{k} = u^*_{-k} \neq u_{-k}$, resulting in a finite quantum distance between the two momenta of a Kramers pair $d_{k,-k} \neq 0$. This finite quantum distance maximizes at $k = \frac{q_G}{2}$. %, consequently, a peak at $q_G$ appears in $\bar{D}^f(q)$ thus $\chi_g^f(q)$, giving rise to the QGFOs.
	In Fig.~\ref{Fig_2}(b), the quantum distance exhibits $D_2$ symmetry as a result of the time-reversal and index-exchange symmetry, i.e., $d_{k_1, k_2} = d_{-k_1, -k_2} = d_{k_2, k_1}$. Two prominent peaks appear at $\pm(\frac{q_G}{2}, -\frac{q_G}{2})$, and further manifest themselves as a peak in the averaged quantum distance $\bar{D}^f(q=q_G)$ tunable via $\delta, \delta'$, as shown in Fig.~\ref{Fig_2}(c). Therefore, a local impurity strongly suppresses scatterings at $q=q_G$, imprinting spatial oscillations with period $\lambda_G = 2\pi/q_G$.
	In contrast, when $\delta=\delta'$, the inversion symmetry is recovered, so the quantum distance between a Kramers pair vanishes and there is no local maximum in $\bar{D}^f(q)$ nor $\chi^f_g(q)$. As a result, QGFOs disappear in this case.
%	We define the real-space density of states modulations as $ \delta\rho\left(x\right) = \sum_{E_{\text{min}}^f<E_i<E_{\text{max}}^f} {f_{E_i}\left|\psi_i(x)\right|^2} - \rho_0$, and present the normalized quantities $\delta\rho_{norm}(x) = \delta\rho\left(x\right)/\rho_0$ and its Fourier transform $\delta\rho_{norm}(q) = \delta\rho(q)/ \operatorname{max}[\delta\rho(q)] $ in Figs.~\ref{Fig_1} and \ref{Fig_2}.
	
%%%%%%%%%%%%%%%%%%%%%%%%%%%%%%%%
\subsection*{Normalized density of states modulations}
We define the real-space density of states modulations as 
\begin{equation}
	\begin{aligned}
		\delta\rho\left(x\right) = \sum_{E_{\text{min}}^f<E_i<E_{\text{max}}^f} {f_{E_i}\left|\psi_i(x)\right|^2} - \rho_0,
	\end{aligned}
	\label{Eq_LDOS_def}
\end{equation}
and normalized density modulations as $\delta\rho_{norm}(x) = \delta\rho\left(x\right)/\rho_0$. The Fourier spectra of the density modulations is normalized as $\delta\rho_{norm}(q) = \delta\rho(q)/ \operatorname{max}[\delta\rho(q)] $ in Figs.~\ref{Fig_1}-\ref{Fig_2}.

	%%%%%%%%%%%%%%%%%%%%%%%%
%	\emph{Spatial spread of QGFOs and quantum metric length.---}
	\subsection*{Spatial spread of QGFOs and quantum metric length}
	Combining Eqs.~\eqref{Eq_delta_rho_k} and \eqref{Eq_chi_tot_High_temperature}, the LDOS modulations at high temperature are:
	\begin{equation}
		\begin{aligned}
			\delta\rho(\bm{r}) 
			= - \frac{U_0}{4TV} \sum_{\bm{q}} [1-\bar{D}^f(\bm{q})] e^{i\bm{q} \cdot \bm{r}}.
		\end{aligned}
		\label{Eq_LDOS_Avearged_QD}
	\end{equation}
	For a local impurity potential with $U_0 \ge 0$, since $(1-\bar{D}^f(\bm{q})) \ge 0$, Fourier transformation yields a negative definite LDOS variation, $\delta \rho(\bm{r}) \le 0$, by Bochner's theorem~\cite{bochner1932vorlesungen}.
	Thus, the amplitude of $\delta\rho (\bm{r})$ defines a probability distribution function $\delta\rho(\bm{r})/\sum_{\bm{r}} \delta\rho(\bm{r})$. Therefore, we define the spatial spread $\Omega_{\mathrm{QGFOs}}$ of the LDOS modulations in the high temperature regime by the variance:
	\begin{equation}
		\Omega_{\mathrm{QGFOs}}
		\equiv \frac{\sum_{\bm{r}} \bm{r}^2\,\delta\rho(\bm{r})}{\sum_{\bm{r}} \delta\rho(\bm{r})} - \left(\frac{ \sum_{\bm{r}} \bm{r}\,\delta\rho(\bm{r})}{\sum_{\bm{r}} \delta\rho(\bm{r})} \right)^2,
		\label{Eq_QS_def}
	\end{equation}
	with the normalization denominator $\sum_{\bm{r}} \delta\rho(\bm{r}) = \delta\rho_{\bm{q}=0} = -U_0/(4T)$. The first moment vanishes because $\sum_{\bm{r}}\bm{r} e^{i\bm{q}\cdot\bm{r}} = -iV \nabla_{\bm{q}}\delta_{\bm{q},0}$. For the second moment, we use $\sum_{\bm{r}} \bm{r}^2 e^{i\bm{q}\cdot\bm{r}} = -V \nabla_{\bm{q}}^2 \delta_{\bm{q},0}$, integrate by parts, and obtain
	\begin{equation}
		\sum_{\bm{r}} \bm{r}^2\,\delta\rho(\bm{r}) 
		= -\frac{U_0}{4T V} \sum_{\bm{k}} 2 \operatorname{Tr}\bigl[\mathcal{Q}^f(\bm{k})\bigr],
		\label{Eq_second_moment}
	\end{equation}
	where $\mathcal{Q}_{\alpha\beta}(\bm{k}) = \bra{\partial_{k_{\alpha}} u_{\bm{k}}} (1 - \ket{u_{\bm{k}}}\bra{u_{\bm{k}}}) \ket{\partial_{k_{\beta}} u_{\bm{k}}}$
	is the quantum geometric tensor. Since $\mathcal{Q}^f = \mathcal{G}^f - \frac{i}{2}\mathcal{B}^f$, its trace purely contains the quantum metric. Therefore
	\begin{equation}
		\Omega_{\mathrm{QGFOs}} = \frac{2}{Na}\sum_{\bm{k}} \,\operatorname{Tr} \bigl[\mathcal{G}^f(\bm{k})\bigr].
		\label{Eq_Omega2}
	\end{equation}
	Defining the quantum metric length as Eq.~\eqref{QML}, in the thermodynamic limit, we recover $\Omega_{\mathrm{QGFOs}} = 2a\,\ell_{\mathrm{QM}}$ [Eq.~\eqref{Eq_QGFOs_extent_QML}].
	
	In two dimensions, the quantum metric is universally bounded by the Berry curvature $\mathcal{B}^f$  through the inequality~\cite{PhysRevB.90.165139}: $\operatorname{Tr}\mathcal{G}^f(\bm{k}) \ge |\mathcal{B}^f(\bm{k})|$. Integrating over the BZ yields 
	\begin{equation}
		\frac{1}{N}\sum_{\bm{k}} \operatorname{Tr} \mathcal{G}^f(\bm{k}) 
		\ge \frac{a^2}{(2\pi)^2}\int |\mathcal{B}^f(\bm{k})|\,d^2k
		\ge \frac{a^2|C|}{2\pi},
	\end{equation}
	where $C$ is the Chern number. Hence, for a (nearly) flat band with finite Chern number, the spatial spread of QGFOs is bounded from below by the topological invariant
	\begin{equation}
		\begin{aligned}
			\Omega_\mathrm{QGFOs}
			\ge \frac{|C|}{\pi} a^2.
		\end{aligned}
		\label{Eq_integrated_QM_Chern_bound_inequality}
	\end{equation}

	\section{Data, Materials, and Software Availability}
	All study data are included in the article and/or SI Appendix.

	\section{Acknowledgment}
	We thank Xilin Feng, Fu-Chun Zhang for inspiring discussions. K. T. L. acknowledges the support of the Ministry of Science and Technology, China, the New Cornerstone Foundation, and the Hong Kong Research Grants Council through Grants No. MOST23SC01-A, No. RFS2021-6S03, No. C6053-23G, No. AoE/P-701/20, AoE/P-604/25R, No. 16309223, No. 16311424 and No. 16300325. PL acknowledges support by DOE (USA) office of Basic Sciences Grant No. DE-FG02-03ER46076.

	\bibliographystyle{apsrev4-2}
	\bibliography{QGFOsNotes}

%\end{document}   % no \documentclass or \begin{document} inside Main.tex
		
		% --- Supplementary section ---
		\clearpage
		\appendix		
		% Reset counters and redefine numbering
		\setcounter{equation}{0}
		\setcounter{section}{0}
		\setcounter{figure}{0}
		\setcounter{table}{0}
		\setcounter{page}{1}
		\renewcommand{\theequation}{S\arabic{equation}}
		\renewcommand{\thesection}{\Roman{section}}
		\renewcommand{\thefigure}{S\arabic{figure}}
		\renewcommand{\thetable}{\arabic{table}}
		\renewcommand{\tablename}{Supplementary Table}
		\renewcommand{\bibnumfmt}[1]{[S#1]}
		\renewcommand{\citenumfont}[1]{S#1}

		% Switch to one column
		\onecolumngrid
		
		\clearpage
%		\begin{center}
%			\textbf{\large Supplemental Material for "Quantum Geometric Friedel Oscillations"}\\[0.5em]
%			Xing-Lei Ma, Jinchao Zhao, Bo-Qing Wu, 
%			Patrick A. Lee \footnote{palee@mit.edu},
%			K. T. Law\footnote{phlaw@ust.hk}\\[0.5em]
%			Department of Physics, Hong Kong University of Science and Technology, Clear Water Bay, Hong Kong, China\\[0.5em]
%			
%			Department of Physics, Massachusetts Institute of Technology, Cambridge, MA 02139, USA
%			\date{\today}
%			\maketitle
%		\end{center}
		
		\begin{center}
			\textbf{\large Supplemental Material for ``Quantum Geometric Friedel Oscillations''}
			\\[0.5em]
			Xing-Lei Ma\textsuperscript{1}, Jinchao Zhao\textsuperscript{1}, Bo-Qing Wu\textsuperscript{1}, Patrick A. Lee\textsuperscript{2}\textsuperscript{*}, K. T. Law\textsuperscript{1}\textsuperscript{\dag} \\[0.5em]
			{\footnotesize\itshape
				\textsuperscript{1}Department of Physics, Hong Kong University of Science and Technology, Clear Water Bay, Hong Kong, China\\
				\textsuperscript{2}Department of Physics, Massachusetts Institute of Technology, Cambridge, MA 02139, USA \\
			}
			{\footnotesize\textnormal
				\textsuperscript{*}palee@mit.edu \quad \textsuperscript{\dag}phlaw@ust.hk
			}
		\end{center}

		\tableofcontents
		\vspace{2em}

\section{Friedel Oscillations with Quantum Geometry}
\label{Sec_Analytical_form_QGFOs}

We now derive the general analytical form of Friedel oscillations when quantum geometric effects are present. The conventional Friedel oscillations have their periods characterized by the Fermi momentum and exhibit a long-range tail determined by the thermal coherence length. We show that nontrivial quantum geometry introduces fundamentally new oscillatory channels with distinct wavevectors and exponential decay profiles.

\subsection{Real-space Susceptibility and Residue Decomposition}
The real-space susceptibility for a flat band $f$ takes the form
\begin{equation}
	\chi^f(\bm{r}, \omega) = -\frac{1}{\pi} \Im \operatorname{Tr} \bigl[ \mathcal{G}^f_{\omega} (\bm{r}) \mathcal{T}_{\omega} \mathcal{G}^f_{\omega} (-\bm{r}) \bigr],
	\label{Eq_real_space_susceptibility}
\end{equation}
where $\mathcal{G}^f_{\omega}(\bm{r})$ is the retarded Green's function projected onto an isolated flat band. In momentum space, it reads:
\begin{equation}
	\mathcal{G}_{\omega}^f(\bm{k}) 
	= \frac{P^f_{\bm{k}}}{\omega^+ - E^f_{\bm{k}}}
	= \frac{\prod_{n\ne f} \left(\mathcal{H}_{\bm{k}}-E^n_{\bm{k}}\right)}{(\omega^+ - E^f_{\bm{k}}) \prod_{n\ne f} \left(E^f_{\bm{k}}-E^n_{\bm{k}} \right) },
	\label{Eq_projected_GF}
\end{equation}
where $\omega^+ = \omega + i0^+$, and $P^f_{\bm{k}} \equiv |u^f_{\bm{k}}\rangle \langle u^f_{\bm{k}}|$ is the flat-band projection operator. 
For a tight-binding Hamiltonian with finite-range hoppings, the numerator on the right side of Eq.~\eqref{Eq_projected_GF} is holomorphic in the finite complex plane. Therefore, the meromorphic structure of the Green's function in the complex $\bm{k}$‑plane is determined by its denominator, which reveals two types of singularities from the solutions of
\begin{subequations}
	\begin{align}
		&\{\bm{k}_{\omega}\}:\; \omega^+-E_{\bm{k}}^{f} = 0, \qquad \\
		&\{\bm{k}_{nf} \}:\; E_{\bm{k}}^{n}-E_{\bm{k}}^{f} = 0 \; (n \neq f).
		\label{Eq_pole_structure_2}
	\end{align}
	\label{Eq_pole_structure}
\end{subequations}
The first set of poles originate from the flat-band dispersion, whereas the second set arises purely from quantum geometry. While both types of poles take complex values in general, the first set lies infinitesimally close to the real axis and we denote those in the upper (lower) half-plane as $\{\bm{k}_{\omega}^{u}\}$ ($\{\bm{k}_{\omega}^{l}\}$)). In contrast, the second type of pole acquires finite imaginary parts due to a local band gap $\Delta_g \neq 0$ (see Sec.\ref{Sec_Pole_band_minima}), and is irremovable when the bands $n$ and $f$ share coupled degrees of freedom.
In addition, symmetry imposes further constraints to the pole structure. Hermiticity of $\mathcal{H}_{\bm{k}}$ guarantees complex‑conjugate pairs, which we denote as $\{\bm{k}_{nf}^{u}\}$ (upper half-plane) and $\{\bm{k}_{nf}^{l}\}$ (lower half-plane). For one such complex‑conjugate pair, we adopt a denotation convention: $\bm{k}_{nf}^{l=\bar{u}} \equiv (\bm{k}_{nf}^{u})^*$.
Moreover, time‑reversal symmetry further relates $\bm{k}_{nf}^{u}$ to $\bm{k}_{nf}^{-u} \equiv -\bm{k}_{nf}^{\bar{u}}$, and we refer to $(\bm{k}_{nf}^{u},\bm{k}_{nf}^{-u})$ as a Kramers pair of poles. Inversion symmetry also pins a pair at $\pm\bm{k}_{nf}^{u}$.

The real‑space Green’s function follows from the Fourier transform:
\begin{equation}
	\mathcal{G}_{\omega}^f(\bm{r}) = \frac{1}{N} \sum_{\bm{k}} e^{i\bm{k}\cdot\bm{r}} \frac{P^f_{\bm{k}}}{\omega^{\dagger} - E^f_{\bm{k}}}.
\end{equation}
For $\bm{r}>0$, the contour is closed in the upper half‑plane, picking up poles with $\Im(\bm{k})\ge 0$. By the residue theorem,
\begin{equation}
	\mathcal{G}_{\omega}^f(\bm{r}) 
	= \sum_{\{\bm{k}^u_{\omega}\}} A_{\bm{k}^u_{\omega}} e^{i\bm{k}_{\omega} \cdot \bm{r}} 
	+ \sum_{n\neq f}\sum_{\{\bm{k}^{u}_{nf}\}} A_{\bm{k}^{u}_{nf}} e^{i\bm{k}^{u}_{nf} \cdot \bm{r}},
	\label{Eq_G_omega_r}
\end{equation}
where the residue coefficients are
%\begin{equation}
%	\begin{aligned}
	%		A_{\bm{k}_{\omega}} &= i \operatorname*{Res}_{\bm{k} = \bm{k}_{\omega}} \left(\frac{1}{\omega^+ - E^f_{\bm{k}}} \right) 
	%		\left. \frac{ \prod_{m\ne f} \bigl[\mathcal{H}_{\bm{k}}-E^m_{\bm{k}}\bigr]}{ \prod_{m\ne f} \bigl[E^f_{\bm{k}}-E^m_{\bm{k}} \bigr] } \right|_{\bm{k}=\bm{k}_{\omega}},\\[4pt]
	%		A_{\bm{k}^{u}_{nf}} &= i \operatorname*{Res}_{\bm{k} = \bm{k}^{u}_{nf}} \left(\frac{1}{E^f_{\bm{k}} - E^n_{\bm{k}}}  \right) 
	%		\left. \frac{\prod_{m\ne f} \bigl[\mathcal{H}_{\bm{k}}-E^m_{\bm{k}}\bigr]}{ (\omega^+ - E^n_{\bm{k}}) \prod_{ m\neq n,f} \bigl[E^n_{\bm{k}}-E^m_{\bm{k}} \bigr] }\right|_{\bm{k}=\bm{k}^{u}_{nf}}.
	%	\end{aligned}
%	\label{Eq_residues}
%\end{equation}
\begin{equation}
	\begin{aligned}
		A_{\bm{k}^u_{\omega}} &= \frac{ V_{cell}}{(2\pi)^d} 2\pi i \operatorname*{Res}_{\bm{k} = \bm{k}^u_{\omega}} \left(\frac{1}{\omega^+ - E^f_{\bm{k}}} \right) 
		P^f_{\bm{k}^u_{\omega}}, \\
		A_{\bm{k}^{u}_{nf}} &=  \frac{ V_{cell}}{(2\pi)^d} 2\pi i \operatorname*{Res}_{\bm{k} = \bm{k}^{u}_{nf}} \left[ P^f_{\bm{k}}  \right] 
		\frac{1}{ (\omega^+ - E^f_{ \bm{k}^{u}_{nf} })}.
	\end{aligned}
	\label{Eq_residues}
\end{equation}
Here, $V_{cell}$ is the volume of a unit cell.
For $\bm{r}<0$, the contour closes in the lower half‑plane, yielding the analogous decomposition with $\{\bm{k}^{l}_{\omega}\}$ and $\{\bm{k}^{l}_{nf}\}$.

\subsection{Three Oscillatory Channels} 
Inserting Eq.~\eqref{Eq_G_omega_r} and its $\bm{r}<0$ counterpart into Eq.~\eqref{Eq_real_space_susceptibility}, we obtain three distinct types of oscillatory modes:
\begin{equation}
	\begin{aligned}
		\chi^f(\bm{r}, \omega) = -\frac{1}{\pi} \Im \Bigg\{ & \sum_{\bm{k}^u_{\omega}, \bm{k}^l_{\omega} } \mathcal{C}_{\bm{k}^u_{\omega},\bm{k}^l_{\omega}} e^{i(\bm{k}^u_{\omega}-\bm{k}^l_{\omega}) \cdot \bm{r}} \\
		& + \sum_n \sum_{\bm{k}^u_{nf}, \bm{k}^{l}_{nf}} \mathcal{C}_{\bm{k}^u_{nf},\bm{k}^l_{nf}} e^{i(\bm{k}^u_{nf}-\bm{k}^{l}_{nf}) \cdot \bm{r}} \\
		& + \sum_n \sum_{\bm{k}^{u/l}_{\omega}, \bm{k}^{u/l}_{nf} } \left[ \mathcal{C}_{\bm{k}^u_{nf},\bm{k}^l_{\omega}} e^{-i(\bm{k}^l_{\omega}-\bm{k}^u_{nf}) \cdot \bm{r}} + \mathcal{C}_{\bm{k}^u_{\omega},\bm{k}^l_{nf}} e^{i(\bm{k}^u_{\omega}-\bm{k}^{l}_{nf}) \cdot \bm{r}} \right] \Bigg\},
	\end{aligned}
	\label{Eq_chi_r_omega_analytic}
\end{equation}
which we denote as $\chi^f_1$, $\chi^f_2$, and $\chi^f_3$, respectively. The coefficients are determined by the residues in Eq.~\eqref{Eq_residues}:
\begin{subequations}
	\begin{align}
		&\mathcal{C}_{\bm{k}^u_{\omega},\bm{k}^l_{\omega}} = \operatorname{Tr} \left[A_{\bm{k}^u_{\omega}} \mathcal{T}_{\omega} A_{\bm{k}^l_{\omega}} \right] = -\frac{ V_{cell}^2}{(2\pi)^{2(d-1)}} \operatorname*{Res}_{\bm{k} = \bm{k}^u_{\omega}} \left(\frac{1}{\omega^+ - E^f_{\bm{k}}} \right) \operatorname*{Res}_{\bm{k} = \bm{k}^l_{\omega}} \left(\frac{1}{\omega^+ - E^f_{\bm{k}}} \right)
		\Lambda^f_{\bm{k}^u_{\omega}, \bm{k}^l_{\omega}} \tilde{\Lambda}^f_{\bm{k}^l_{\omega},\bm{k}^u_{\omega}},
		%		\left|\braket{u^f_{\bm{k}^u_{\omega}}|\mathcal{T}_{\omega}|u^f_{\bm{k}^l_{\omega}} } \right|^2,
		\label{Eq_chi_coefficients1} \\
		&\mathcal{C}_{\bm{k}^u_{nf},\bm{k}^l_{nf}} = \operatorname{Tr} \left[ A_{\bm{k}^u_{nf}} \mathcal{T}_{\omega} A_{\bm{k}^{l}_{nf}}  \right] = - \frac{ V_{cell}^2}{(2\pi)^{2(d-1)}} \operatorname{Tr} \left[\operatorname*{Res}_{\bm{k} = \bm{k}^{u}_{nf}} \left( P^f_{\bm{k}}  \right) \mathcal{T}_{\omega} \operatorname*{Res}_{\bm{k} = \bm{k}^{l}_{nf}} \left( P^f_{\bm{k}}  \right) \right]
		\frac{1}{ (\omega^+ - E^f_{ \bm{k}^{u}_{nf} }) (\omega^+ - E^f_{ \bm{k}^{l}_{nf} }) },  \label{Eq_chi_coefficients2} \\
		&\mathcal{C}_{\bm{k}^u_{nf},\bm{k}^l_{\omega}} = \operatorname{Tr} \left[ A_{\bm{k}^u_{nf}} \mathcal{T}_{\omega} A_{\bm{k}^l_{\omega}}   \right] = - \frac{ V_{cell}^2}{(2\pi)^{2(d-1)}} \operatorname{Tr} \left[ \operatorname*{Res}_{\bm{k} = \bm{k}^{u}_{nf}} \left( P^f_{\bm{k}}  \right) \mathcal{T}_{\omega} P^f_{\bm{k}^l_{\omega}}  \right] 
		\operatorname*{Res}_{\bm{k} = \bm{k}^l_{\omega}} \left(\frac{1}{\omega^+ - E^f_{\bm{k}}} \right)
		\frac{1}{ \omega^+ - E^f_{ \bm{k}^{u}_{nf} } },  \label{Eq_chi_coefficients3} \\
		&\mathcal{C}_{\bm{k}^u_{\omega},\bm{k}^l_{nf}} = \operatorname{Tr} \left[ A_{\bm{k}^u_{\omega}} \mathcal{T}_{\omega} A_{\bm{k}^{l}_{nf}}  \right] = - \frac{ V_{cell}^2}{(2\pi)^{2(d-1)}} \operatorname{Tr} \left[  P^f_{\bm{k}^u_{\omega}} \mathcal{T}_{\omega} \operatorname*{Res}_{\bm{k} = \bm{k}^{l}_{nf}} \left( P^f_{\bm{k}}  \right) \right] 
		\operatorname*{Res}_{\bm{k} = \bm{k}^u_{\omega}} \left(\frac{1}{\omega^+ - E^f_{\bm{k}}} \right)
		\frac{1}{ \omega^+ - E^f_{ \bm{k}^{l}_{nf} } } , \label{Eq_chi_coefficients4} 
	\end{align}
	\label{Eq_chi_coefficients}
\end{subequations}
where $\Lambda^f_{\bm{k}, \bm{k}'} = \braket{u^f_{\bm{k}}|u^f_{\bm{k}'}}$ and $\tilde{\Lambda}^f_{\bm{k}, \bm{k}'} =  \braket{u^f_{\bm{k}}|\mathcal{T}_{\omega}|u^f_{\bm{k}'}}$. 

For a parabolic dispersion in a 1D flat band $E^f_{{k}} = \frac{{k}^2}{2m} - \mu$ (lattice constant is set as $a=1$), we can simplify the component from band dispersion. Note that at energy $\omega$, the residue value at the pole ${k}^{u/l}_{\omega} = \pm \sqrt{2m(\omega^+ + \mu)}$ produces $\operatorname*{Res}_{\bm{k} = {k}^{u/l}_{\omega}} \left(\frac{1}{\omega^+ - E^f_{{k}}} \right) = \frac{m}{{k}^{u/l}_{\omega}}$. Then the first term in Eq.~\eqref{Eq_chi_r_omega_analytic} reduces to
\begin{equation}
	\begin{aligned}
		\chi_1^f(r, \omega) = -\frac{1}{\pi} \left[\frac{m^2}{|k^u_{\omega}|^2} \sin(2\sqrt{2|k_{\omega}| r}) -\pi \delta(\omega+\mu) \frac{m}{2}   \right]
		\Lambda^f_{\bm{k}^u_{\omega}, \bm{k}^l_{\omega}} \tilde{\Lambda}^f_{\bm{k}^l_{\omega},\bm{k}^u_{\omega}} .
	\end{aligned}
	\label{chi_f_omega_1}
\end{equation}
For a weak impurity, we approximate $\mathcal{T}_{\omega} \approx \mathcal{I}$, then integrate over $\omega$ weighted by the Fermi-Dirac distribution $f_{T=0}(\omega) = \Theta(E_F - \omega)$. In the limit $\Lambda_{k,k'} = 1$, we obtain the well-established Friedel oscillations:
\begin{equation}
	\begin{aligned}
		\chi_1^f(r) = \int  f(\omega) \chi_1^f(r, \omega) \diff\omega = - \frac{\cos(2k_Fr)}{2k_Fr} + \mathcal{O}(\frac{1}{r^2}) .
	\end{aligned}
\end{equation}
%More generically, Eq.~\eqref{chi_f_omega_1} suggests that the zero-temperature Friedel oscillations are also shaped by the quantum geometry below the Fermi surface, that is, the quantum distance between any two degenerate states (in energy-conserving scatterings) contributes to the susceptibility.

To analyze the remaining parts of the flat-band susceptibility in Eq.~\eqref{Eq_chi_r_omega_analytic}, we define the Hermitian and anti-Hermitian part of the $\mathcal{T}$-matrix as:
\begin{equation}
	\begin{aligned}
		\mathcal{T}^h_{\omega} \equiv \frac{\mathcal{T}_{\omega} + \mathcal{T}^{\dagger}_{\omega}}{2}, \quad
		\mathcal{T}^a_{\omega} \equiv \frac{\mathcal{T}_{\omega} - \mathcal{T}^{\dagger}_{\omega}}{2}.
	\end{aligned}
\end{equation}
Similarly, we define $\mathcal{C}^h \equiv \operatorname{Tr} \left[ A_{\bm{k}^u} \mathcal{T}^h_{\omega} A_{\bm{k}^l}   \right]$ and $\mathcal{C}^a \equiv \operatorname{Tr} \left[ A_{\bm{k}^u} \mathcal{T}^a_{\omega} A_{\bm{k}^l}  \right]$ for each coefficients in Eq.~\eqref{Eq_chi_coefficients}.

Now we look at the quantum geometric component $\chi_2^f$ in Eq.~\eqref{Eq_chi_r_omega_analytic}.
As mentioned earlier, hermiticity leads to poles at both $\bm{k}^{u}_{nf}$ and $\bm{k}^{l=\bar{u}}_{nf}$ in the flat-band projection operator. Therefore, for each pair of such complex conjugate poles, flat-band dispersion and the residues of the flat-band projection operator in Eq.~\eqref{Eq_chi_coefficients2} satisfy 
\begin{equation}
	\begin{aligned}
		E^f_{\bm{k}^{u}_{nf}} = \left( E^f_{\bm{k}^{\bar{u} }_{nf}} \right)^*, \quad \operatorname*{Res}_{\bm{k} = \bm{k}^{u}_{nf}} \left( P^f_{\bm{k}}  \right) =  \left[ \operatorname*{Res}_{\bm{k} = \bm{k}^{\bar{u}}_{nf}} \left( P^f_{\bm{k}}  \right)\right]^{\dagger}.
	\end{aligned}
\end{equation} 
For such a pair, the coefficients satisfy $\mathcal{C}^h_{\bm{k}^u_{nf},\bm{k}^{\bar{u}}_{nf}} = \left[ \mathcal{C}^h_{\bm{k}^u_{nf},\bm{k}^{\bar{u}}_{nf}}\right]^*$ and $\mathcal{C}^a_{\bm{k}^u_{nf},\bm{k}^{\bar{u}}_{nf}} = -\left[ \mathcal{C}^a_{\bm{k}^u_{nf},\bm{k}^{\bar{u}}_{nf}}\right]^*$, thus being real and imaginary, respectively. Hence, their contribution reads:
\begin{equation}
	\begin{aligned}
		\frac{1}{\pi}\Im \left[   \mathcal{C}_{\bm{k}^u_{nf},\bm{k}^{\bar{u}}_{nf}} e^{i(\bm{k}^u_{nf}-\bm{k}^{\bar{u}}_{nf}) \cdot \bm{r}} \right]
		%		&= \frac{1}{\pi}  \operatorname{Tr} \left[\operatorname*{Res}_{\bm{k} = \bm{k}^{u}_{nf}} \left( P^f_{\bm{k}}  \right) \mathcal{T}^a_{\omega} \operatorname*{Res}_{\bm{k} = \bm{k}^{\bar{u}}_{nf}} \left( P^f_{\bm{k}}  \right) \right]		\frac{e^{- 2\Im(\bm{k}^u_{nf}) \cdot \bm{r}} }{ (\omega - E^f_{ \bm{k}^{u}_{nf} }) (\omega - E^f_{ \bm{k}^{\bar{u}}_{nf} }) } \\
		&= \frac{1}{\pi} |\mathcal{C}^a_{\bm{k}^u_{nf},\bm{k}^{\bar{u}}_{nf}}| e^{- 2\Im(\bm{k}^u_{nf}) \cdot \bm{r}} .
	\end{aligned}
	\label{Eq_cc_pair_contribution}
\end{equation}

Above, Eq.~\eqref{Eq_cc_pair_contribution} suggests that when there is only one pair of poles in $\{\bm{k}_{nf}\}$, such as the Lieb lattice~\cite{lieb1989two}, the flat-band susceptibility will exhibit a purely exponential decay, which has also been revealed in the RKKY interactions of exact flat bands~\cite{PhysRevB.108.155429,PhysRevB.104.155151,luo2024influence}. To verify this, we compute the charge‑density response of a Lieb lattice with a single pair of poles in the flat‑band projection operator (hence a single quantum metric hot spot) and find no oscillatory component: the response is purely exponential, as shown in Fig.~\ref{Fig_Lieb_single_QMhotspot}(b,c).
In contrast, with multiple pairs of poles the susceptibility will acquire oscillatory features in addition to the exponential decay in Eq.~\eqref{Eq_cc_pair_contribution}. For example, when there are two distinct poles in the upper complex plane $\bm{k}^{u, u'\neq u}_{nf}$,%and their complex conjugate partners $\bm{k}^{\bar{u}, \bar{u'}}_{nf}$, 
their cross contribution to $\chi_2^f$ is:
\begin{equation}
	\begin{aligned}
		&\frac{1}{\pi}\Im \left[ 
		%\sum_n \sum_{\bm{k}^u_{nf}, \bm{k}^{u'}_{nf}}
		\left( \mathcal{C}_{\bm{k}^u_{nf},\bm{k}^{\bar{u'}}_{nf}} e^{i(\bm{k}^u_{nf}-\bm{k}^{\bar{u'}}_{nf}) \cdot \bm{r}}  + \mathcal{C}_{\bm{k}^{u'}_{nf},\bm{k}^{\bar{u}}_{nf}} e^{i(\bm{k}^{{u}'}_{nf}-\bm{k}^{\bar{u}}_{nf}) \cdot \bm{r}} \right) \right] \\
		=& \frac{1}{\pi}\Im \left[ \left( \mathcal{C}^h_{\bm{k}^u_{nf},\bm{k}^{\bar{u'}}_{nf}} e^{i(\bm{k}^u_{nf}-\bm{k}^{\bar{u'}}_{nf}) \cdot \bm{r}}  + c.c. \right)  + \left( \mathcal{C}^a_{\bm{k}^u_{nf},\bm{k}^{\bar{u'}}_{nf}} e^{i(\bm{k}^u_{nf}-\bm{k}^{\bar{u'}}_{nf}) \cdot \bm{r}} - c.c. \right) \right] \\
		=& \frac{2}{\pi} \left\{\Re\left( \mathcal{C}^a_{\bm{k}^u_{nf},\bm{k}^{\bar{u'}}_{nf}} \right) \sin{ \left[ \Re(\bm{k}^u_{nf}-\bm{k}^{\bar{u'}}_{nf}) \cdot \bm{r} \right]}
		+ \Im\left( \mathcal{C}^a_{\bm{k}^u_{nf},\bm{k}^{\bar{u'}}_{nf}} \right)  \cos{ \left[ \Re(\bm{k}^u_{nf}-\bm{k}^{\bar{u'}}_{nf}) \cdot \bm{r} \right]}  \right\}
		e^{- \Im(\bm{k}^u_{nf}-\bm{k}^{\bar{u'}}_{nf}) \cdot \bm{r}} \\
		=& \frac{2}{\pi } |\mathcal{C}^a_{\bm{k}^u_{nf},\bm{k}^{\bar{u'}}_{nf}}|\sin{ \left[ \Re(\bm{k}^{{u}}_{nf}-\bm{k}^{\bar{u'}}_{nf}) \cdot \bm{r} + \phi \right]}
		e^{- \Im(\bm{k}^u_{nf}-\bm{k}^{\bar{u'}}_{nf}) \cdot \bm{r}},
	\end{aligned}
	\label{Eq_two_pairs_contribution}
\end{equation}
where in the second step, we have used the relation $\mathcal{C}^a_{\bm{k}^u_{nf},\bm{k}^{\bar{u'}}_{nf}} = [\mathcal{C}_{\bm{k}^{u'}_{nf},\bm{k}^{\bar{u}}_{nf}}] ^*$, and in the last step we have defined a phase shift which satisfies $\tan{\phi} \equiv \Im\left( \mathcal{C}^a_{\bm{k}^u_{nf},\bm{k}^{\bar{u'}}_{nf}} \right) / \Re\left( \mathcal{C}^a_{\bm{k}^u_{nf},\bm{k}^{\bar{u'}}_{nf}} \right)$.
%%%%%%%%% f-sum rule of the phase shift?

\begin{figure}[tbp]
	\centering
	\includegraphics[width=0.8\linewidth]{ 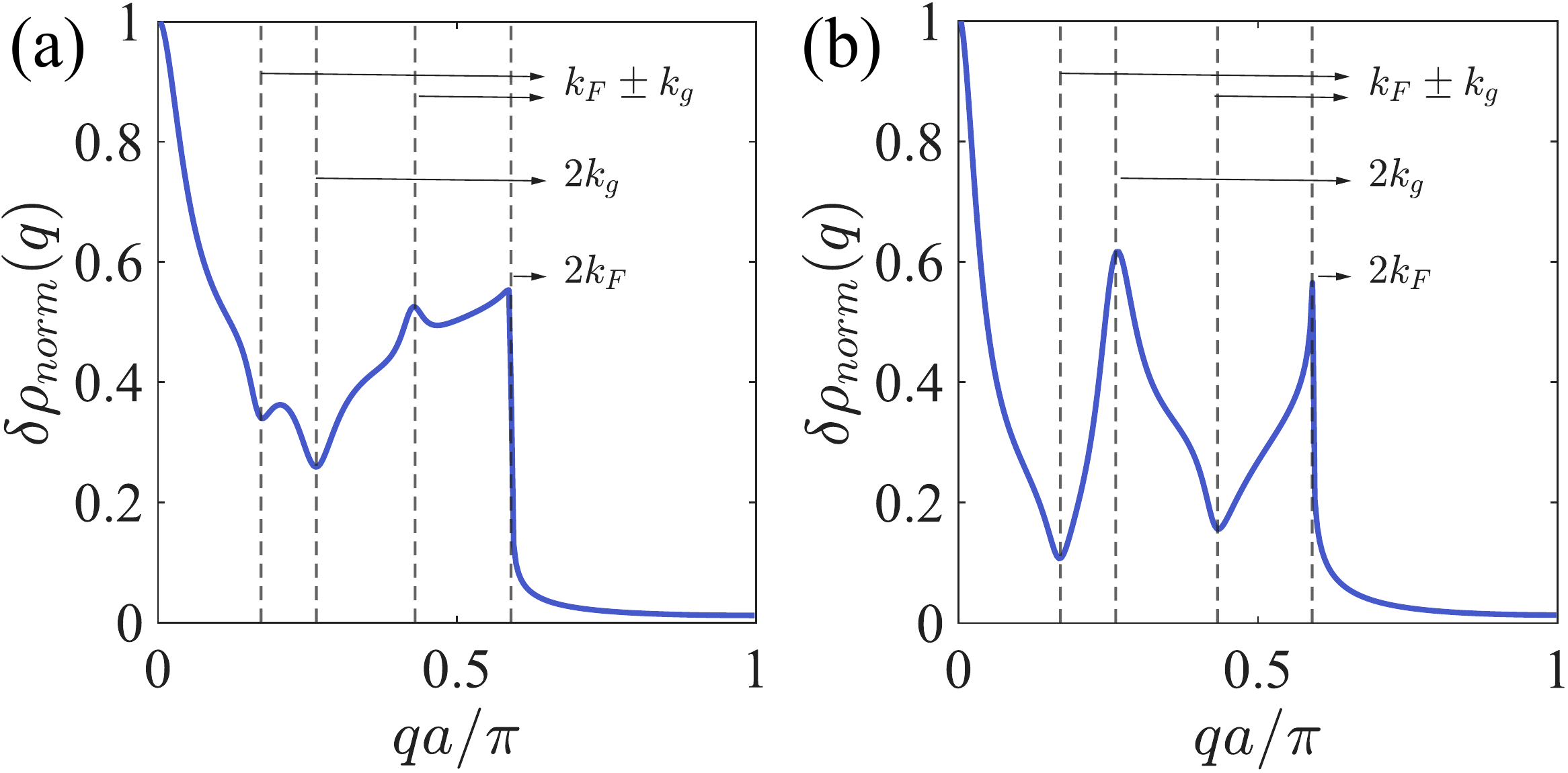}
	\caption{Three types of oscillatory components in the zero-temperature Friedel oscillations for (a) the 1D three-band model, and (b) a 1D two-band model (detailed in Section~\ref{Sec_Two_Band_Model}). Common parameters: $J=1$, $t=10^{-4}J$, $\mu = -1.2t$, and $U_0 = 2.2 \times 10^{-3}J$. Special parameters for (a) $\delta=0.03$, $\delta'=5\delta$; (b)  $\beta=0.03$, $\beta'= 5\beta$. Vertical dashed lines denote the three types of components from Fermi momentum $2k_F$, quantum geometric wavevector $2k_g$, and their cross components $k_F \pm k_g$.}
	\label{Fig_Three_types_of_components}
\end{figure}
Taking both Eq.~\eqref{Eq_cc_pair_contribution} and Eq.~\eqref{Eq_two_pairs_contribution} into account, the total contributions of two pairs of poles read
\begin{equation}
	\begin{aligned}
		\chi^f_{nf, \text{pair}} (\bm{r}, \omega) =& -\frac{1}{\pi} %\sum_n \sum_{\bm{k}^u_{nf}, \bm{k}^{u'}_{nf}} 
		\left\{ |\mathcal{C}^a_{\bm{k}^u_{nf},\bm{k}^{\bar{u}}_{nf}} | + |\mathcal{C}^a_{\bm{k}^{u'}_{nf},\bm{k}^{\bar{u'}}_{nf}}| + 2  |\mathcal{C}^a_{\bm{k}^u_{nf},\bm{k}^{\bar{u'}}_{nf}}|  \sin{ \left[ \bm{q}_G \cdot \bm{r} + \phi \right]} \right\}
		e^{- \bm{\xi}_G^{-1} \cdot \bm{r}}.
	\end{aligned}
	\label{Eq_Two_pairs_total_contribution}
\end{equation}
Here, the oscillation wavevector $\bm{q}_G$ and decay length\footnote{The decay length is written as a vector as the system can be anisotropic in general; In 1D, it is a scalar.} $\bm{\xi}_G$ are determined by
\begin{subequations}
	\begin{align}
		\bm{q}_G &= \Re\bigl(\bm{k}^{u}_{nf} - \bm{k}^{u'}_{nf}\bigr) \label{Eq_q_G}, \\
		\bm{\xi}^{-1}_G   &=  \Im\bigl(\bm{k}^u_{nf} - \bm{k}^{\bar{u'}}_{nf}\bigr) 
		\label{Eq_xi_G}.
	\end{align}
	\label{Eq_q_G_xi_G}
\end{subequations}
The discussion of Eq.~\eqref{Eq_q_G_xi_G} and how the decay length and QGFO wavevector are related to quantum metric length and quantum metric hot spots can be found in Sec.\ref{Sec_Robustness} and Sec.\ref{Sec_q_G__QM_hotspots_separation}, respectively.
More generically, there can be multiple pairs of poles in the flat-band projection operator, and every two pairs contribute as Eq.~\eqref{Eq_Two_pairs_total_contribution}. Therefore, the total contributions in $\chi^f_2$ are
\begin{equation}
	\begin{aligned}
		\chi^f_{2}(\bm{r}) =& -\frac{1}{\pi} \sum_n \sum_{\{\text{pair}\}}  \int f(\omega)  \chi^f_{nf, \text{pair}} (\bm{r}, \omega) \diff\omega.
	\end{aligned}
	\label{Eq_total_QG_contribution}
\end{equation}

\begin{figure}[tbp]
	\centering
	\includegraphics[width=0.8\linewidth]{ 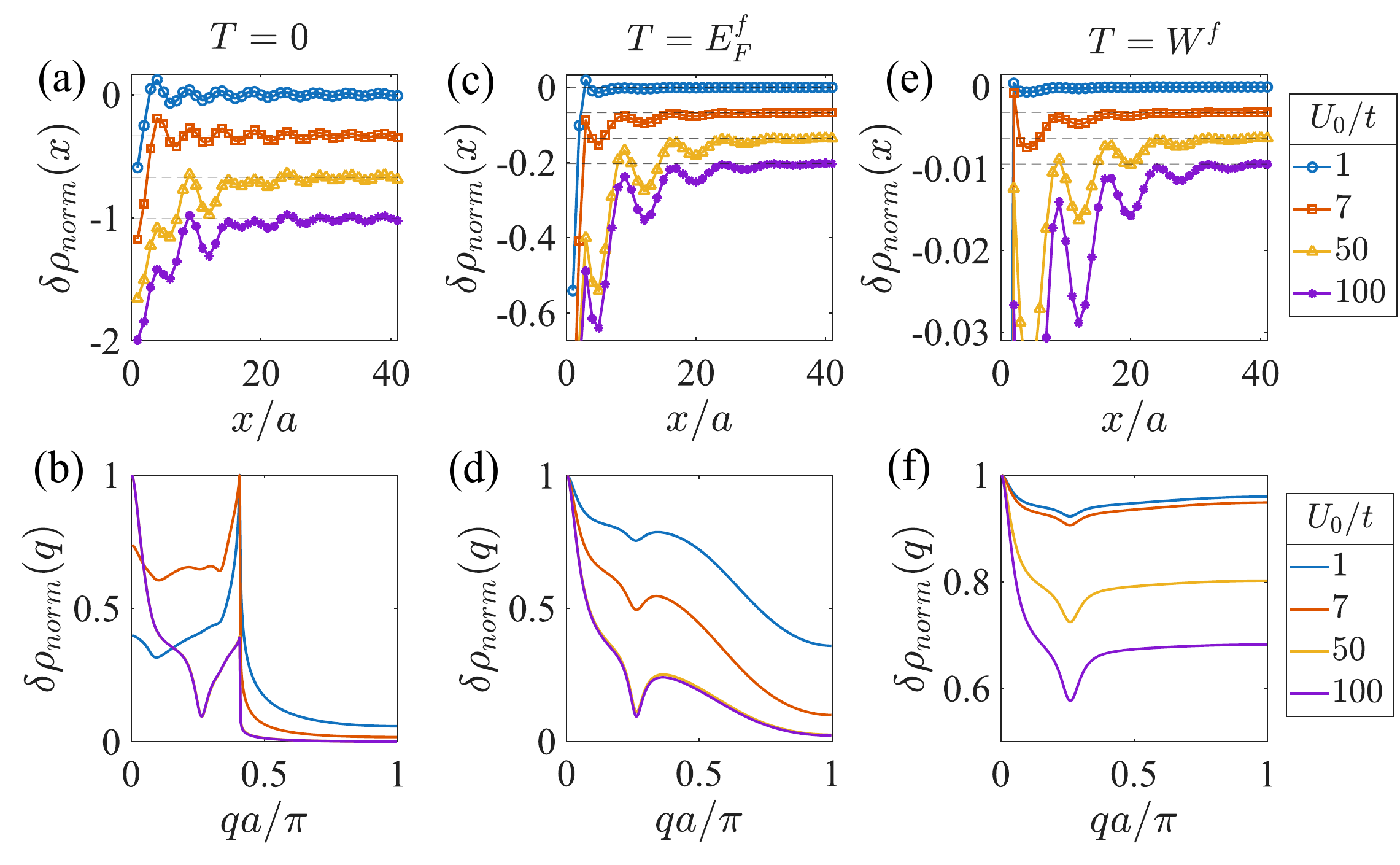}
	\caption{QGFOs dependence on impurity strength $U_0$ in the 1D three-band model, at temperature (a-b) $T=E_F^f$ and (c-d) $T = W^f$ (We take $k_B = 1$ throughout the Supplemental Materials). (a,c) Real‑space density modulations. (b,d) Fourier spectra. The parameters: $J=1$, $t=10^{-4}J$, $\delta=0.04$, $\delta'=5\delta$, $U_0 = 5\times 10^{-3}J$. Horizontal dashed lines in (a) and (c) mark the asymptotic bulk background. Vertical dashed lines in (b) and (d) indicate the quantum geometric wavevector $q_G$.}
	\label{Fig_U0_dependence}
\end{figure}

Furthermore, when $\mathcal{T}$-matrix depends weakly on energy, the integrand $\chi^f_{nf, \text{pair}} (\bm{r}, \omega) $ in Eq.~\eqref{Eq_total_QG_contribution} can be factored out, and we obtain a general analytical form for QGFOs:
\begin{equation}
	\chi^f_2 (\bm{r}) \;\sim\; -\bigl[A + B \sin\!\bigl( \bm{q}_G \cdot \bm{r} + \phi \bigr)\bigr] e^{ - \bm{\xi}_G^{-1} \cdot \bm{r}},
	\label{Eq_general_QGFOs_form}
\end{equation}
where $A$ and $B$ are shorthand denotations of the $\mathcal{C}$ coefficients in Eq.~\eqref{Eq_Two_pairs_total_contribution}.
In time-reversal symmetric systems, the poles appear as Kramers pairs $(\bm{k}_{nf}^{u},\bm{k}_{nf}^{u'=-u})$, and then the oscillation wavevector and decay length in Eq.~\eqref{Eq_q_G_xi_G} yield $\bm{q}_G = 2 \Re(\bm{k}_{nf}^{u})$ and $\xi_G = 1/(2\Im (\bm{k}_{nf}^{u}))$. Plugging this into Eq.~\eqref{Eq_general_QGFOs_form}, it well captures the QGFOs observed in Fig.~1 in the main text due to the presence of a nearly touched Kramers pair in the 1D three-band model.
Notably, the right hand side of Eq.~\eqref{Eq_total_QG_contribution} has $\omega$ dependence implicitly in the coefficients $\mathcal{C}^a$ through $\mathcal{T}_{\omega}$. 
For a strong impurity, the $\mathcal{T}$-matrix acquires a significant non-Hermitian part, thus a larger $\mathcal{C}^a$, which amplifies the quantum geometric components. Thus the associated dip in the density modulation spectrum becomes more prominent for a stronger impurity, as revealed in Fig.~\ref{Fig_U0_dependence}(a,b). Therefore, we come to the conclusion that the QGFOs persist beyond the Born approximation.
%For weak impurity with reasonable Born approximations, the $\mathcal{T}$-matrix becomes hermitian, which means $\mathcal{T}^a_{\omega} = 0$ thus $\mathcal{C}^a = 0$. As a result, the second type of oscillations in Eq.~\eqref{Eq_chi_r_omega_analytic} vanishes.
%Therefore, beyond weak impurity limit, $\chi^f_2$ emerges in general. 

As for the third, cross term in Eq.~\eqref{Eq_chi_r_omega_analytic} $\chi_3^f$, this channel originate from the scatterings between Fermi momentum $\bm{k}_F$ and the quantum geometric components $\Re{(\bm{k}_{nf})}$. We omit the detailed analysis and only present some numerical results in Fig.~\ref{Fig_Three_types_of_components}.

In summary, the three oscillatory channels identified in Eq.~\eqref{Eq_chi_r_omega_analytic} are:

\begin{enumerate}
	\item \textbf{Fermi-surface scattering}: $2\bm{k}_F$ from ($\bm{k}_{F}$,$-\bm{k}_{F}$) pairs [$\chi^f_1$],
	\item \textbf{Purely quantum-geometric}: $\Re(\bm{k}^u_{nf} - \bm{k}^{l}_{nf})$ with exponential envelope $\Im(\bm{k}^u_{nf} - \bm{k}^{l}_{nf})$ [$\chi^f_2$],
	\item \textbf{Mixed terms}: $\bm{k}_F - \Re(\bm{k}^{u}_{nf})$ [$\chi^f_3$].
\end{enumerate}

At low temperature $T \ll W^f$, in principle, all three channels can coexist.
In Fig.~\ref{Fig_Three_types_of_components}, we numerically identify all three types of oscillatory components at zero temperature: $2k_F$, $2k_g$, and $k_F \pm k_g$. They are present both in the 1D three-band model and a two-band model detailed in Sec.\ref{Sec_Two_Band_Model}.
At $T\gtrsim W^f$, the Fermi‑surface‑dependent components including $\chi^f_1$ and $\chi^f_3$ in Eq.~\eqref{Eq_chi_r_omega_analytic} are washed out, while the quantum-geometric component $\chi^f_2$ is sustained, as demonstrated in Fig.~\ref{Fig_U0_dependence}(e) and (f).

%\subsection{\label{Sec_Robustness}Universal Properties and Robustness}
\subsection{Robustness of QGFOs}
%\subsection{\label{Sec_Robustness}Robustness Against Impurity Variations}
\subsubsection{\label{Sec_Robustness}Impurity Variations}
\begin{figure}[tbp]
	\centering
	\includegraphics[width=0.8\linewidth]{ 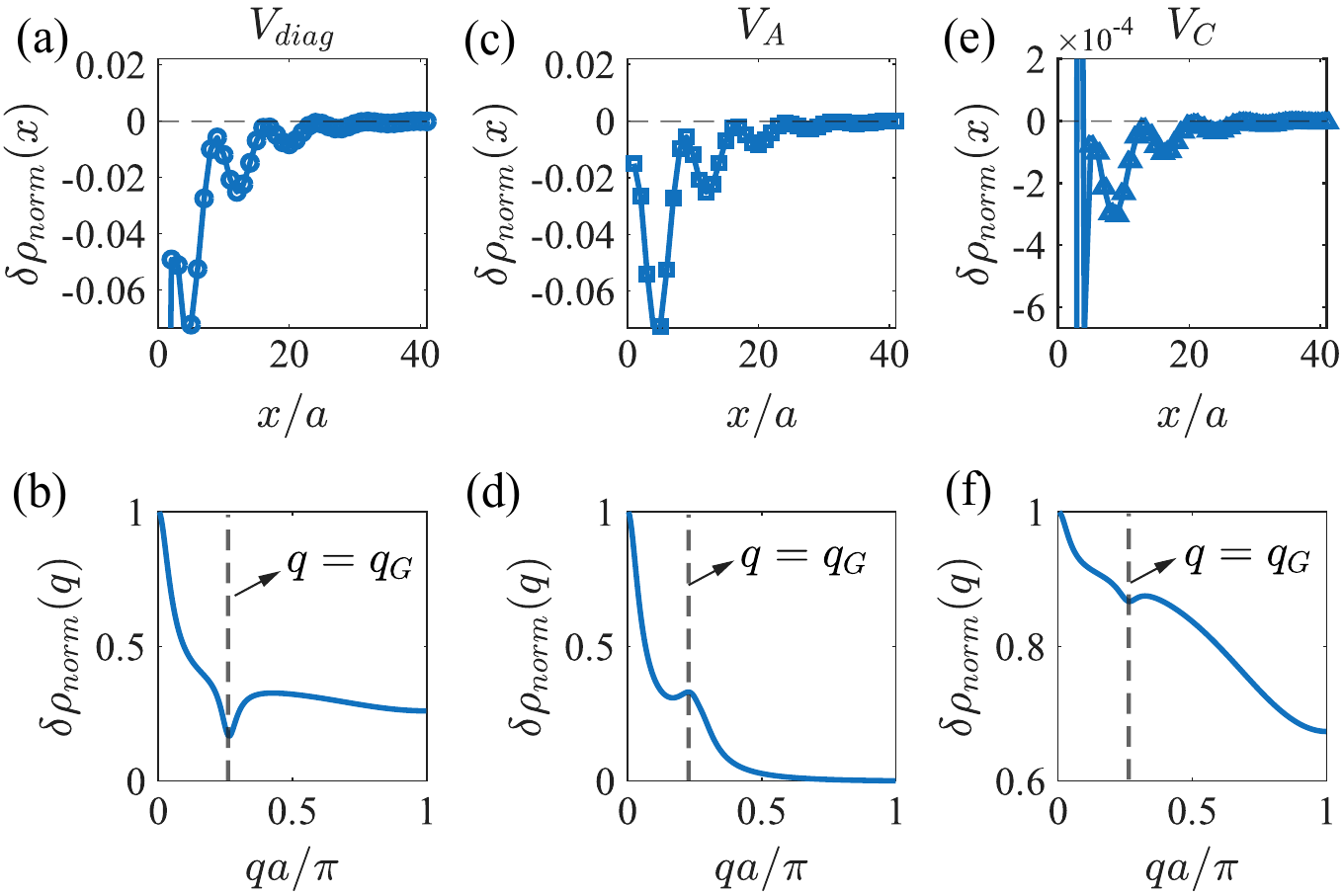}
	\caption{QGFOs dependence on different impurity forms in the 1D three-band model. (a,b) An impurity with orbital-diagonal form $V_{\text{diag}}$. (c,d) An impurity introduced at A-orbital $V_{A}$. (e,f) An impurity introduced at C-orbital $V_{C}$. The parameters: $J=1$, $t=10^{-4}J$, $\delta=0.03$, $\delta'=5\delta$, $U_0 = 1\times 10^{-2}J$, and $k_B T = t$. Vertical dashed lines in (b,d,f) mark the quantum geometric wavevector $q_G$.}
	\label{Fig_imp_form_dependence}
\end{figure}

The QGFOs exhibit universal features that are insensitive to variations in the details of the impurity potential. %The characteristic wavevector $\bm{q}_G$ and decay length $\bm{\xi}_G$ are determined solely by the pole structure of the flat-band projection operator, independent of impurity details. The imaginary parts $\Im(\bm{k}_{nf})$ are fundamentally bounded by the integrated quantum metric length $\ell_{\mathrm{QM}}$ as established in Ref.~\cite{zhao2026quantummetricboundstate}, ensuring that the spatial decay length $\xi_G \sim 1/\Im(\bm{k})$ cannot vanish.
Because Eq.~\eqref{Eq_general_QGFOs_form} follows from the flat-band pole structure, which is a bulk property, the characteristic wavevector $\bm{q}_G$ and the spatial decay $\bm{\xi}_G$ are fixed by the band projector and do not depend on impurity strength or orbital form; only the overall amplitude is affected by the $\mathcal{T}$‑matrix. This robustness is demonstrated in Fig.~\ref{Fig_U0_dependence}, where the Fourier dip at $q_G$ remains unchanged as $U_0$ and $T$ vary, and in Fig.~\ref{Fig_imp_form_dependence}, where different orbital dependence [including and orbital-diagonal impurity $V_{\text{diag}} = \text{diag}(1,1,1)$, A-orbital impurity $V_A = \text{diag}(1,0,0) $ and C-orbital impurity $V_C = \text{diag}(0,0,1) $] alters the short‑range behavior but leaves the oscillation period intact.
This confirms the universal character of the QGFOs.

Crucially, this robustness is deeply rooted in the flat-band quantum geometry. First, the real parts of the complex pole, $\Re(\bm{k})$, are set by the nearly touching momenta $\bm{k}_g$, which correspond to the quantum metric hot spots separation (see Sec.\ref{Sec_q_G__QM_hotspots_separation}). This alignment defines the characteristic wavevector of the resulting real-space oscillations. Second, we invoke the framework established in~\cite{zhao2026quantummetricboundstate}, where it is proven that the imaginary part of the pole, $\Im(\bm{k})$, is fundamentally bounded by the bulk quantum geometry. Specifically, the integrated quantum metric length $\ell_{\mathrm{QM}}$ restricts how far these complex singularities can migrate from the real momentum axis, effectively imposing a strict upper bound on $\Im(\bm{k})$. Because the spatial decay length of the oscillations scales as $\xi_G \sim 1/\Im(\bm{k})$, this geometric constraint explicitly bridges the Hilbert-space properties to the real-space dynamics, guaranteeing that the spatial extent of the QGFOs cannot vanish. 
In the three-band model, we find that in the limit $0 < \delta \ll \lambda \ll 1$, the decay length $\xi_G \simeq \frac{1}{2\sqrt{\lambda}}a$, which is proportional to the quantum metric length $\ell_{\mathrm{QM}} \simeq \frac{\pi}{2\sqrt{\lambda}}a$. Therefore, we find the asymptotic relation between the decay length and quantum metric length: $\xi_G \simeq \frac{1}{\pi} \ell_{\mathrm{QM}}$ in the large $\ell_{\mathrm{QM}}$ limit.

\iffalse
Firstly, the real parts of the poles are set by the nearly touching momenta $\bm{k}_g$ with peaking quantum metric. Near the band bottom of the adjacent dispersive band $n$, let the effective mass be $m_n^*(\bm{k}_g)$ and local gap $\Delta_g(\bm{k}_g)$, then Eq.~\eqref{Eq_pole_structure} reduces to $(\bm{k}^u_{nf}-\bm{k}_g)^2/2m_n^* + \Delta_g(\bm{k}_g)=0$, giving
\begin{equation}
	\Re(\bm{k}^u_{nf}) \simeq \bm{k}_g,\qquad \Im(\bm{k}^u_{nf}) \simeq \sqrt{2m_n^*\Delta_g(\bm{k}_g)} \quad (m_n^*\Delta_g>0).
	\label{Eq_pole_parabolic}
\end{equation}
\fi

%In the low-energy regime with parabolic band approximations, these modes correspond to: $2\bm{k}_F$, $\bm{k}_F - \bm{k}_{nf}$, and $2\bm{k}_{nf}$. These three distinct contributions are all present and can be observed at zero temperature (see SM~\cite{supp}), though usually the Fermi momentum contributions would dominate over the others.

%%Note: Even Eq.~\eqref{Eq_pole_structure} yields solutions with non-zero real part, it may not always give rise to QGFOs (An example can be found in the SM). In this case, one needs to look at the averaged quantum distance, which would always apply.

%\subsection{\label{Sec_filling} Robustness of QGFOs against Filling fluctuations}
%\subsection{\label{Sec_filling} Robustness Against Filling Fraction Variations}
\subsubsection{\label{Sec_filling}Chemical potential variations}

\begin{figure}[htbp]
	\centering
	\includegraphics[width=0.6\linewidth]{ 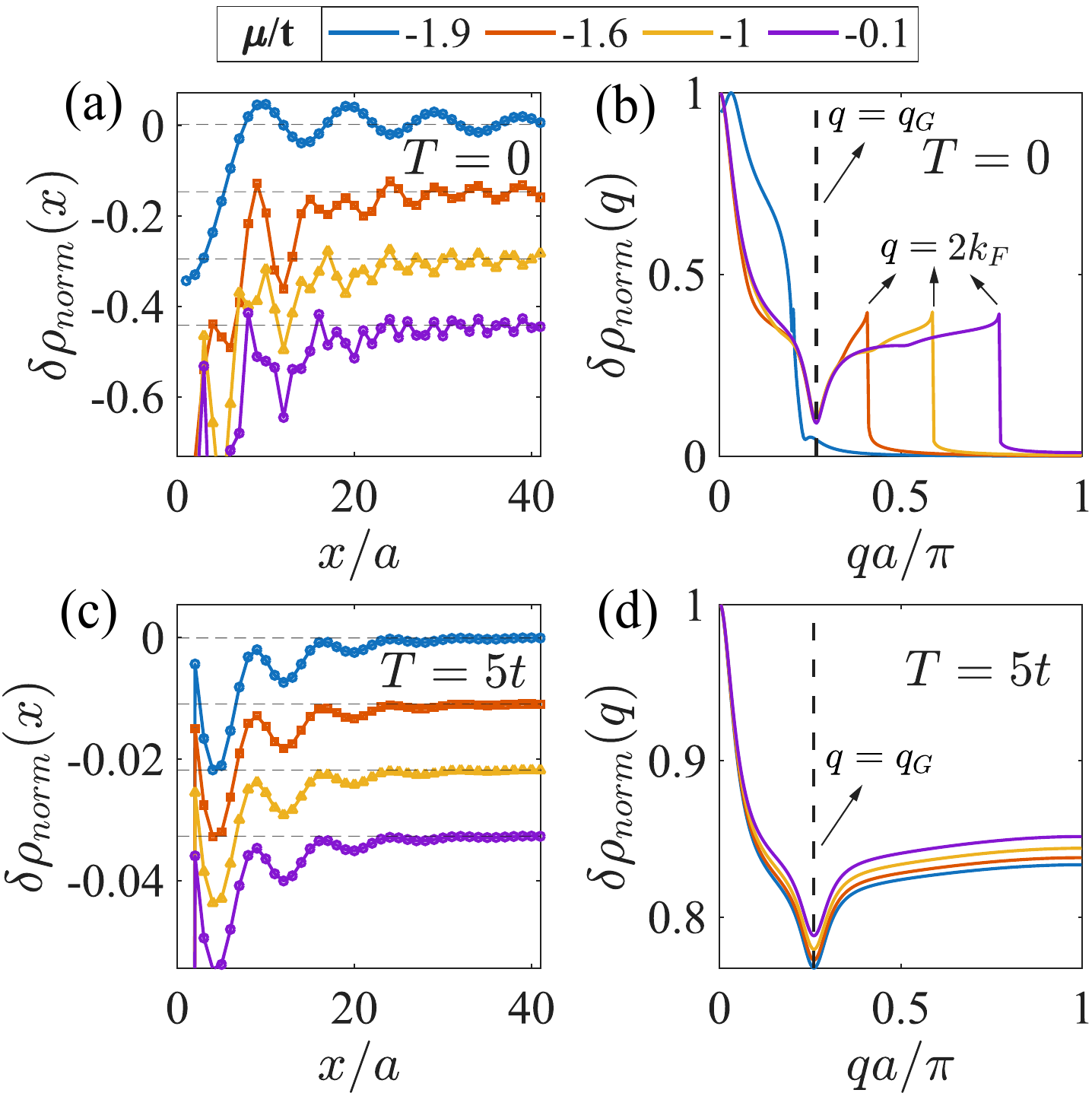}
	\caption{Filling dependence of the Friedel oscillations (a-b) at zero temperature, and (c-d) at $T=5t$. (a,c) Normalized real‑space density modulations $\delta\rho_{\text{norm}}(x)$. Horizontal dashed lines in (a) and (c) mark the vanishing bulk background. (b,d) Corresponding Fourier spectra. Vertical dashed lines indicate the quantum geometric wavevector $q_G$ and the Fermi-surface nesting $2k_F$. The $2k_F$ peak shifts with filling at $T=0$ but is absent at $T>W^f$, while the $q_G$ dip remains unchanged. Parameters:  $J=1$, $t=10^{-4}J$, $\delta=0.03$, $\delta'=5\delta$, $U_0 = 5\times 10^{-3}J$. }
	\label{Fig_QGFOs_fillings_Temperature}
\end{figure}
In flat bands, the quenched bandwidth renders the precise filling fraction or chemical potential rather difficult to control experimentally. It is thus necessary to investigate the dependence of the QGFOs on chemical potential fluctuations. In Fig.~\ref{Fig_QGFOs_fillings_Temperature}, we present numerical results on the Friedel oscillations with varying chemical potential $\mu$ at temperature $T=0$ and $T=5t>W^f$, using the 1D three-band model.
At $T=0$, both the conventional $2k_F$ oscillations and the QGFOs are present [panel (a)]. Their Fourier spectra in (b) contain a peak at $2k_F$ and a dip at $q_G$. As $\mu$ varies, the $2k_F$ peak shifts because it tracks the Fermi surface, whereas the $q_G$ dip remains unchanged. At $T=5t>W^f$, the conventional oscillations are washed out entirely, and only the $q_G$ dip survives (c,d). Moreover, the $q_G$ feature does not shift with $\mu$, confirming that the QGFOs are insensitive to the Fermi level. This robustness against filling fluctuations as well as thermal smearing, stems from the origin of the QGFOs in the averaged quantum distance $\bar{D}^f(\bm{q})$, which is a property of the flat-band Hilbert space rather than a particular Fermi surface. %As long as $\mu$ stays within the flat band, $\bar{D}^f(\bm{q})$—and therefore $q_G$—does not change. Thus, the QGFOs are immune to both thermal smearing and filling fluctuations, in stark contrast to the conventional $2k_F$ oscillations.

%%%%%%%%%%%%%%%%%%%%%%%%%%%%%%
\section{Linking QGFO Wavevectors to Quantum Metric Hot Spot Separations}
\label{Sec_q_G__QM_hotspots_separation}

In the previous section, we established the generic form for QGFOs as Eq.~\eqref{Eq_general_QGFOs_form}, with their characteristic wavevector and decay length determined by Eq.~\eqref{Eq_q_G_xi_G}. In the main text, using a three-band model, we showed that the QGFO wavevector $\bm{q}_G$ is determined by the quantum metric hot spots separation, and the spatial decay length is governed by the quantum metric length.
Here, we aim to rigorously link $\bm{q}_G$ to the momentum-space separation of quantum metric hot spots.
In the following subsections, we first show that near local minima of a dispersive band that is adjacent to the flat band, the meromorphic structure of the flat-band projection operator in Eq.~\eqref{Eq_pole_structure} yields poles whose real parts coincide with these minima. We then show that quantum metric hot spots arise at the same locations. The separation between these quantum metric hot spots therefore directly sets the QGFO wavevectors.

%\subsection{\label{Sec_Pole_band_minima}Pole structure of flat-band projection operator near dispersive band minima}
\subsection{\label{Sec_Pole_band_minima}QGFO Wavevectors Near Dispersive Band Minima}
We examine the second type of singularity of the flat-band Green's function, solving Eq.~\eqref{Eq_pole_structure_2}. For simplicity, we consider a flat band $E^f_{\bm{k}} \simeq E^f_0$ and its nearest dispersive neighbor $E^d_{\bm{k}} $. Suppose the dispersive band has a local minimum at $\bm{k}_0$, the local band gap at this minimum is $\Delta_{\bm{k}_0} \equiv E^d_{\bm{k}_0} - E^f_0$. Near this minimum, we can obtain poles contributed by these two bands $\bm{k}_{nf}$ by solving:
\begin{equation}
	\begin{aligned}
		E^d_{\bm{k}_{nf}} - E^d_{\bm{k}_0} = - \Delta_{\bm{k}_0}.
	\end{aligned}
	\label{Eq_two_band_pole_equation}
\end{equation}
When the local gap is sufficiently small compared to other energy scales, we adopt a quadratic expansion of $E^d(\bm{k})$ around $\bm{k}_0$ in the complex plane:
\begin{equation}
	\begin{aligned}
		E^d_{\bm{k}_0+\bm{\eta}} 
		= E^d_{\bm{k}_0} + \frac{1}{2} \sum_{ij} \left(\frac{1}{M}\right)_{ij} \eta_i \eta_j 
		+ \mathcal{O}(|\bm{\eta}|^3).
	\end{aligned}
	\label{Eq_E_d_expansion}
\end{equation}
Here, we have defined the effective mass using the Hessian $\left(\frac{1}{M}\right)_{ij}  \equiv \left. \frac{\partial^2 E_n}{\partial k_i \partial k_j} \right|_{\bm{k}_0}$ and $\eta_i \equiv (\bm{k}_{nf}-\bm{k}_0)_i$.
Diagonalizing the effective mass with eigenvalues $m_{\alpha}>0$ and orthonormal eigenvectors $\hat{\bm{e}}_\alpha$, we obtain
\begin{equation}
	\begin{aligned}
		\sum_{\alpha} \frac{\eta_{\alpha}^2}{2m_\alpha} \simeq -\Delta_{\bm{k}_0},
	\end{aligned}
	\label{Eq_ellipsoid}
\end{equation}
In $d$-dimensions, Eq.~\eqref{Eq_ellipsoid} indicates that the imaginary part of $\bm{\eta} = (\eta_1, \eta_2, ..., \eta_d)$ defines a $d$-dimensional ellipsoid, and it can be described by $d-1$ free parameters. For example, in 2D, the poles (in the upper complex plane) from Eq.~\eqref{Eq_two_band_pole_equation} can be parameterized by a single parameter $\theta$ as:
\begin{equation}
	\begin{aligned}
		\bm{k}^{u/\bar{u}}_{nf} &= \bm{k}_0 + \bm{\eta}^{u/\bar{u}} \\
		&\simeq \bm{k}_0 \pm  i ( \sqrt{ 2m_{1} \Delta_{\bm{k}_0} } \cos\theta \, \hat{\bm{e}}_1 + \sqrt{ 2m_{2} \Delta_{\bm{k}_0} } \sin\theta \, \hat{\bm{e}}_2).
	\end{aligned}
	\label{Eq_k_nf_sol}
\end{equation}
Notably, for an isotropic system, or near a high symmetry point such as the $\Gamma$ point, the effective mass $m_\alpha = m_0$, and Eq.~\eqref{Eq_k_nf_sol} reduces to $\bm{k}_{nf} = \bm{k}_0 + i\sqrt{2m_0\Delta_{\bm{k}_0}} \, \hat{\bm{e}}_{\bm{r}}$ with $\hat{\bm{e}}_{\bm{r}}$ the radial unit vector.

When there are multiple local minima in the band $D^d_{\bm{k}}$, each of them produces a pole as Eq.~\eqref{Eq_ellipsoid} and \eqref{Eq_k_nf_sol}, as long as the quadratic expansions in Eq.~\eqref{Eq_E_d_expansion} remain valid approximations. As a result, for any two such local minima $\bm{k}_{0,p}$ and $\bm{k}_{0,q}$ with poles they generate $ \bm{k}_{nf, p} = \bm{k}_{0,p} + \bm{\eta}_{p}$ and $ \bm{k}_{nf,q} = \bm{k}_{0,q} + \bm{\eta}_{q}$, one obtains the characteristic wavevector and decay length for QGFOs following Eq.~\eqref{Eq_q_G_xi_G}:
\begin{subequations}
	\begin{align}
		\bm{q}_G &\simeq \bm{k}_{0,p} - \bm{k}_{0,q} \label{Eq_q_G_pair}, \\
		\bm{\xi}^{-1}_G   &\simeq -i\bigl( \bm{\eta}^u_p - \bm{\eta}^{\bar{u}}_q \bigr)
		\label{Eq_xi_G_pair}.
	\end{align}
	\label{Eq_q_G_xi_G_pair}
\end{subequations}
For isotropic local minima, the decay length reduces to a scalar $\xi_G \simeq \bigl(\sqrt{2m_0\Delta_{\bm{k}_{0,p}}} + \sqrt{2m_0\Delta_{\bm{k}_{0,q}}} \bigr) $. 

In conclusion, the local minima of a dispersive band $E^d_{\bm{k}}$ generate pole structures of the flat-band projection operator, and the real (imaginary) parts of the poles are given by the momenta location (dispersive band effective mass and band gap). Importantly, the QGFO wavevectors are determined by the separation of these local minima. We reiterate that this conclusion relies on the validity of the quadratic expansion in Eq.~\eqref{Eq_E_d_expansion}, and for well separated local minima with small band gaps, it is a reasonably good approximation.

\subsection{\label{Sec_QM_hotspots_band_minima}Emergence of Quantum Metric Hot Spots Near Dispersive Band Minima}

In this subsection, we provide a detailed derivation of the conditions under which the flat-band quantum metric exhibits a hot spot at a local minimum of an adjacent dispersive band. We show that, while the hot spot is strictly pinned to the band minimum only when symmetry enforces a stationary numerator in the quantum metric tensor in Eq.~\eqref{Eq_QM_definition}, the actual displacement of the peak for a generic low-symmetry minimum can be arbitrarily small for a sufficient suppressed local gap.
% is parametrically smaller than the Brillouin zone (BZ) scale and hence negligible in practice.
Consider the dominant component of the quantum metric tensor from the inter-band contribution between the flat band $f$ and an adjacent dispersive band $d$:
\begin{equation}
	g^{ab}_{\bm{k}} \simeq \frac{ N^{ab}_{\bm{k}} }{ \Delta^2_{\bm{k}} },
	\label{Eq_QM_definition}
\end{equation}
where $N^{ab}_{\bm{k}}$ is related to the inter-band velocity operator $\hat{v}^a_{\bm{k}} \equiv \partial_{k_a} H_{\bm{k}}$ and the Bloch vectors of the two bands $u^f_{\bm{k}}$ and $u^d_{\bm{k}}$ by
\begin{equation}
	N^{ab}_{\bm{k}} = \Re \big\{ \langle u^f_{\bm{k}} | \hat{v}^a_{\bm{k}} | u^d_{\bm{k}} \rangle \langle u^d_{\bm{k}} | \hat{v}^b_{\bm{k}} | u^f_{\bm{k}} \rangle \big\}.
\end{equation}

Because the denominator in Eq.~\eqref{Eq_QM_definition} is squared, the quantum metric is resonantly enhanced wherever $\Delta(\bm{k})$ is locally minimized\footnote{To reduce higher order effects, we assume the local minima considered here are well approximated by quadratic expansions.}, provided $N^{ab}_{\bm{k}}$ does not vanish identically. Generically, a quantum metric hot spot which satisfies
\begin{equation}
	\bm{\nabla}_{\bm{k}} g^{ab}_{\bm{k}} = 0,
	\label{Eq_stationary_QM_condition}
\end{equation}
can deviate from this local minimum. We now examine how close the quantum metric hot spot can approach the dispersive band minimum $\bm{k}_0$. Eqs.~\eqref{Eq_QM_definition} and \eqref{Eq_stationary_QM_condition} lead to
\begin{equation}
	\frac{\bm{\nabla}_{\bm{k}} N^{ab}_{\bm{k}}}{\Delta^2} - \frac{2N^{ab}_{\bm{k}}\bm{\nabla}_{\bm{k}} N^{ab}_{\bm{k}}}{\Delta^3}  = 0.
	\label{Eq_stationary_QM_condition_expanded}
\end{equation}
Around the local minimum $\bm{k}_0$, we expand $N^{ab}_{\bm{k}}$ and $\Delta_{\bm{k}}$ to second order in $\bm{\delta}$, and obtain their gradient as
\begin{align}
	\bm{\nabla} N_{\bm{k}} &= \mathbf{D}^N_{\bm{k_0}}+ \mathbf{H}^N_{\bm{k_0}}\, \bm{\delta} + \mathcal{O}(\bm{\delta}^2), \\
	\bm{\nabla} \Delta_{\bm{k}} &= \mathbf{H}^{\Delta}_{\bm{k_0}} \, \bm{\delta} + \mathcal{O}(\bm{\delta}^2) ,
\end{align}
where the gradient is defined as $\mathbf{D}^N_{\bm{k_0}}\equiv \left. \bm{\nabla} N^{ab}_{\bm{k}} \right|_{\bm{k}_0}$,  and the Hessian matrices for $N^{ab}$ and $\Delta$ are defined as $\mathbf{H}^N_{\bm{k_0}} \equiv \frac{1}{2} \bm{\nabla} \bm{\nabla} N^{ab}_{\bm{k}}$ and $\mathbf{H}^{\Delta}_{\bm{k_0}} \equiv \frac{1}{2} \bm{\nabla} \bm{\nabla} \Delta_{\bm{k}}$, respectively. Then, Eqs.~\eqref{Eq_stationary_QM_condition} and \eqref{Eq_stationary_QM_condition_expanded} yield
\begin{equation}
	\Delta_{\bm{k}_0} \left( \mathbf{D}^N_{\bm{k_0}}+ \mathbf{H}^N_{\bm{k_0}}\bm{\delta} \right) - 2 N_0 \mathbf{H}^{\Delta}_{\bm{k_0}} \bm{\delta} = 0.
\end{equation}
Thus, solving $\bm{\delta}$ gives the displacement of the actual quantum metric peak from the local minimum $\bm{k}_0$:
\begin{equation}
	\bm{\delta} = - \Delta_{\bm{k}_0} \left( \Delta_{\bm{k}_0} \mathbf{H}^N_{\bm{k_0}}- 2 N_0 \mathbf{H}^{\Delta}_{\bm{k_0}} \right)^{-1} \mathbf{D}^N_{\bm{k_0}}.
	\label{Eq_QM_hot_spot_displacement}
\end{equation}

In general, the matrix in parentheses in Eq.~\eqref{Eq_QM_hot_spot_displacement} is invertible, so how close $\bm{\delta}$ approaches zero would depend on $\Delta_{\bm{k}_0}$ and $\mathbf{D}^N_{\bm{k_0}}$. At high-symmetry momenta such as $\Gamma$, $M$ or $K$, where crystal symmetries enforce an even function $N^{ab}_{\bm{k}} $ of $\bm{k}$, one finds
\begin{equation}
	\mathbf{D}^N_{\bm{k_0}}= 0.
	\label{Eq_D_N_vanish_condition}
\end{equation}
Thus a quantum metric hot spot will be pinned at such high symmetry points. For generic low-symmetry momenta $\bm{k}_0$, while $\mathbf{D}^N_{\bm{k_0}}\neq 0$ in general, we argue that for a sufficiently small local gap $\Delta_{\bm{k}_0}$, the quantum metric hot spot coincides with $\bm{k}_0$.
Therefore, we arrive at the conclusion that near dispersive band minima with a sufficiently small band gap, quantum metric hot spots arise.

\subsection{Conclusion}
Combining the results of the preceding subsections, we establish the bridge linking the oscillation periods of QGFOs to momentum-space separation of quantum metric hot spots.
In Sec.\ref{Sec_Pole_band_minima}, we demonstrated that the meromorphic structure of the flat-band projection operator yields complex poles near dispersive band minima. Crucially, to quadratic order, the real part of such a pole is pinned to the band minimum, while its imaginary part is governed by the local band gap and effective mass. Consequently, for a pair of minima, the QGFO wavevectors are determined by the differences of these complex poles, as given in Eq.~\eqref{Eq_q_G_pair}.
In Sec.\ref{Sec_QM_hotspots_band_minima}, we analyzed the behavior of the quantum metric tensor near a dispersive band minimum. We found that while the location of a quantum metric hot spot can be displaced from the dispersive band minimum by [Eq.~\eqref{Eq_QM_hot_spot_displacement}], this displacement vanishes identically at high-symmetry momenta. Meanwhile, for generic low-symmetry minima, it will approach zero in the limit of sufficiently small local gap---the regime in which the hot spot is most pronounced.

\begin{figure}[tbp]
	\centering
	\includegraphics[width=0.8\linewidth]{ 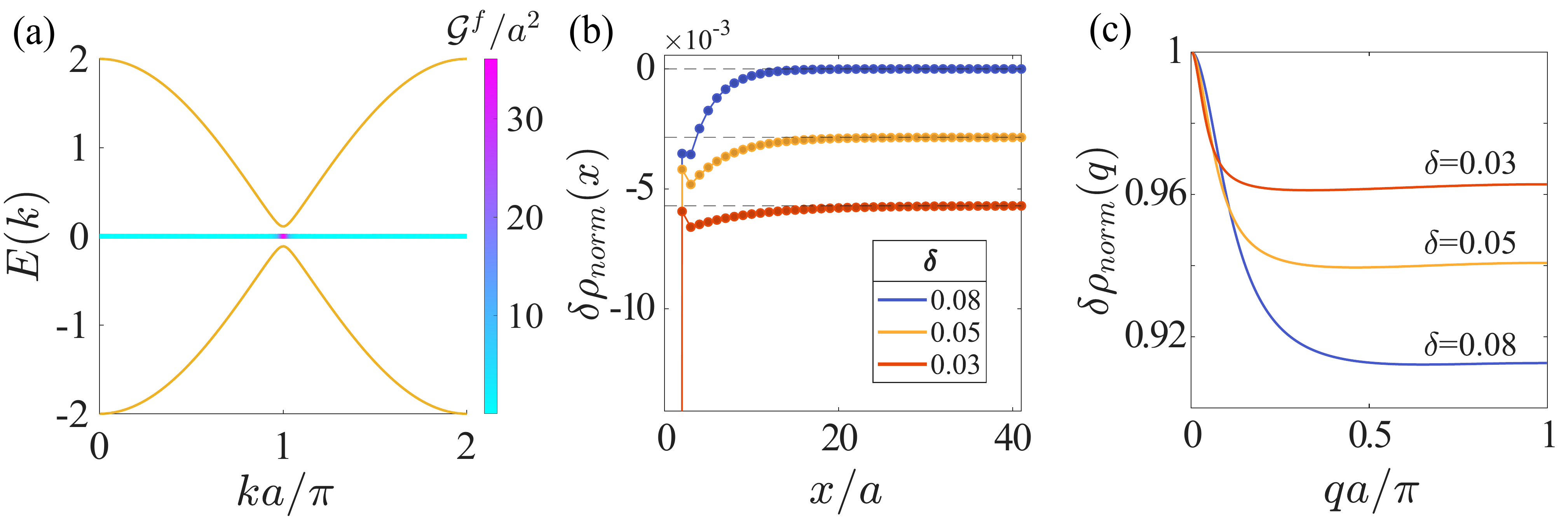}
	\caption{Charge density modulations for 1D Lieb lattice. (a) The flat band in the middle exhibits a single quantum metric hot spot in the flat band. (b) The charge density modulations around a single impurity exhibit a purely exponential decay with no oscillations. (c) The corresponding Fourier spectra of the density modulations show no local peaks or dips at finite $q$. Parameters: $J=1$, $t=10^{-4}J$, $\mu = -1.6t$, $U_0 = 1 \times 10^{-2}J$, and $k_BT = 10t$.}
	\label{Fig_Lieb_single_QMhotspot}
\end{figure}

Therefore, we conclude that, near dispersive band minima, the poles of the flat-band projection operator coincide with quantum metric hot spots, and the QGFOs wavevectors $\bm{q}_G$ are determined by the momentum-space separation of the quantum metric hot spots.
In other words, the presence of QGFOs requires multiple quantum metric hot spots, and the momentum-space separation of any two of them will give rise to an oscillatory component in the charge density responses (i.e. a peak or a dip in the flat-band susceptibility). Contrastingly, if there is only a single quantum metric hot spot in the flat band, such as the 1D Lieb lattice [Fig.~\ref{Fig_Lieb_single_QMhotspot}(a) shows the band structure as well as the quantum metric distribution in the flat band], the charge density response to a local impurity will exhibit a purely exponential decay without any spatial oscillations [Fig.~\ref{Fig_Lieb_single_QMhotspot}(b)].
This confirms the three-band model analysis in the main text and grounds the generic QGFO form [Eq.~\eqref{Eq_general_QGFOs_form}] in the underlying quantum geometry.

\section{Inter-band Susceptibilities}
\label{Sec_DispBand_contribution}

In the main text we kept only the flat‑band contribution to the susceptibility. Here, we further analyze the inter‑band contributions to the susceptibility, and show that they are negligible when $T,U_0\ll\Delta_g$, but become relevant at higher temperatures when $T\simeq \Delta_g$. This justifies the flat‑band projection invoked in the main text.

%In the main text, through the flat-band projection, we have dropped the inter-band contributions between flat band and higher bands. We now comment on the conditions for this treatment to be justified. As we show in the following, the magnitude of inter-band susceptibility $\chi^{\text{inter}}(\bm{q})$ between flat band and dispersive bands is bounded by $\frac{1}{\Delta_g}\left[1-\bar{D}^{\text{inter}}(\bm{q})\right]$ from above, which, compared with the flat band contribution, suggests that the intra-band contribution $\chi^f$ dominates when $T \ll \Delta_g$. Going beyond this regime, that is, when $T \lesssim \Delta_g$, the averaged inter-band quantum distance would also contribute to the QGFOs. For instance, if $\bar{D}^{\text{inter}}$ has features such as dips at certain momentum, they will also be manifested as oscillation periods in real space. %More discussions on this are expanded in the later sections on flat band RKKY interactions and Supplemental Materials\cite{}.
%\subsection{Dispersive-band contributions}

We work within the Born approximation, $\mathcal{T}_{\omega}=\mathcal{I}$, and expand the full-band susceptibility as
\begin{equation}
	\begin{aligned}
		\chi(\bm{q}) 
		= & -\frac{1}{\pi} \Im \left[ \frac{1}{N} \sum_{\bm{k}}  \int_\omega f(\omega)  \operatorname*{Tr} \left[\mathcal{G}_{\omega}(\bm{k}+\bm{q}) \mathcal{G}_{\omega}(\bm{k}) \right] \right] \\
		= & \frac{1}{N} \sum_{\bm{k}}  \int_\omega f(\omega) \Im \left[ -\frac{1}{\pi} \sum_{n,m}  \frac{|\braket{u_n(\bm{k}+\bm{q}) | u_m(\bm{k})} |^2}{(\omega^{\dagger} - E^n_{\bm{k}+\bm{q}})(\omega^{\dagger} - E^m_{\bm{k}})} \right]  
		%		= & \chi^f(\bm{q}) + \chi^l(\bm{q}) +\chi^h(\bm{q}) + \chi^{l-f}(\bm{q}) + \chi^{h-f}(\bm{q})  + \chi^{h-l}(\bm{q}) 
		.
	\end{aligned}
	\label{Eq_chi_q_FullBand_expansion}
\end{equation}
We consider three sets of bands that are well separated in energy: flat band $f$, higher dispersive bands $h$ and lower dispersive bands $l$. The full-band susceptibility decomposes into intra-set ($\chi^f,\chi^l,\chi^h$) and inter-set ($\chi^{l-f},\chi^{h-f},\chi^{l-h}$) contributions. For simplicity, we assume all bands are well-isolated so that the only relevant bands are those nearby the flat band. As such, we consider one band in each dispersive set that is closest to the flat band. In the regime $W^f \ll T\ll \Delta_g$, we estimate each of the dispersive‑band terms, which can be categorized into following three parts:

1. Intra‑band dispersive contributions:
\begin{equation}
	\begin{aligned}
		\chi^{l}(\bm{q}) %&\ \ =  -\frac{1}{\pi}		\Im \left[ \frac{1}{N} \sum_{\bm{k}} \int_\omega f(\omega) \frac{1-d^{l}_{\bm{k}, \bm{k}+\bm{q}}}{(\omega^+-E^l_{\bm{k}+\bm{q}}) (\omega^+-E^l_{\bm{k}}) } \right] \\
		&\ \ = \frac{1}{N} \sum_{\bm{k}} \frac{f(E^l_{\bm{k}+\bm{q}}) - f(E^l_{\bm{k}})}{E^d_{\bm{k}+\bm{q}} - E^l_{\bm{k}}} (1-d^{l}_{\bm{k}, \bm{k}+\bm{q}}) \\
		&\overset{T\ll \Delta_g}{\simeq}  \frac{1}{N} \sum_{\bm{k}} \frac{e^{E^l_{\bm{k}}/T} - e^{E^l_{\bm{k}+\bm{q}}/T}}{E^l_{\bm{k}+\bm{q}} - E^l_{\bm{k}}} (1-d^{l}_{\bm{k}, \bm{k}+\bm{q}}) \\
		&\ \  \simeq 0,
	\end{aligned}
	\label{Eq_intra_l}
\end{equation}
and
\begin{equation}
	\begin{aligned}
		\chi^{h}(\bm{q}) &\overset{T\ll \Delta_g}{\simeq}  0.
	\end{aligned}
\end{equation}
%\begin{equation}
%	\begin{aligned}
	%		\chi^{h}(\bm{q}) %&\ \ =  -\frac{1}{\pi}		\Im \left[ \frac{1}{N} \sum_{\bm{k}} \int_\omega f(\omega) \frac{1-d^{h}_{\bm{k}, \bm{k}+\bm{q}}}{(\omega^+-E^h_{\bm{k}+\bm{q}}) (\omega^+-E^h_{\bm{k}}) } \right] \\
	%		&\ \ = \frac{1}{N} \sum_{\bm{k}} \frac{f(E^h_{\bm{k}+\bm{q}}) - f(E^h_{\bm{k}})}{E^h_{\bm{k}+\bm{q}} - E^h_{\bm{k}}} (1-d^{h}_{\bm{k}, \bm{k}+\bm{q}}) \\
	%		&\overset{T\ll \Delta_g}{\simeq}  \frac{1}{N} \sum_{\bm{k}} \frac{e^{-E^h_{\bm{k}+\bm{q}}/T} - e^{-E^h_{\bm{k}}/T}}{E^h_{\bm{k}+\bm{q}} - E^h_{\bm{k}}} (1-d^{h}_{\bm{k}, \bm{k}+\bm{q}}) \\
	%		&\ \  \simeq 0.
	%	\end{aligned}
%\end{equation}

%1. Intra‑f

2. Inter-band contributions between the flat and lower dispersive bands $\chi^{l-f}$:
\begin{equation}
	\begin{aligned}
		\chi^{l-f}(\bm{q})  %&\ \ =  -\frac{1}{\pi}		\Im \left[ \frac{1}{N} \sum_{\bm{k}} \int_\omega f(\omega) \frac{1-d^{l-f}_{\bm{k}, \bm{k}+\bm{q}}}{(\omega^+-E^h_{\bm{k}+\bm{q}}) (\omega^+-E^h_{\bm{k}}) } \right] \\
		&\ \ = \ \frac{1}{N} \sum_{\bm{k}} \frac{f(E^l_{\bm{k}+\bm{q}}) - f(E^f_{\bm{k}})}{E^l_{\bm{k}+\bm{q}} - E^f_{\bm{k}}} (1-d^{l-f}_{\bm{k}, \bm{k}+\bm{q}}) \\
		&\overset{W^f \ll T\ll \Delta_g}{\simeq}  \frac{1}{N} \sum_{\bm{k}} \frac{1-e^{E^l_{\bm{k}+\bm{q}}/T} - \frac{1}{2}(1-\frac{E^f_{\bm{k}}}{2T})}{E^l_{\bm{k}+\bm{q}} - E^f_{\bm{k}}} (1-d^{l-f}_{\bm{k}, \bm{k}+\bm{q}}) \\
		&\ \overset{W^f \ll \Delta_g}{\simeq} \frac{1}{N} \sum_{\bm{k}} \frac{1}{2E^l_{\bm{k}+\bm{q}}} (1-d^{l-f}_{\bm{k}, \bm{k}+\bm{q}})  \\
		&\ \ \ \  \ge -\frac{1}{2\Delta_g} [1-\bar{D}^{l-f}(\bm{q})],
	\end{aligned}
	\label{Eq_chi_lf}
\end{equation}
and similarly, between the flat and higher dispersive bands $\chi^{h-f}$:
\begin{equation}
	\begin{aligned}
		\chi^{h-f}(\bm{q}) %&\ \ =  -\frac{1}{\pi}		\Im \left[ \frac{1}{N} \sum_{\bm{k}} \int_\omega f(\omega) \frac{1-d^{h-f}_{\bm{k}, \bm{k}+\bm{q}}}{(\omega^+-E^h_{\bm{k}+\bm{q}}) (\omega^+-E^h_{\bm{k}}) } \right] \\
		%		&\ \ \ = \frac{1}{N} \sum_{\bm{k}} \frac{f(E^h_{\bm{k}+\bm{q}}) - f(E^f_{\bm{k}})}{E^h_{\bm{k}+\bm{q}} - E^f_{\bm{k}}} (1-d^{h-f}_{\bm{k}, \bm{k}+\bm{q}}) \\
		%		&\overset{W^f \ll T\ll \Delta_g}{\simeq}  \frac{1}{N} \sum_{\bm{k}} \frac{e^{-E^h_{\bm{k}+\bm{q}}/T} - \frac{1}{2}(1-\frac{E^f_{\bm{k}}}{2T})}{E^h_{\bm{k}+\bm{q}} - E^f_{\bm{k}}} (1-d^{h-f}_{\bm{k}, \bm{k}+\bm{q}}) \\
		%		&\ \overset{W^f \ll \Delta_g}{\simeq} - \frac{1}{N} \sum_{\bm{k}} \frac{1}{2E^h_{\bm{k}+\bm{q}}} (1-d^{h-f}_{\bm{k}, \bm{k}+\bm{q}})  \\
		\ge - \frac{1}{2\Delta_g} [1-\bar{D}^{h-f}(\bm{q})] .
	\end{aligned}
	\label{Eq_chi_hf}
\end{equation}

3. Inter‑band contribution between the two dispersive bands:
\begin{equation}
	\begin{aligned}
		\chi^{l-h}(\bm{q}) %&\ \ =  -\frac{1}{\pi}		\Im \left[ \frac{1}{N} \sum_{\bm{k}} \int_\omega f(\omega) \frac{1-d^{l-h}_{\bm{k}, \bm{k}+\bm{q}}}{(\omega^+-E^l_{\bm{k}+\bm{q}}) (\omega^+-E^h_{\bm{k}}) } \right] \\
		&\ \ = \ \ \frac{1}{N} \sum_{\bm{k}} \frac{f(E^l_{\bm{k}+\bm{q}}) - f(E^h_{\bm{k}})}{E^l_{\bm{k}+\bm{q}} - E^h_{\bm{k}}} (1-d^{l-h}_{\bm{k}, \bm{k}+\bm{q}}) \\
		&\overset{T\ll \Delta_g}{\simeq}  \frac{1}{N} \sum_{\bm{k}} \frac{1-e^{E^l_{\bm{k}+\bm{q}}/T} - e^{-E^h_{\bm{k}}/T}}{E^l_{\bm{k}+\bm{q}} - E^h_{\bm{k}}} (1-d^{l-h}_{\bm{k}, \bm{k}+\bm{q}}) \\
		&\ \  \geq -\frac{1}{2\Delta_g} [1-\bar{D}^{l-h}(\bm{q})].
	\end{aligned}
	\label{Eq_chi_lh}
\end{equation}

\begin{figure}[tbp]
	\centering
	\includegraphics[width=1\linewidth]{ 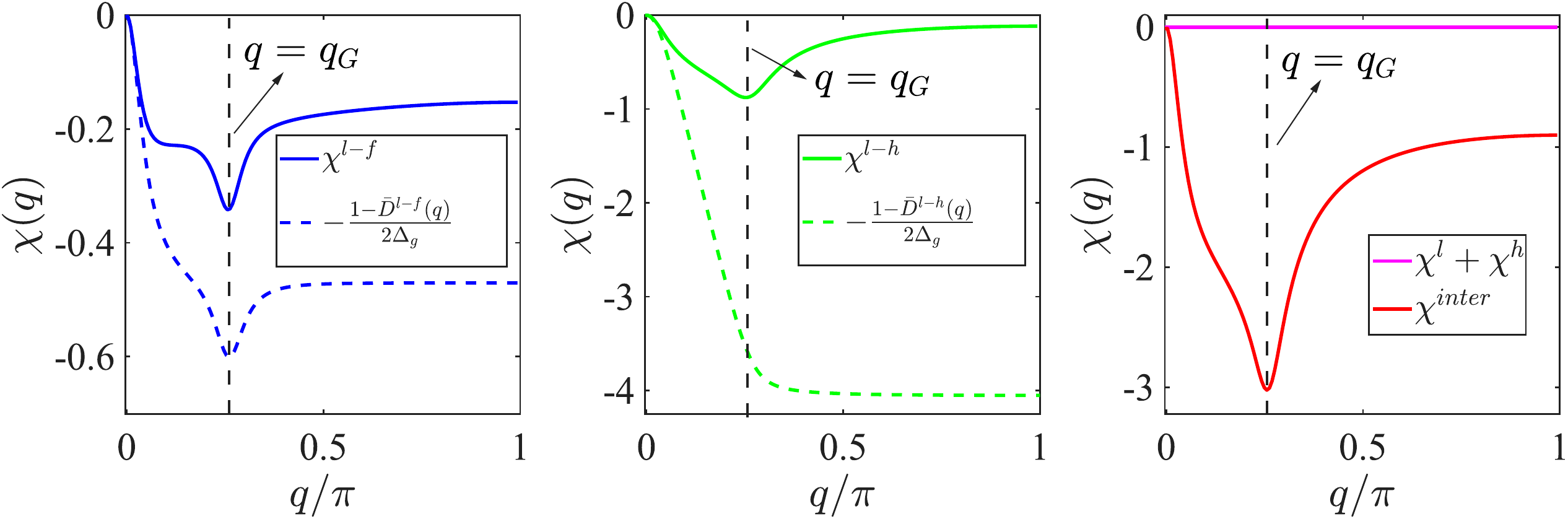}
	\caption{Inter-band susceptibilities for the 1D three-band model. (a) $\chi^{l-f}$ and its lower bound (amplitude upper bound) from inter-band averaged quantum distance (dashed line) according to Eq.~\eqref{Eq_chi_lf}. (b) $\chi^{l-h}$ and its lower bound from Eq.~\eqref{Eq_chi_lh}. (c) The total inter-band susceptibilities $\chi^{\text{inter}}$ vs. the intra-band susceptibilities from dispersive bands $\chi^{l} + \chi^{h}$. Vertical dashed lines denote the dipping feature at momentum $q_G$. Parameters:  $J=1$, $t=10^{-4}J$, $\delta=0.03$, $\delta'=5\delta$, and $T = 10t$.}
	\label{Fig_Inter_band_susceptibilities}
\end{figure}

Above, we have used the approximations $f(E_k^{l})  \simeq 1-e^{E_k^l/T}$, $f(E_k^{h}) \simeq e^{-E_k^h/T}$, and  $f(E_k^{h}) \simeq \frac{1}{2}(1- E_k^f/2T)$ in the regime $W^f \ll T \ll \Delta_g$.
Collecting all inter‑band terms (including the reciprocal partners $f$‑$l$, $f$‑$h$, and $h$‑$l$), the total inter‑band susceptibility is $\chi^{\text{inter}} (\bm{q}) \equiv  2(\chi^{l-f}(\bm{q}) + \chi^{h-f}(\bm{q}) + \chi^{l-h}(\bm{q}) )$. Since the inter-band susceptibilities are all negative, the above inequalities yield an upper bound to the total inter-band susceptibilites:
\begin{equation}
	\begin{aligned}
		\left|\chi^{\text{inter}} (\bm{q}) \right| \leq \frac{1}{\Delta_g} [3-\bar{D}^{l-f}(\bm{q}) -\bar{D}^{h-f}(\bm{q}) - \bar{D}^{l-h}(\bm{q})]. 
		%		\\ &\gg - \frac{1}{\Delta_g},
	\end{aligned}
	\label{Eq_chi_disp_tot}
\end{equation}
This bound saturates when $W^f,W^l,W^h\ll\Delta_g$. Eqs.~\eqref{Eq_chi_lf}--\eqref{Eq_chi_disp_tot} make it explicit that the inter‑band averaged quantum distances $\bar{D}^{l-f},\bar{D}^{h-f},\bar{D}^{l-h}$ control the magnitude of the inter‑band susceptibility. Since they are in the order of $\sim \mathcal{O}(1)$, the right‑hand side of~\eqref{Eq_chi_disp_tot} reduces to $\sim 1/\Delta_g$. Compared with the flat‑band contribution $\chi^f\sim 1/(4T) \gg 1/\Delta_g$ for $T\ll\Delta_g$, the inter‑band terms are negligible, validating the flat-band projection.  Only when $T$ approaches $\Delta_g$ do the inter‑band contributions become competitive, and any structure in the inter‑band quantum distances—peaks, dips, kinks—will then imprint additional oscillatory components in the LDOS, modifying the flat-band Friedel oscillations. This is precisely the origin of the alternating FM/AFM crossover in the RKKY interaction discussed later in Section~\ref{Sec_application}.

Eqs.~\eqref{Eq_intra_l}--\eqref{Eq_chi_disp_tot} are confirmed in Fig.~\ref{Fig_Inter_band_susceptibilities}. Panel (a) shows $\chi^{l-f}$ together with the lower bound $-\frac{1}{2\Delta_g}[1-\bar{D}^{l-f}]$ (dashed line) from Eq.~\eqref{Eq_chi_lf}; panel (b) shows the same for $\chi^{l-h}$. Both inter‑band susceptibilities exhibit a dip at the quantum geometric wavevector $q_G$,
which, from the lower bounds in (a,b), suggests a quantum geometric origin: the inter-band quantum distance between $f$ and $l$ shows up a minimum at $q_G$; the inter-band quantum distance between dispersive band $l$ and $h$ exhibits a kink near $q_G$, which is dressed by the inter-band Lindhard weight and gives rise to the dip in $\chi^{lh}$.
Panel (c) displays the total inter‑band susceptibility $\chi^{\text{inter}}$ (curve in red) and the (negligible) intra‑band dispersive parts $\chi^l+\chi^h$ (curve in magenta).

\section{QGFOs at Finite Temperatures}

In the main text, we have focused on the high-temperature regime $k_B T \gg W^f$, in which the conventional Friedel oscillations are washed out completely, leaving the QGFOs solely governed by the averaged quantum distance.
However, the presence of dominant QGFOs is not contingent on extremely high temperature. Given that the spatial extent of conventional Friedel oscillations is governed by the thermal coherence length $\xi_T$, the oscillations are strongly suppressed when $\xi_T$ falls below the oscillation wavelength, i.e., $\xi_T \lesssim \pi/k_F$. For a parabolic flat band, this suppression occurs at temperatures $ k_BT \simeq E_F^f \ll W^f$, where $E_F^f \equiv E_F - E^f_{\text{min}}$ is the Fermi energy measured from the flat-band bottom. Such temperatures are much lower than the bandwidth but high enough to completely smear out the Fermi surface. For realistic flat-band platforms such as magic-angle twisted bilayer graphene, the bandwidth is $\sim 10$ meV~\cite{cao2018unconventional}, so with a low filling fraction, e.g., $\nu^f \sim 0.1$, QGFOs in such systems---if present---can be predominantly observable at temperatures as low as $10$ K.
\iffalse
Fig.~\ref{Fig_1}(d) suggests that the thermal length $\xi_T$ can already be reduced to sub-lattice-constant scale when the thermal energy $k_BT$ is one order of magnitude smaller than the flat-band bandwidth. On the other hand, the bandwidth in realistic flat-band platforms such as magic-angle twisted bilayer graphene is of $\sim 10$ meV~\cite{cao2018unconventional}, so we expect the QGFOs, if present, should be observable at temperature $\sim 10$ K. So we now examine an intermediate temperature regime in which the temperature is much lower than the bandwidth but high enough to completely smear out the Fermi surface. 
Taking a low filling fraction in the flat band $\nu^f \ll 1$, a temperature $ T \simeq E_F^f \ll W^f$ --- where $E_F^f \equiv E_F - E^f_{\text{min}}$ is the Fermi energy measured from the flat-band bottom %($E^f_{\text{min}/\text{max}} \equiv \operatorname{min/max} (E^f_{\bm{k}})$) 
--- suffices to wash out the conventional Friedel oscillations. 
Because states well above $E_F$ are thermally frozen, both scattering among those states and large momentum transfers $|\bm{q}| \gg 2k_F$ are suppressed.
\fi
To investigate this regime, let us examine the quantum geometric susceptibility at finite temperature:
\begin{equation}
	\begin{aligned}
		\chi^f_g(\bm{q}) = - \frac{1}{N} \sum_{\bm{k}} \int_{\epsilon} f'(\epsilon) C^f_{\bm{k}, \bm{q},0}(\epsilon)
		%\frac{\Theta(\epsilon - E^f_{\bm{k}+\bm{q}}) - \Theta(\epsilon - E^f_{\bm{k}})}{E^f_{\bm{k}+\bm{q}} - E^f_{\bm{k}}} 
		d^f_{\bm{k}, \bm{k}+\bm{q}},
	\end{aligned}
	\label{Eq_chi_g_finite_T}
\end{equation}
where $C^f_{\bm{k}, \bm{q}, 0}(\epsilon)$ is the zero-temperature Lindhard weight with Fermi energy taken at $\epsilon$. Eq.~\eqref{Eq_chi_g_finite_T} indicates that the dominant quantum geometry arises from the region near the Fermi energy $E^f_F$ within an energy window $\sim k_B T$. Since the states well above and below $E_F$ are thermally frozen, scattering processes within each frozen sector are suppressed, as are large energy transfers $\Delta E = |E_{\bm{k}+\bm{q}} - E_{\bm{k}}| \gg k_B T$ between two sectors. Contributions of the associated quantum distance to $\chi^f_g(\bm{q})$ are suppressed exponentially for the former processes and algebraically ($\sim 1/\Delta E$) for the latter. Therefore, for temperature $ k_B T \simeq E_F^f$, the relevant quantum geometry is primarily captured by the partially averaged quantum distance
\begin{equation}
	\begin{aligned}
		\bar{D}_p^{f}(\bm{q}) = \frac{1}{\mathcal{N}} \sum_{E^f_{\text{min}} < E^f_{\bm{k}},E^f_{\bm{k}+\bm{q}} < \Lambda_c} d^f_{\bm{k}, \bm{k}+\bm{q}},
	\end{aligned}
	\label{Eq_PAQD}
\end{equation}
%$\bar{D}_p^{f}(\bm{q}) = \frac{1}{N} \sum_{E^f_{\text{min}} < E^f_{\bm{k}},E^f_{\bm{k}+\bm{q}} < \Lambda_c} d^f_{\bm{k}, \bm{k}+\bm{q}}$, 
%$\bar{D}_p^{f}(\bm{q}) = \frac{1}{N} \sum_{0<|\bm{k}|,|\bm{k}+\bm{q}|<\Lambda_c} d^f_{\bm{k}, \bm{k}+\bm{q}}$,
where $\Lambda_c = \kappa k_BT < E^f_{\text{max}}$ (with $\kappa \sim \mathcal{O}(1)$, practically $\kappa \approx 3$) is an energy cutoff chosen such that $\bar{D}_p^{f}(\bm{q})$ contains all substantial quantum geometric information within the thermally accessible window. The normalization factor $\mathcal{N} = \sum_{E^f_{\text{min}} < E^f_{\bm{k}},E^f_{\bm{k}+\bm{q}} < \Lambda_c} 1$.

\begin{figure}[tbp]
	\centering
	\includegraphics[width=0.6\linewidth]{ 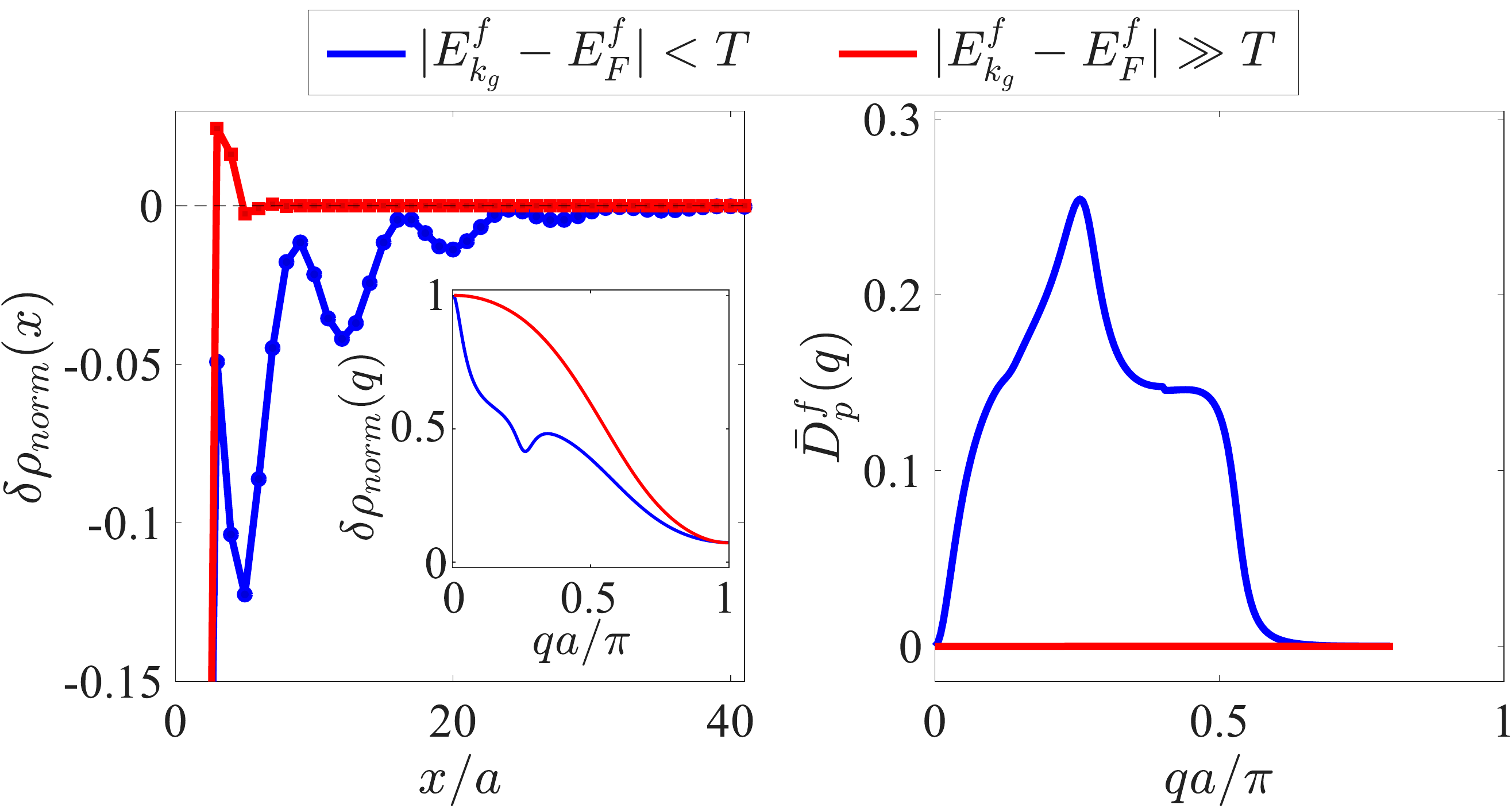}
	\caption{(a) Comparison of QGFOs with distinct low-energy quantum geometric structures: Quantum metric hot spots living within (blue curves) versus far above (red curves) the smeared Fermi surface. Inset: corresponding Fourier spectrum. (b) Partially averaged quantum distance $\bar{D}_p^{f}(q)$ for the two regimes. Shared parameters for the two cases are: $J=1$, $\mu = -1.6t$, $\lambda=0.02$, $\delta = 0.03$, $\delta' = 5\delta$, $U_0 = 0.01J$, $k_BT=E_F^f=0.4t$, $U_0=10t$, and $\Lambda_c = 3.5k_BT$. The hopping strength is $t=10^{-4}J$ for the blue curve, and $t=-10^{-4}J$ for the red curve.}
	\label{Fig_Band_Flip}
\end{figure}

%==================================================
\subsection{Thermal Accessibility of Quantum-Metric Hot Spots}
As an example, we examine the 1D flat-band model in this regime and compare the QGFOs before and after altering the quantum geometry of the thermally accessible low-energy states. Fixing the flat-band filling fraction $\nu^f \ll 1$ and temperature $k_BT = E_F^f$, the quantum metric hot spots in the flat-band lie within the smeared Fermi surface, i.e., $|E^f_{k_g} - E^f_F| < k_BT$, giving rise to prominent QGFOs with characteristic wavevector $q_G$ [blue curves in Fig.~\ref{Fig_Band_Flip}(a)]. In contrast, reversing the hopping sign $t \to -t$ shifts the band bottom by half a BZ while leaving the dispersive bands intact, rendering the quantum metric hot spots thermally inaccessible, i.e., $|E^f_{k_g} - E^f_F| \gg k_BT$. As depicted in the red curves in Fig.~\ref{Fig_Band_Flip}(a), the oscillation pattern fades after the shift and the spectral dip at $q_G$ disappears, signaling a trivial low-energy quantum geometric structure. Correspondingly, panel (b) confirms that the partially averaged quantum distance exhibits a pronounced peak at $q=q_G$ only before the shift, whereas the shifted case is featurelessly trivial. In conclusion, in the regime $ k_BT \simeq E_F^f$, thermal broadening confines the response to a smeared Fermi window, which completely washes out the conventional Friedel oscillations, and the QGFOs are governed by $\bar{D}_p^{f}(\bm{q})$. We further corroborate this behavior in a 2D flat-band model in the End Matter.

%%%%%%%%%%%%%%%%%%%%%%%%%%%%%
%	\emph{2D model with QGFOs.}
\subsection{2D model with QGFOs}
\begin{figure}[bbp]
	\centering
	\includegraphics[width=0.6\linewidth]{ 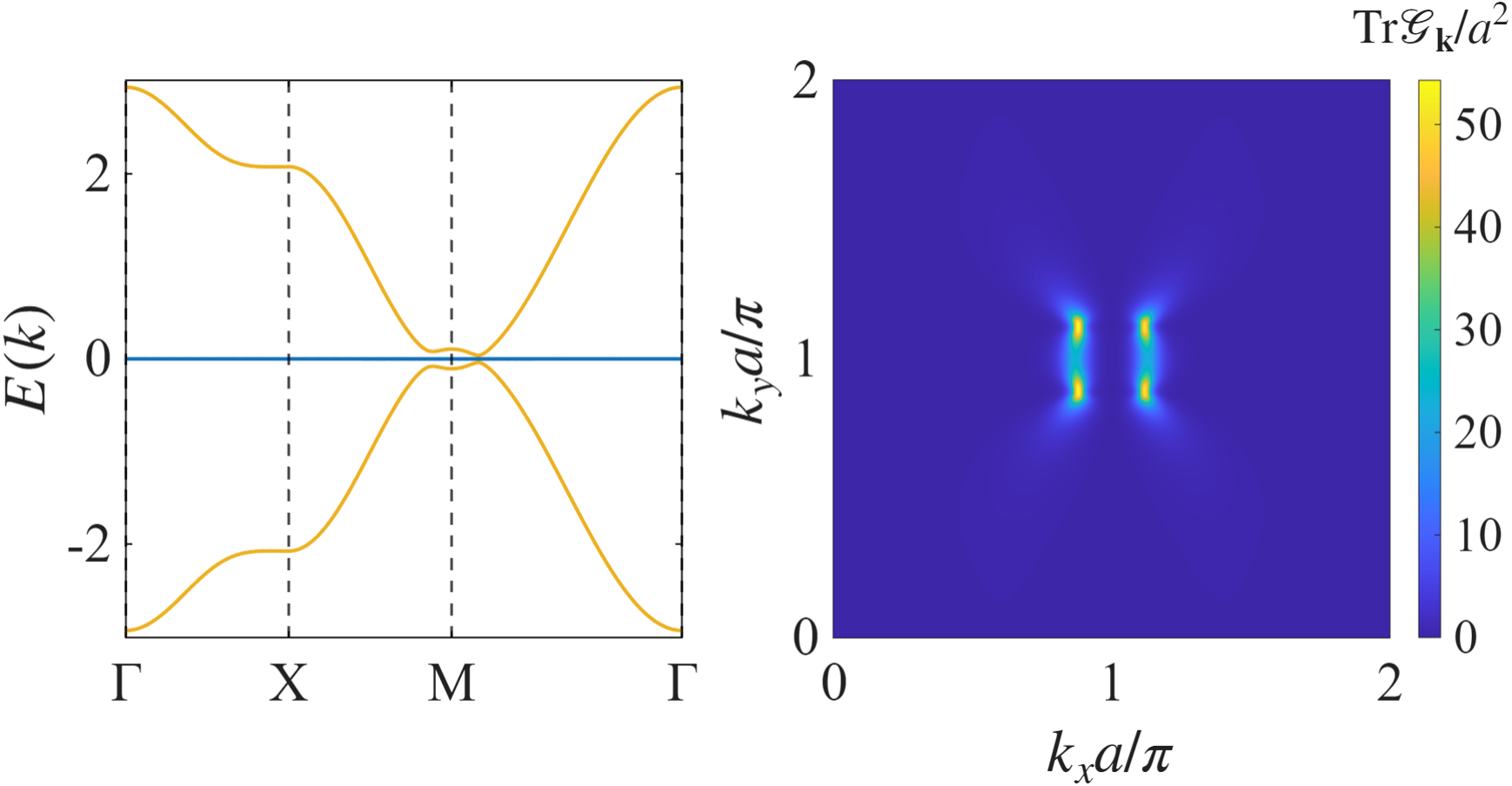}
	\caption{(a) Band structure of the 2D model. (b) Brillouin Zone distribution of traced quantum metric $\operatorname{Tr}\mathcal{G}_{\bm{k}}$. Parameters: $J = 1$, $t=10^{-4}J$, $\delta' = 5\delta$, $\mu = -0.48W^f$ with bandwidth $W^f = 8t$.
	}
	\label{Fig_2DModel_QGFOs_Band_structure}
\end{figure}

\begin{figure}[tbp]
	\centering
	\includegraphics[width=0.6\linewidth]{ 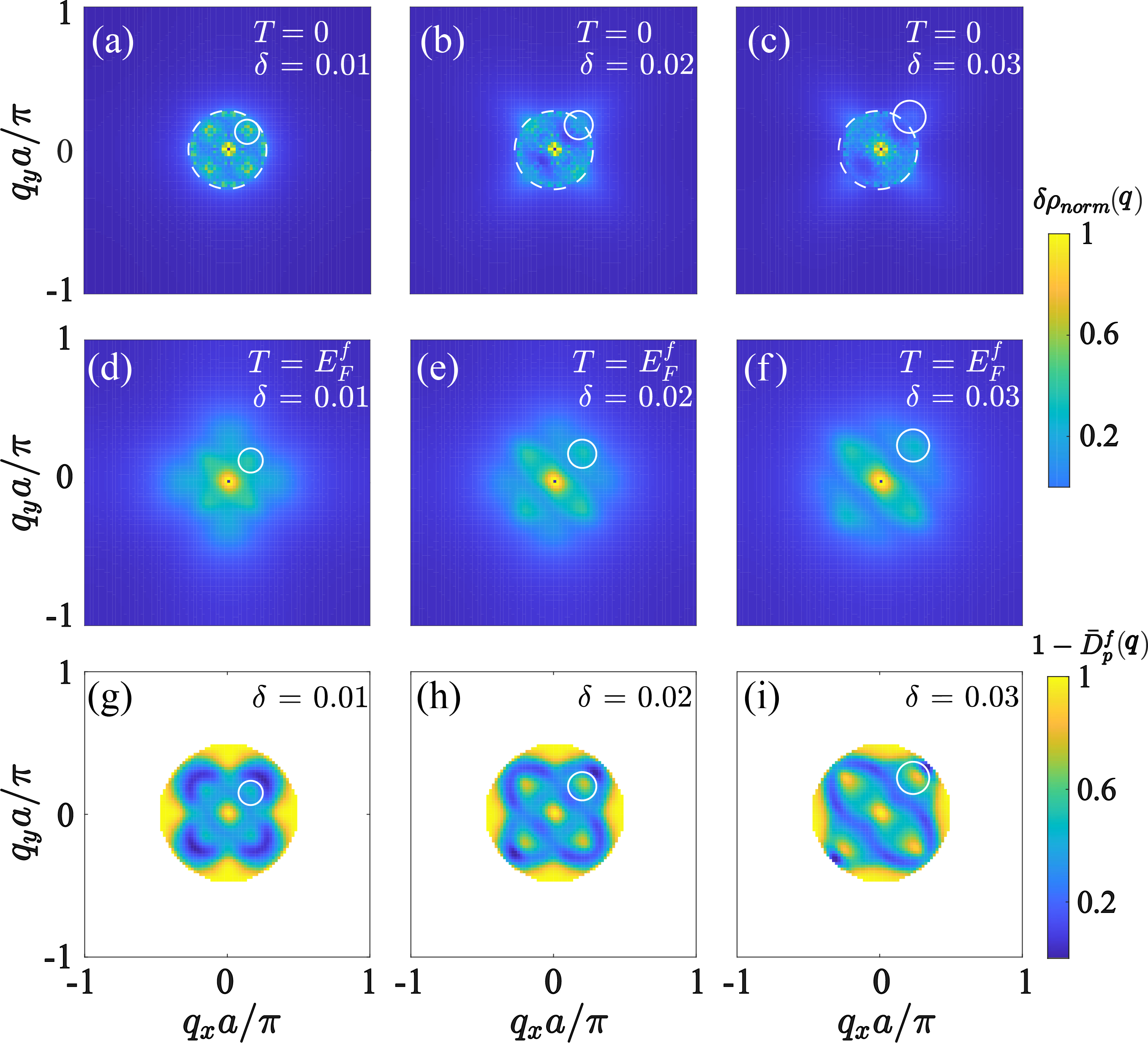}
	\caption{2D flat-band QGFOs. (a-c) Normalized LDOS variations in Fourier-space at $T=0$ for three values of $\delta$. Dashed circles mark the conventional $2k_F$ oscillations; solid circles mark the QGFOs. (d-f) Same as (a-c) at $T=E_F^f$, where the conventional oscillations are washed out and only QGFOs peaks remain. (g-i) Partially averaged quantum distance, rearranged as $1-\bar{D}^f_p(\bm{q})$; the peak locations match the QGFO positions in (d-f). Parameters: $J = 1$, $t=10^{-4}J$, $\delta' = 5\delta$, $\mu = -0.48W^f$, $E_F^f=0.02W^f$ with bandwidth $W^f = 8t$, $U_0 = t$, and the cutoff $\Lambda_c = 3.5T$. 
	}
	\label{Fig_2DModel_QGFOs_Quantum_distance}
\end{figure}

Now we further consider a two-dimensional flat-band model and examine the QGFOs in the finite temperature regime. The model shares the same $3\times3$ structure as the 1D three-band Hamiltonian with the following substitutions:
\begin{equation}
	\begin{aligned}
		%		H^{\text{2D}}(\bm{k}) &= \begin{pmatrix}
			%			0	 	& a_{\bm{k}} & b_{\bm{k}} \\
			%			a_{\bm{k}}^{\dagger}  & 0  & 0\\
			%			b_{\bm{k}}^{\dagger}  &0   & 0
			%		\end{pmatrix} + (E_f(\bm{k})-\mu) \mathcal{I}_3 \\
		%		\text{with} \quad 
		\quad a_{\bm{k}} &= [ 1+(1+\delta)e^{ik_x}/2+(1+\delta')e^{-ik_x}/2]J, \\
		b_{\bm{k}} &= [ 1+(1+\delta)e^{ik_y}/2+(1+\delta')e^{-ik_y}/2]J,
	\end{aligned}
	\label{Eq_2D_Model_Hamilt}
\end{equation}
and $ E^f_{\bm{k}} = 2t [\cos(k_x)+\cos(k_y)] -\mu$, where $0 < t,|\mu| \ll J$ again guarantees a nearly flat band. The band structure along high-symmetry points is shown in Fig.~\ref{Fig_2DModel_QGFOs_Band_structure}(a), where the flat band in the middle has a parabolic dispersion at low energy. It nearly touches the dispersive bands at the momenta of two Kramers pairs $(\pm k_g^x, \pm k_g^y)$, and the traced quantum metric $\operatorname{Tr}\mathcal{G}_{\bm{k}}$ exhibits hot spots at these momenta, as shown in panel (b). Consequently, the quantum distance between each Kramers pair exhibits a local dip, giving rise to the QGFOs when a local impurity is introduced. Fig.~\ref{Fig_2DModel_QGFOs_Quantum_distance}(a--c) show the zero-temperature Fourier-space LDOS variations normalized as $\delta\rho_{norm}(\bm{q}) = \delta\rho(\bm{q}) / \operatorname{max}[\delta\rho(\bm{q})]$ for three values of parameter $\delta$. The conventional $2k_F$ oscillations (white dashed circles) are independent of $\delta$ and disappear when temperature increases to $T = E_F^f$ [panels (d--f)], where $E_F^f \equiv E_F - E^f_{\text{min}}$ is the Fermi energy measured from the flat-band bottom [$E^f_{\text{min}/\text{max}} \equiv \operatorname{min/max} (E^f_{\bm{k}})$]. The QGFOs components (white solid circles) are $\delta$-tunable and persist at $T=E_F^f$. 
%In this temperature regime, we show in the SM~\cite{supp} that the presence of QGFOs is governed by the partially averaged quantum distance
%\begin{equation}
%	\begin{aligned}
%		\bar{D}_p^{f}(\bm{q}) = \frac{1}{\mathcal{N}} \sum_{E^f_{\text{min}} < E^f_{\bm{k}},E^f_{\bm{k}+\bm{q}} < \Lambda_c} d^f_{\bm{k}, \bm{k}+\bm{q}},
%	\end{aligned}
%	\label{Eq_PAQD}
%\end{equation}
%where $\Lambda_c = \kappa k_BT < E^f_{\text{max}}$ (with $\kappa \sim \mathcal{O}(1)$, practically $\kappa \approx 3$) is an energy cutoff chosen such that $\bar{D}_p^{f}(\bm{q})$ contains the substantially relevant low-energy quantum geometric information within the thermally accessible window ($\sim k_BT$)~\cite{supp}. The normalization factor $\mathcal{N} = \sum_{E^f_{\text{min}} < E^f_{\bm{k}},E^f_{\bm{k}+\bm{q}} < \Lambda_c} 1$. 
The results are presented in Fig.~\ref{Fig_2DModel_QGFOs_Quantum_distance}.
Panels (g--i) show $1-\bar{D}_p^{f}(\bm{q})$ for different values of parameter $\delta$, which exhibits pronounced peaking features that coincide with the QGFO locations in (d--f), confirming that QGFOs at this intermediate temperature regime are governed by the low-energy quantum geometry. Notably, unlike the 1D model, the QGFOs in the 2D model persist when both time-reversal and inversion symmetry are present, suggesting that inversion breaking is not generically required to have QGFOs.

%%%%%%%%%%%%%%%%%%%%%%%%%%%%%
%%%%%%%%%%%%%%%%%%%%%%%%%%%%%
\section{\label{Sec_Two_Band_Model}Two-band Model}

%%%%%%%%%%%%%%%%%%%%%%%%%%%%%
\begin{figure}[htbp]
	\centering
	\includegraphics[width=1\linewidth]{ 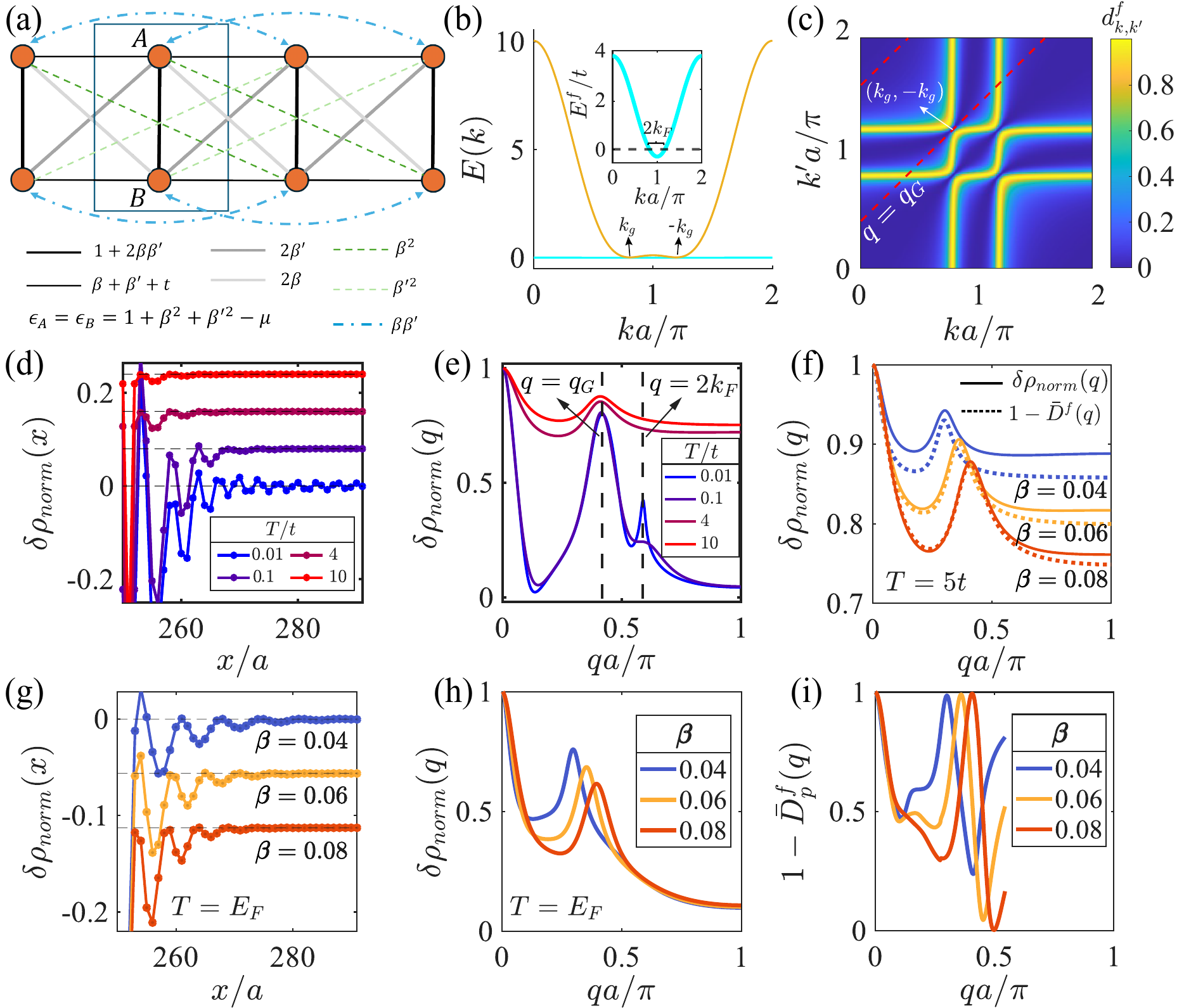}
	\caption{Two-band model with QGFOs. (a) Schematic illustration of the Lattice structure and hoppings; the solid rectangle represents a unit cell. (b) Band structure. Inset: Zoom-in of the flat band. The dashed line is the Fermi level; $\pm k_g$ denote the nearly touched Kramers pair. (c) Quantum distance $d_{k,k'}$ in the $k$-$k'$ plane. Red dashed lines highlight the cut $q=q_G$ corresponding to the peak of $\bar{D}(q)$; the white arrow marks $(k_g, -k_g)$ which connects the nearly touched Kramers pair in (a). (d) Normalized density modulations $\delta\rho_{norm}(x) \equiv \delta\rho(x)/ \rho_0$ ($\rho_0$ total density of states) near a single impurity of strength $U_0 = 10t$ at varying temperature. (e) Fourier spectra $\delta\rho_{norm}(q)$ at the same temperatures; dashed lines mark the peaking momenta at $q_G$ and $2k_F$. (f) Comparison between the QGFOs spectra (solid) and the averged quantum distance $\bar{D}^f(q)$ (dotted) for different $\beta$ at temperature $T=5t$. (g-h) QGFOs in the intermediate-temperature regime $T=E_F=0.2t$ for: (g) real-space $\delta\rho_{norm}(x)$, (h) Fourier spectra, for varying $\beta$. (i) Partially averaged quantum distance $1-\bar{D}_p^f(q)$ with energy cutoff $\Lambda_c=3.5T$. Other Parameters:  $\beta = 0.08$ for (b-e); $J=1$, $t=10^{-4}J$, $\mu = -1.8t$, $\beta' = 5\beta$ for all panels, except $\mu = -1.2t$ for (d-e); impurity strength $U_0=t$ for (f-h).}
	\label{Fig_2Band_model_QGFOs}
\end{figure}

We now present a two-band model that hosts QGFOs with a minimal set of bands, demonstrating the universality of the phenomena. The Hamiltonian is $H_{\text{2-band}} = \sum_{k,\sigma=\pm} \Psi_{k,\sigma}^{\dagger} \mathcal{H}_{\sigma}(k) \Psi_{k,\sigma}$ where $\Psi_{k,\sigma} = (c_{k,\sigma,A}, c_{k,\sigma,B})^{\operatorname{T}}$ and
\begin{equation}
	\begin{aligned}
		\mathcal{H}_{\text{2-band}}(k) = \begin{pmatrix}
			a_k + E^f_k	 	& b_{k} \\
			b_{k}^{\dagger} & a_k + E^f_k
		\end{pmatrix},
	\end{aligned}
\end{equation}
with 
\begin{equation}
	\begin{aligned}
		\quad a_{k} &  = J \left| 1+ (1+\beta)e^{-ika}/2  + (1+\beta')e^{ika}/2    \right|^2, \\
		\quad  b_{k} & = J \left( 1+ (1+\beta)e^{-ika}/2  + (1+\beta')e^{ika}/2    \right)^2, \\
		%		\quad \text{and} \quad  
		E^f_k  &= 2t \cos(ka) -\mu .
		\label{Eq_Model_Hamilt_2band}
	\end{aligned}
\end{equation}
The real-space lattice is shown in  Fig.~\ref{Fig_2Band_model_QGFOs}(a); the hoppings extend up to next-nearest neighbors and decay as a power-law with distance, controlled by the parameters $\beta,\beta' \ll 1$.
The model features a flat band with dispersion $E^f_k$ and a dispersive band $E^d_k = 2|a_k|$. Taking $t, |\mu| \ll J$ ensures the flatness of the lower band. The dimensionless parameters $\beta$ and $\beta'$ shape the quantum geometry of the flat band.
As shown in Fig.~\ref{Fig_2Band_model_QGFOs}(b), the two bands nearly touch at a Kramers pair $k = \pm k_g$. This near-degeneracy creates a sharp peak in the quantum metric, which means the flat-band Bloch vectors at these points dramatically differ from neighboring states. The quantum distance $d_{k,k'}$ [panel (c)] exhibits a set of highlighted curves connecting $k=\pm k_g$ to other momenta. In contrast to the three-band model discussed in the main text, the quantum distance between the two nearly touched points is a local minimum, resulting in a dip at $q_G \equiv 2k_g$ in the averaged quantum distance [or, a peak in $1-\bar{D}^f(q)$, as shown by the dashed lines in Fig.~\ref{Fig_2Band_model_QGFOs}(f).

Following the same method as in the main text, we compute the flat-band Friedel oscillations of this two-band model. Fig.~\ref{Fig_2Band_model_QGFOs}(d,e) shows the temperature evolution. At $T \ll W^f$, short-range oscillations exhibit two independent and modulating components, i.e., $q_G$ from the quantum geometry and $2k_F$ from band dispersion. As temperature increases, the Fermi-momentum component diminishes, while the $q_G$ feature persists even beyond the bandwidth, leaving purely quantum geometric oscillations. At $T=5t>W^f$, the Fourier spectrum $\delta\rho_{\text{norm}}(q)$ [panel (f)] agrees well with the averaged quantum distance $1-\bar{D}^f(q)$, confirming the high‑temperature relation $\delta\rho(q) \propto [1-\bar{D}^f(q)]$. Small discrepancies, most pronounced at smallest $\beta$, arise from minor deviations from the Born approximation; they diminish as $\beta$ (thus the band gap) increases. Crucially, the peak position $q_G$ remains fixed for all parameter variations (including temperature, impurity strength or fillings), as it is governed solely by the flat-band quantum geometry.
%%% we may introduce $\alpha$ to tune the polar angle of the flat-band Bloch vectors on the Bloch sphere, which, changes the quantum metric but band dispersions remain unchanged (indpendent of $\alpha$). In this case, the peaking features of the QGFOs spectra are unchanged.
In Fig.~\ref{Fig_2Band_model_QGFOs}(g--i), we further validate the QGFOs in the intermediate-temperature regime, i.e., $T=E_F \ll W^f$. The partially averaged quantum distance $1-\bar{D}_p^f(q)$ [panel (i)] accurately predicts the peak positions of the QGFO spectra in (h), although the full $q$‑space profile still carries a fingerprint of the band dispersion. This two‑band model thus confirms that QGFOs are a generic feature of flat‑band systems with nontrivial quantum geometry.

\section{\label{Sec_application}Application -- Quantum Geometric RKKY Interaction}

Magnetic impurities in a metal interact indirectly via conduction electrons, producing the RKKY interaction~\cite{ruderman1954indirect,kasuya1956theory, yosida1957magnetic}. This coupling oscillates between ferromagnetic (FM) and antiferromagnetic (AFM) with impurity separation, reflecting the underlying Friedel oscillations. When temperature dominates over electronic kinetics, the conventional RKKY interaction is washed out, particularly in flat bands~\cite{PhysRevB.108.155429,PhysRevB.104.155151,luo2024influence,PhysRevB.99.205430,oriekhov2020rkky}, allowing quantum geometry to take over. This naturally leads to a quantum geometric RKKY interaction, governed by the QGFOs introduced above.
We consider localized magnetic moments at $\bm{R}_{\text{i}} \ (i=1,2,...)$ with spin $\bm{S} = S\hat{\bm{z}}$, with a local exchange coupling to itinerant described by~\cite{PhysRevB.81.205416}:
\begin{equation}
	\begin{aligned}
		H_{\text{imp}}(\bm{R}) &= J_K \sum_{i,\alpha} \delta(\bm{R} - \bm{R}_{\text{i},\alpha}) \bm{S} \cdot \bm{\sigma}, %_{\bm{R}} ,
		\label{Eq_magnetic_imp_interaction}
	\end{aligned}
\end{equation}
where $J_K$ is the Kondo coupling and $\bm{\sigma}$ is the vector of Pauli matrices acting on the conduction-electron spin. Any two magnetic moments, say $\bm{S}_1$ at $\bm{R}$ and $\bm{S}_2$ at $\bm{R} + \bm{r}$, couple indirectly through the conduction electrons:
\begin{equation}
	\begin{aligned}
		H_{\text{RKKY}} &= \sum_{\bm{R}, \bm{r}} J_{\text{RKKY}}(\bm{r}) \bm{S}_1(\bm{R}) \cdot \bm{S}_2(\bm{R}+\bm{r}).
		\label{Eq_RKKY_interaction}
	\end{aligned}
\end{equation}
The RKKY coupling $J_{\text{RKKY}}$ is proportional to the static susceptibility~\cite{coleman2015introduction}, and therefore its spatial dependence is encoded in the susceptibilities of Eqs.~(2)--(6) in the main text%~\eqref{Eq_chi_q_FullBand}--\eqref{Eq_chi_tot_High_temperature}
, inheriting both conventional and quantum geometric oscillations.

\begin{figure}[htbp]
	\centering
	\includegraphics[width=0.4\linewidth]{ 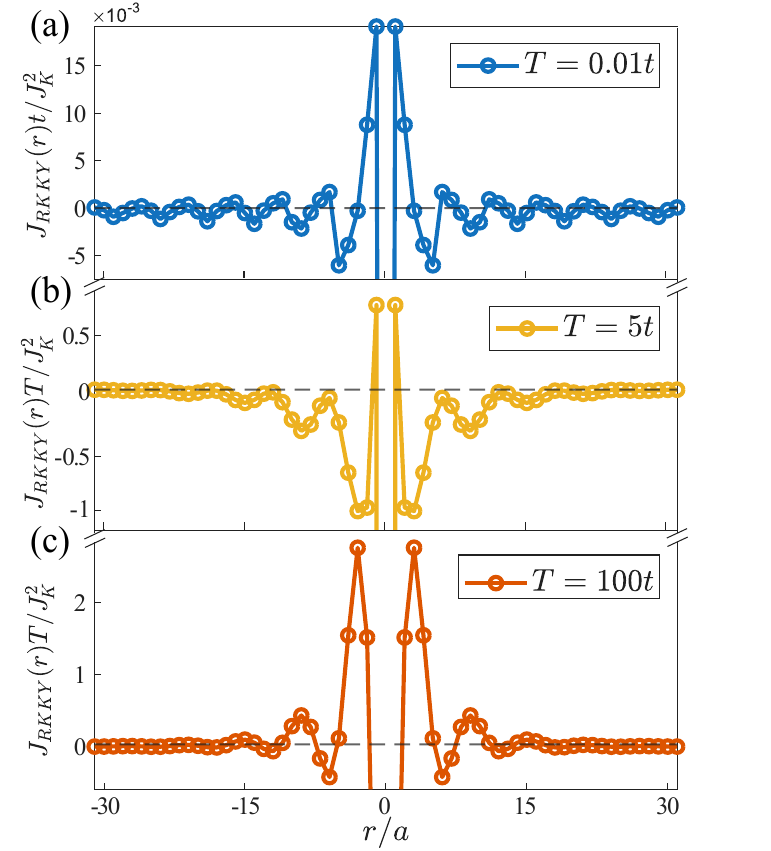}
	\caption{RKKY coupling $J_{RKKY}(r)$ as a function of distance for three temperature regimes. (a) $T\ll W^f$: conventional RKKY, power-law decay, alternating FM/AFM. (b) $W^f \lesssim T \ll   \Delta_g$: Quantum geometric RKKY dominated by flat band, exponential decay, purely FM in discrete regions. (c)$ W^f \ll T \lesssim \Delta_g$: Quantum geometric RKKY with significant interband contributions, returning alternating FM/AFM. All panels: $J = 1$, $t=10^{-4}J$, $\delta=0.05$, $\delta' = 5\delta$, $\mu = -1.6t$, and the magnetic impurities are orbital-diagonal with strength $U_0 = 2t$. $J_{RKKY}$ is normalized by $J_K^2/t$ for (a), and $J_K^2/T$ for (b) and (c).}
	%		are different for (a) and (b-c) due to distinct magnetization regimes.}
\label{Fig_J_RKKY_r}
\end{figure}

In the 1D flat-band model, we compute $J_{\text{RKKY}}$ by exact diagonalization using  $J_{\text{RKKY}} = (F_{\text{FM}}-F_{\text{AFM}})/2S^2$, where $F_{\text{FM}}$ ($F_{\text{AFM}}$) is the free energies for the FM (AFM) alignment of the two moments~\cite{deaven1991simple,PhysRevB.81.205416}. Negative (positive) $J_{\text{RKKY}}$ signifies FM (AFM) coupling.
Fig.~\ref{Fig_J_RKKY_r} shows the distance dependence of the flat band RKKY coupling strength $J_{RKKY}(r)$ at three temperature scales. At $T \ll W^f$ [Fig.~\ref{Fig_J_RKKY_r}(a)], a sharp Fermi surface gives the conventional RKKY interaction with power-law decay and alternating FM/AFM sign. When the temperature increases to be comparable with the bandwidth $T \gtrsim W^f$ [Fig.~\ref{Fig_J_RKKY_r}(b)], this conventional contribution is washed out, leaving the quantum geometric RKKY coupling. It shows exponential decay -- consistent with the QGFOs -- and a purely FM alignment in discrete spatial regions. Further increasing the temperature to the scale of the band gap $ W^f \ll T \lesssim \Delta_g$ [Fig.~\ref{Fig_J_RKKY_r}(c)], interband scattering between flat and dispersive bands becomes significant (see Sec.~\ref{Sec_DispBand_contribution}), and the oscillations revert to alternating FM/AFM with an exponentially decaying profile.
The thermal crossover highlights the decisive role of QGFOs in shaping the flat band RKKY interaction.

\bibliography{SM_Friedel_QG_Notes}

\begin{thebibliography}{80}%
\makeatletter
\providecommand \@ifxundefined [1]{%
 \@ifx{#1\undefined}
}%
\providecommand \@ifnum [1]{%
 \ifnum #1\expandafter \@firstoftwo
 \else \expandafter \@secondoftwo
 \fi
}%
\providecommand \@ifx [1]{%
 \ifx #1\expandafter \@firstoftwo
 \else \expandafter \@secondoftwo
 \fi
}%
\providecommand \natexlab [1]{#1}%
\providecommand \enquote  [1]{``#1''}%
\providecommand \bibnamefont  [1]{#1}%
\providecommand \bibfnamefont [1]{#1}%
\providecommand \citenamefont [1]{#1}%
\providecommand \href@noop [0]{\@secondoftwo}%
\providecommand \href [0]{\begingroup \@sanitize@url \@href}%
\providecommand \@href[1]{\@@startlink{#1}\@@href}%
\providecommand \@@href[1]{\endgroup#1\@@endlink}%
\providecommand \@sanitize@url [0]{\catcode `\\12\catcode `\$12\catcode
  `\&12\catcode `\#12\catcode `\^12\catcode `\_12\catcode `\%12\relax}%
\providecommand \@@startlink[1]{}%
\providecommand \@@endlink[0]{}%
\providecommand \url  [0]{\begingroup\@sanitize@url \@url }%
\providecommand \@url [1]{\endgroup\@href {#1}{\urlprefix }}%
\providecommand \urlprefix  [0]{URL }%
\providecommand \Eprint [0]{\href }%
\providecommand \doibase [0]{https://doi.org/}%
\providecommand \selectlanguage [0]{\@gobble}%
\providecommand \bibinfo  [0]{\@secondoftwo}%
\providecommand \bibfield  [0]{\@secondoftwo}%
\providecommand \translation [1]{[#1]}%
\providecommand \BibitemOpen [0]{}%
\providecommand \bibitemStop [0]{}%
\providecommand \bibitemNoStop [0]{.\EOS\space}%
\providecommand \EOS [0]{\spacefactor3000\relax}%
\providecommand \BibitemShut  [1]{\csname bibitem#1\endcsname}%
\let\auto@bib@innerbib\@empty
%</preamble>
\bibitem [{\citenamefont {Friedel}(1958)}]{friedel1958metallic}%
  \BibitemOpen
  \bibfield  {author} {\bibinfo {author} {\bibfnamefont {J.}~\bibnamefont
  {Friedel}},\ }\href@noop {} {\bibfield  {journal} {\bibinfo  {journal} {Il
  Nuovo Cimento (1955-1965)}\ }\textbf {\bibinfo {volume} {7}},\ \bibinfo
  {pages} {287} (\bibinfo {year} {1958})}\BibitemShut {NoStop}%
\bibitem [{\citenamefont {Bena}(2016)}]{bena2016friedel}%
  \BibitemOpen
  \bibfield  {author} {\bibinfo {author} {\bibfnamefont {C.}~\bibnamefont
  {Bena}},\ }\href@noop {} {\bibfield  {journal} {\bibinfo  {journal} {Comptes
  Rendus. Physique}\ }\textbf {\bibinfo {volume} {17}},\ \bibinfo {pages} {302}
  (\bibinfo {year} {2016})}\BibitemShut {NoStop}%
\bibitem [{\citenamefont {Ruderman}\ and\ \citenamefont
  {Kittel}(1954)}]{ruderman1954indirect}%
  \BibitemOpen
  \bibfield  {author} {\bibinfo {author} {\bibfnamefont {M.~A.}\ \bibnamefont
  {Ruderman}}\ and\ \bibinfo {author} {\bibfnamefont {C.}~\bibnamefont
  {Kittel}},\ }\href@noop {} {\bibfield  {journal} {\bibinfo  {journal}
  {Physical Review}\ }\textbf {\bibinfo {volume} {96}},\ \bibinfo {pages} {99}
  (\bibinfo {year} {1954})}\BibitemShut {NoStop}%
\bibitem [{\citenamefont {Kasuya}(1956)}]{kasuya1956theory}%
  \BibitemOpen
  \bibfield  {author} {\bibinfo {author} {\bibfnamefont {T.}~\bibnamefont
  {Kasuya}},\ }\href@noop {} {\bibfield  {journal} {\bibinfo  {journal}
  {Progress of theoretical physics}\ }\textbf {\bibinfo {volume} {16}},\
  \bibinfo {pages} {45} (\bibinfo {year} {1956})}\BibitemShut {NoStop}%
\bibitem [{\citenamefont {Yosida}(1957)}]{yosida1957magnetic}%
  \BibitemOpen
  \bibfield  {author} {\bibinfo {author} {\bibfnamefont {K.}~\bibnamefont
  {Yosida}},\ }\href@noop {} {\bibfield  {journal} {\bibinfo  {journal}
  {Physical Review}\ }\textbf {\bibinfo {volume} {106}},\ \bibinfo {pages}
  {893} (\bibinfo {year} {1957})}\BibitemShut {NoStop}%
\bibitem [{\citenamefont {Ashcroft}\ \emph {et~al.}(1976)\citenamefont
  {Ashcroft}, \citenamefont {Mermin} \emph {et~al.}}]{ashcroft1976solid}%
  \BibitemOpen
  \bibfield  {author} {\bibinfo {author} {\bibfnamefont {N.~W.}\ \bibnamefont
  {Ashcroft}}, \bibinfo {author} {\bibfnamefont {N.~D.}\ \bibnamefont
  {Mermin}}, \emph {et~al.},\ }\href@noop {} {\bibinfo {title} {Solid state
  physics}} (\bibinfo {year} {1976})\BibitemShut {NoStop}%
\bibitem [{\citenamefont {Kohn}\ and\ \citenamefont
  {Luttinger}(1965)}]{kohn1965new}%
  \BibitemOpen
  \bibfield  {author} {\bibinfo {author} {\bibfnamefont {W.}~\bibnamefont
  {Kohn}}\ and\ \bibinfo {author} {\bibfnamefont {J.}~\bibnamefont
  {Luttinger}},\ }\href@noop {} {\bibfield  {journal} {\bibinfo  {journal}
  {Physical Review Letters}\ }\textbf {\bibinfo {volume} {15}},\ \bibinfo
  {pages} {524} (\bibinfo {year} {1965})}\BibitemShut {NoStop}%
\bibitem [{\citenamefont {Simion}\ and\ \citenamefont
  {Giuliani}(2005)}]{PhysRevB.72.045127}%
  \BibitemOpen
  \bibfield  {author} {\bibinfo {author} {\bibfnamefont {G.~E.}\ \bibnamefont
  {Simion}}\ and\ \bibinfo {author} {\bibfnamefont {G.~F.}\ \bibnamefont
  {Giuliani}},\ }\href@noop {} {\bibfield  {journal} {\bibinfo  {journal}
  {Physical Review B}\ }\textbf {\bibinfo {volume} {72}},\ \bibinfo {pages}
  {045127} (\bibinfo {year} {2005})}\BibitemShut {NoStop}%
\bibitem [{\citenamefont {Affleck}\ \emph {et~al.}(2008)\citenamefont
  {Affleck}, \citenamefont {Borda},\ and\ \citenamefont
  {Saleur}}]{affleck2008friedel}%
  \BibitemOpen
  \bibfield  {author} {\bibinfo {author} {\bibfnamefont {I.}~\bibnamefont
  {Affleck}}, \bibinfo {author} {\bibfnamefont {L.}~\bibnamefont {Borda}},\
  and\ \bibinfo {author} {\bibfnamefont {H.}~\bibnamefont {Saleur}},\
  }\href@noop {} {\bibfield  {journal} {\bibinfo  {journal} {Physical Review
  B}\ }\textbf {\bibinfo {volume} {77}},\ \bibinfo {pages} {180404} (\bibinfo
  {year} {2008})}\BibitemShut {NoStop}%
\bibitem [{\citenamefont {Grassme}\ and\ \citenamefont
  {Bussemer}(1993)}]{grassme1993friedel}%
  \BibitemOpen
  \bibfield  {author} {\bibinfo {author} {\bibfnamefont {R.}~\bibnamefont
  {Grassme}}\ and\ \bibinfo {author} {\bibfnamefont {P.}~\bibnamefont
  {Bussemer}},\ }\href@noop {} {\bibfield  {journal} {\bibinfo  {journal}
  {Physics Letters A}\ }\textbf {\bibinfo {volume} {175}},\ \bibinfo {pages}
  {441} (\bibinfo {year} {1993})}\BibitemShut {NoStop}%
\bibitem [{\citenamefont {Cavaliere}\ \emph {et~al.}(2014)\citenamefont
  {Cavaliere}, \citenamefont {Ziani}, \citenamefont {Negro},\ and\
  \citenamefont {Sassetti}}]{cavaliere2014thermally}%
  \BibitemOpen
  \bibfield  {author} {\bibinfo {author} {\bibfnamefont {F.}~\bibnamefont
  {Cavaliere}}, \bibinfo {author} {\bibfnamefont {N.~T.}\ \bibnamefont
  {Ziani}}, \bibinfo {author} {\bibfnamefont {F.}~\bibnamefont {Negro}},\ and\
  \bibinfo {author} {\bibfnamefont {M.}~\bibnamefont {Sassetti}},\ }\href@noop
  {} {\bibfield  {journal} {\bibinfo  {journal} {Journal of Physics: Condensed
  Matter}\ }\textbf {\bibinfo {volume} {26}},\ \bibinfo {pages} {505301}
  (\bibinfo {year} {2014})}\BibitemShut {NoStop}%
\bibitem [{\citenamefont {Egger}\ and\ \citenamefont
  {Grabert}(1996)}]{egger1996friedel}%
  \BibitemOpen
  \bibfield  {author} {\bibinfo {author} {\bibfnamefont {R.}~\bibnamefont
  {Egger}}\ and\ \bibinfo {author} {\bibfnamefont {H.}~\bibnamefont
  {Grabert}},\ }in\ \href@noop {} {\emph {\bibinfo {booktitle} {Quantum
  transport in semiconductor submicron structures}}}\ (\bibinfo  {publisher}
  {Springer},\ \bibinfo {year} {1996})\ pp.\ \bibinfo {pages}
  {133--158}\BibitemShut {NoStop}%
\bibitem [{\citenamefont {Geim}\ and\ \citenamefont
  {Grigorieva}(2013)}]{geim2013van}%
  \BibitemOpen
  \bibfield  {author} {\bibinfo {author} {\bibfnamefont {A.~K.}\ \bibnamefont
  {Geim}}\ and\ \bibinfo {author} {\bibfnamefont {I.~V.}\ \bibnamefont
  {Grigorieva}},\ }\href@noop {} {\bibfield  {journal} {\bibinfo  {journal}
  {Nature}\ }\textbf {\bibinfo {volume} {499}},\ \bibinfo {pages} {419}
  (\bibinfo {year} {2013})}\BibitemShut {NoStop}%
\bibitem [{\citenamefont {Cao}\ \emph {et~al.}(2018)\citenamefont {Cao},
  \citenamefont {Fatemi}, \citenamefont {Fang}, \citenamefont {Watanabe},
  \citenamefont {Taniguchi}, \citenamefont {Kaxiras},\ and\ \citenamefont
  {Jarillo-Herrero}}]{cao2018unconventional}%
  \BibitemOpen
  \bibfield  {author} {\bibinfo {author} {\bibfnamefont {Y.}~\bibnamefont
  {Cao}}, \bibinfo {author} {\bibfnamefont {V.}~\bibnamefont {Fatemi}},
  \bibinfo {author} {\bibfnamefont {S.}~\bibnamefont {Fang}}, \bibinfo {author}
  {\bibfnamefont {K.}~\bibnamefont {Watanabe}}, \bibinfo {author}
  {\bibfnamefont {T.}~\bibnamefont {Taniguchi}}, \bibinfo {author}
  {\bibfnamefont {E.}~\bibnamefont {Kaxiras}},\ and\ \bibinfo {author}
  {\bibfnamefont {P.}~\bibnamefont {Jarillo-Herrero}},\ }\href@noop {}
  {\bibfield  {journal} {\bibinfo  {journal} {Nature}\ }\textbf {\bibinfo
  {volume} {556}},\ \bibinfo {pages} {43} (\bibinfo {year} {2018})}\BibitemShut
  {NoStop}%
\bibitem [{\citenamefont {Cai}\ \emph {et~al.}(2023)\citenamefont {Cai},
  \citenamefont {Anderson}, \citenamefont {Wang}, \citenamefont {Zhang},
  \citenamefont {Liu}, \citenamefont {Holtzmann}, \citenamefont {Zhang},
  \citenamefont {Fan}, \citenamefont {Taniguchi}, \citenamefont {Watanabe}
  \emph {et~al.}}]{cai2023signatures}%
  \BibitemOpen
  \bibfield  {author} {\bibinfo {author} {\bibfnamefont {J.}~\bibnamefont
  {Cai}}, \bibinfo {author} {\bibfnamefont {E.}~\bibnamefont {Anderson}},
  \bibinfo {author} {\bibfnamefont {C.}~\bibnamefont {Wang}}, \bibinfo {author}
  {\bibfnamefont {X.}~\bibnamefont {Zhang}}, \bibinfo {author} {\bibfnamefont
  {X.}~\bibnamefont {Liu}}, \bibinfo {author} {\bibfnamefont {W.}~\bibnamefont
  {Holtzmann}}, \bibinfo {author} {\bibfnamefont {Y.}~\bibnamefont {Zhang}},
  \bibinfo {author} {\bibfnamefont {F.}~\bibnamefont {Fan}}, \bibinfo {author}
  {\bibfnamefont {T.}~\bibnamefont {Taniguchi}}, \bibinfo {author}
  {\bibfnamefont {K.}~\bibnamefont {Watanabe}}, \emph {et~al.},\ }\href@noop {}
  {\bibfield  {journal} {\bibinfo  {journal} {Nature}\ }\textbf {\bibinfo
  {volume} {622}},\ \bibinfo {pages} {63} (\bibinfo {year} {2023})}\BibitemShut
  {NoStop}%
\bibitem [{\citenamefont {Leykam}\ \emph {et~al.}(2018)\citenamefont {Leykam},
  \citenamefont {Andreanov},\ and\ \citenamefont
  {Flach}}]{leykam2018artificial}%
  \BibitemOpen
  \bibfield  {author} {\bibinfo {author} {\bibfnamefont {D.}~\bibnamefont
  {Leykam}}, \bibinfo {author} {\bibfnamefont {A.}~\bibnamefont {Andreanov}},\
  and\ \bibinfo {author} {\bibfnamefont {S.}~\bibnamefont {Flach}},\
  }\href@noop {} {\bibfield  {journal} {\bibinfo  {journal} {Advances in
  Physics: X}\ }\textbf {\bibinfo {volume} {3}},\ \bibinfo {pages} {1473052}
  (\bibinfo {year} {2018})}\BibitemShut {NoStop}%
\bibitem [{\citenamefont {Neves}\ \emph {et~al.}(2024)\citenamefont {Neves},
  \citenamefont {Wakefield}, \citenamefont {Fang}, \citenamefont {Nguyen},
  \citenamefont {Ye},\ and\ \citenamefont {Checkelsky}}]{neves2024crystal}%
  \BibitemOpen
  \bibfield  {author} {\bibinfo {author} {\bibfnamefont {P.~M.}\ \bibnamefont
  {Neves}}, \bibinfo {author} {\bibfnamefont {J.~P.}\ \bibnamefont
  {Wakefield}}, \bibinfo {author} {\bibfnamefont {S.}~\bibnamefont {Fang}},
  \bibinfo {author} {\bibfnamefont {H.}~\bibnamefont {Nguyen}}, \bibinfo
  {author} {\bibfnamefont {L.}~\bibnamefont {Ye}},\ and\ \bibinfo {author}
  {\bibfnamefont {J.~G.}\ \bibnamefont {Checkelsky}},\ }\href@noop {}
  {\bibfield  {journal} {\bibinfo  {journal} {npj Computational Materials}\
  }\textbf {\bibinfo {volume} {10}},\ \bibinfo {pages} {39} (\bibinfo {year}
  {2024})}\BibitemShut {NoStop}%
\bibitem [{\citenamefont {Ye}\ \emph {et~al.}(2024)\citenamefont {Ye},
  \citenamefont {Fang}, \citenamefont {Kang}, \citenamefont {Kaufmann},
  \citenamefont {Lee}, \citenamefont {John}, \citenamefont {Neves},
  \citenamefont {Zhao}, \citenamefont {Denlinger}, \citenamefont {Jozwiak}
  \emph {et~al.}}]{ye2024hopping}%
  \BibitemOpen
  \bibfield  {author} {\bibinfo {author} {\bibfnamefont {L.}~\bibnamefont
  {Ye}}, \bibinfo {author} {\bibfnamefont {S.}~\bibnamefont {Fang}}, \bibinfo
  {author} {\bibfnamefont {M.}~\bibnamefont {Kang}}, \bibinfo {author}
  {\bibfnamefont {J.}~\bibnamefont {Kaufmann}}, \bibinfo {author}
  {\bibfnamefont {Y.}~\bibnamefont {Lee}}, \bibinfo {author} {\bibfnamefont
  {C.}~\bibnamefont {John}}, \bibinfo {author} {\bibfnamefont {P.~M.}\
  \bibnamefont {Neves}}, \bibinfo {author} {\bibfnamefont {S.~F.}\ \bibnamefont
  {Zhao}}, \bibinfo {author} {\bibfnamefont {J.}~\bibnamefont {Denlinger}},
  \bibinfo {author} {\bibfnamefont {C.}~\bibnamefont {Jozwiak}}, \emph
  {et~al.},\ }\href@noop {} {\bibfield  {journal} {\bibinfo  {journal} {Nature
  Physics}\ }\textbf {\bibinfo {volume} {20}},\ \bibinfo {pages} {610}
  (\bibinfo {year} {2024})}\BibitemShut {NoStop}%
\bibitem [{\citenamefont {Provost}\ and\ \citenamefont
  {Vallee}(1980)}]{provost1980riemannian}%
  \BibitemOpen
  \bibfield  {author} {\bibinfo {author} {\bibfnamefont {J.}~\bibnamefont
  {Provost}}\ and\ \bibinfo {author} {\bibfnamefont {G.}~\bibnamefont
  {Vallee}},\ }\href@noop {} {\bibfield  {journal} {\bibinfo  {journal}
  {Communications in Mathematical Physics}\ }\textbf {\bibinfo {volume} {76}},\
  \bibinfo {pages} {289} (\bibinfo {year} {1980})}\BibitemShut {NoStop}%
\bibitem [{\citenamefont {Resta}(2011)}]{resta2011insulating}%
  \BibitemOpen
  \bibfield  {author} {\bibinfo {author} {\bibfnamefont {R.}~\bibnamefont
  {Resta}},\ }\href@noop {} {\bibfield  {journal} {\bibinfo  {journal} {The
  European Physical Journal B}\ }\textbf {\bibinfo {volume} {79}},\ \bibinfo
  {pages} {121} (\bibinfo {year} {2011})}\BibitemShut {NoStop}%
\bibitem [{\citenamefont {T{\"o}rm{\"a}}(2023)}]{torma2023essay}%
  \BibitemOpen
  \bibfield  {author} {\bibinfo {author} {\bibfnamefont {P.}~\bibnamefont
  {T{\"o}rm{\"a}}},\ }\href@noop {} {\bibfield  {journal} {\bibinfo  {journal}
  {Physical Review Letters}\ }\textbf {\bibinfo {volume} {131}},\ \bibinfo
  {pages} {240001} (\bibinfo {year} {2023})}\BibitemShut {NoStop}%
\bibitem [{\citenamefont {Yu}\ \emph {et~al.}(2025)\citenamefont {Yu},
  \citenamefont {Bernevig}, \citenamefont {Queiroz}, \citenamefont {Rossi},
  \citenamefont {T{\"o}rm{\"a}},\ and\ \citenamefont {Yang}}]{yu2025quantum}%
  \BibitemOpen
  \bibfield  {author} {\bibinfo {author} {\bibfnamefont {J.}~\bibnamefont
  {Yu}}, \bibinfo {author} {\bibfnamefont {B.~A.}\ \bibnamefont {Bernevig}},
  \bibinfo {author} {\bibfnamefont {R.}~\bibnamefont {Queiroz}}, \bibinfo
  {author} {\bibfnamefont {E.}~\bibnamefont {Rossi}}, \bibinfo {author}
  {\bibfnamefont {P.}~\bibnamefont {T{\"o}rm{\"a}}},\ and\ \bibinfo {author}
  {\bibfnamefont {B.-J.}\ \bibnamefont {Yang}},\ }\href@noop {} {\bibfield
  {journal} {\bibinfo  {journal} {npj Quantum Materials}\ }\textbf {\bibinfo
  {volume} {10}},\ \bibinfo {pages} {101} (\bibinfo {year} {2025})}\BibitemShut
  {NoStop}%
\bibitem [{\citenamefont {Liu}\ \emph {et~al.}(2025{\natexlab{a}})\citenamefont
  {Liu}, \citenamefont {Qiang}, \citenamefont {Lu},\ and\ \citenamefont
  {Xie}}]{liu2025quantum}%
  \BibitemOpen
  \bibfield  {author} {\bibinfo {author} {\bibfnamefont {T.}~\bibnamefont
  {Liu}}, \bibinfo {author} {\bibfnamefont {X.-B.}\ \bibnamefont {Qiang}},
  \bibinfo {author} {\bibfnamefont {H.-Z.}\ \bibnamefont {Lu}},\ and\ \bibinfo
  {author} {\bibfnamefont {X.}~\bibnamefont {Xie}},\ }\href@noop {} {\bibfield
  {journal} {\bibinfo  {journal} {National Science Review}\ }\textbf {\bibinfo
  {volume} {12}},\ \bibinfo {pages} {nwae334} (\bibinfo {year}
  {2025}{\natexlab{a}})}\BibitemShut {NoStop}%
\bibitem [{\citenamefont {Verma}\ \emph {et~al.}(2026)\citenamefont {Verma},
  \citenamefont {Moll}, \citenamefont {Holder},\ and\ \citenamefont
  {Queiroz}}]{verma2026quantum}%
  \BibitemOpen
  \bibfield  {author} {\bibinfo {author} {\bibfnamefont {N.}~\bibnamefont
  {Verma}}, \bibinfo {author} {\bibfnamefont {P.~J.}\ \bibnamefont {Moll}},
  \bibinfo {author} {\bibfnamefont {T.}~\bibnamefont {Holder}},\ and\ \bibinfo
  {author} {\bibfnamefont {R.}~\bibnamefont {Queiroz}},\ }\href@noop {}
  {\bibfield  {journal} {\bibinfo  {journal} {Nature Reviews Physics}\ ,\
  \bibinfo {pages} {1}} (\bibinfo {year} {2026})}\BibitemShut {NoStop}%
\bibitem [{\citenamefont {Peotta}\ and\ \citenamefont
  {T{\"o}rm{\"a}}(2015)}]{peotta2015superfluidity}%
  \BibitemOpen
  \bibfield  {author} {\bibinfo {author} {\bibfnamefont {S.}~\bibnamefont
  {Peotta}}\ and\ \bibinfo {author} {\bibfnamefont {P.}~\bibnamefont
  {T{\"o}rm{\"a}}},\ }\href@noop {} {\bibfield  {journal} {\bibinfo  {journal}
  {Nature Communications}\ }\textbf {\bibinfo {volume} {6}},\ \bibinfo {pages}
  {8944} (\bibinfo {year} {2015})}\BibitemShut {NoStop}%
\bibitem [{\citenamefont {Julku}\ \emph {et~al.}(2016)\citenamefont {Julku},
  \citenamefont {Peotta}, \citenamefont {Vanhala}, \citenamefont {Kim},\ and\
  \citenamefont {T\"orm\"a}}]{PhysRevLett117_045303}%
  \BibitemOpen
  \bibfield  {author} {\bibinfo {author} {\bibfnamefont {A.}~\bibnamefont
  {Julku}}, \bibinfo {author} {\bibfnamefont {S.}~\bibnamefont {Peotta}},
  \bibinfo {author} {\bibfnamefont {T.~I.}\ \bibnamefont {Vanhala}}, \bibinfo
  {author} {\bibfnamefont {D.-H.}\ \bibnamefont {Kim}},\ and\ \bibinfo {author}
  {\bibfnamefont {P.}~\bibnamefont {T\"orm\"a}},\ }\href
  {https://doi.org/10.1103/PhysRevLett.117.045303} {\bibfield  {journal}
  {\bibinfo  {journal} {Physical Review Letters}\ }\textbf {\bibinfo {volume}
  {117}},\ \bibinfo {pages} {045303} (\bibinfo {year} {2016})}\BibitemShut
  {NoStop}%
\bibitem [{\citenamefont {Hu}\ \emph {et~al.}(2019)\citenamefont {Hu},
  \citenamefont {Hyart}, \citenamefont {Pikulin},\ and\ \citenamefont
  {Rossi}}]{PhysRevLett.123.237002}%
  \BibitemOpen
  \bibfield  {author} {\bibinfo {author} {\bibfnamefont {X.}~\bibnamefont
  {Hu}}, \bibinfo {author} {\bibfnamefont {T.}~\bibnamefont {Hyart}}, \bibinfo
  {author} {\bibfnamefont {D.~I.}\ \bibnamefont {Pikulin}},\ and\ \bibinfo
  {author} {\bibfnamefont {E.}~\bibnamefont {Rossi}},\ }\href@noop {}
  {\bibfield  {journal} {\bibinfo  {journal} {Physical Review Letters}\
  }\textbf {\bibinfo {volume} {123}},\ \bibinfo {pages} {237002} (\bibinfo
  {year} {2019})}\BibitemShut {NoStop}%
\bibitem [{\citenamefont {Xie}\ \emph {et~al.}(2020)\citenamefont {Xie},
  \citenamefont {Song}, \citenamefont {Lian},\ and\ \citenamefont
  {Bernevig}}]{xie2020topology}%
  \BibitemOpen
  \bibfield  {author} {\bibinfo {author} {\bibfnamefont {F.}~\bibnamefont
  {Xie}}, \bibinfo {author} {\bibfnamefont {Z.}~\bibnamefont {Song}}, \bibinfo
  {author} {\bibfnamefont {B.}~\bibnamefont {Lian}},\ and\ \bibinfo {author}
  {\bibfnamefont {B.~A.}\ \bibnamefont {Bernevig}},\ }\href@noop {} {\bibfield
  {journal} {\bibinfo  {journal} {Physical Review Letters}\ }\textbf {\bibinfo
  {volume} {124}},\ \bibinfo {pages} {167002} (\bibinfo {year}
  {2020})}\BibitemShut {NoStop}%
\bibitem [{\citenamefont {Herzog-Arbeitman}\ \emph {et~al.}(2022)\citenamefont
  {Herzog-Arbeitman}, \citenamefont {Peri}, \citenamefont {Schindler},
  \citenamefont {Huber},\ and\ \citenamefont
  {Bernevig}}]{herzog2022superfluid}%
  \BibitemOpen
  \bibfield  {author} {\bibinfo {author} {\bibfnamefont {J.}~\bibnamefont
  {Herzog-Arbeitman}}, \bibinfo {author} {\bibfnamefont {V.}~\bibnamefont
  {Peri}}, \bibinfo {author} {\bibfnamefont {F.}~\bibnamefont {Schindler}},
  \bibinfo {author} {\bibfnamefont {S.~D.}\ \bibnamefont {Huber}},\ and\
  \bibinfo {author} {\bibfnamefont {B.~A.}\ \bibnamefont {Bernevig}},\
  }\href@noop {} {\bibfield  {journal} {\bibinfo  {journal} {Physical Review
  Letters}\ }\textbf {\bibinfo {volume} {128}},\ \bibinfo {pages} {087002}
  (\bibinfo {year} {2022})}\BibitemShut {NoStop}%
\bibitem [{\citenamefont {T{\"o}rm{\"a}}\ \emph {et~al.}(2022)\citenamefont
  {T{\"o}rm{\"a}}, \citenamefont {Peotta},\ and\ \citenamefont
  {Bernevig}}]{torma2022superconductivity}%
  \BibitemOpen
  \bibfield  {author} {\bibinfo {author} {\bibfnamefont {P.}~\bibnamefont
  {T{\"o}rm{\"a}}}, \bibinfo {author} {\bibfnamefont {S.}~\bibnamefont
  {Peotta}},\ and\ \bibinfo {author} {\bibfnamefont {B.~A.}\ \bibnamefont
  {Bernevig}},\ }\href@noop {} {\bibfield  {journal} {\bibinfo  {journal}
  {Nature Reviews Physics}\ }\textbf {\bibinfo {volume} {4}},\ \bibinfo {pages}
  {528} (\bibinfo {year} {2022})}\BibitemShut {NoStop}%
\bibitem [{\citenamefont {Chen}\ and\ \citenamefont
  {Law}(2024)}]{PhysRevLett.132.026002}%
  \BibitemOpen
  \bibfield  {author} {\bibinfo {author} {\bibfnamefont {S.~A.}\ \bibnamefont
  {Chen}}\ and\ \bibinfo {author} {\bibfnamefont {K.}~\bibnamefont {Law}},\
  }\href@noop {} {\bibfield  {journal} {\bibinfo  {journal} {Physical Review
  Letters}\ }\textbf {\bibinfo {volume} {132}},\ \bibinfo {pages} {026002}
  (\bibinfo {year} {2024})}\BibitemShut {NoStop}%
\bibitem [{\citenamefont {Fang}\ \emph {et~al.}(2024)\citenamefont {Fang},
  \citenamefont {Cano},\ and\ \citenamefont
  {Ghorashi}}]{PhysRevLett133_106701}%
  \BibitemOpen
  \bibfield  {author} {\bibinfo {author} {\bibfnamefont {Y.}~\bibnamefont
  {Fang}}, \bibinfo {author} {\bibfnamefont {J.}~\bibnamefont {Cano}},\ and\
  \bibinfo {author} {\bibfnamefont {S.~A.~A.}\ \bibnamefont {Ghorashi}},\
  }\href {https://doi.org/10.1103/PhysRevLett.133.106701} {\bibfield  {journal}
  {\bibinfo  {journal} {Physical Review Letters}\ }\textbf {\bibinfo {volume}
  {133}},\ \bibinfo {pages} {106701} (\bibinfo {year} {2024})}\BibitemShut
  {NoStop}%
\bibitem [{\citenamefont {Du}\ \emph {et~al.}(2021)\citenamefont {Du},
  \citenamefont {Wang}, \citenamefont {Sun}, \citenamefont {Lu},\ and\
  \citenamefont {Xie}}]{du2021quantum}%
  \BibitemOpen
  \bibfield  {author} {\bibinfo {author} {\bibfnamefont {Z.}~\bibnamefont
  {Du}}, \bibinfo {author} {\bibfnamefont {C.}~\bibnamefont {Wang}}, \bibinfo
  {author} {\bibfnamefont {H.-P.}\ \bibnamefont {Sun}}, \bibinfo {author}
  {\bibfnamefont {H.-Z.}\ \bibnamefont {Lu}},\ and\ \bibinfo {author}
  {\bibfnamefont {X.}~\bibnamefont {Xie}},\ }\href@noop {} {\bibfield
  {journal} {\bibinfo  {journal} {Nature Communications}\ }\textbf {\bibinfo
  {volume} {12}},\ \bibinfo {pages} {5038} (\bibinfo {year}
  {2021})}\BibitemShut {NoStop}%
\bibitem [{\citenamefont {Jiang}\ \emph {et~al.}(2025)\citenamefont {Jiang},
  \citenamefont {Holder},\ and\ \citenamefont {Yan}}]{jiang2025revealing}%
  \BibitemOpen
  \bibfield  {author} {\bibinfo {author} {\bibfnamefont {Y.}~\bibnamefont
  {Jiang}}, \bibinfo {author} {\bibfnamefont {T.}~\bibnamefont {Holder}},\ and\
  \bibinfo {author} {\bibfnamefont {B.}~\bibnamefont {Yan}},\ }\href@noop {}
  {\bibfield  {journal} {\bibinfo  {journal} {Reports on Progress in Physics}\
  }\textbf {\bibinfo {volume} {88}},\ \bibinfo {pages} {076502} (\bibinfo
  {year} {2025})}\BibitemShut {NoStop}%
\bibitem [{\citenamefont {Souza}\ \emph {et~al.}(2000)\citenamefont {Souza},
  \citenamefont {Wilkens},\ and\ \citenamefont
  {Martin}}]{souza2000polarization}%
  \BibitemOpen
  \bibfield  {author} {\bibinfo {author} {\bibfnamefont {I.}~\bibnamefont
  {Souza}}, \bibinfo {author} {\bibfnamefont {T.}~\bibnamefont {Wilkens}},\
  and\ \bibinfo {author} {\bibfnamefont {R.~M.}\ \bibnamefont {Martin}},\
  }\href@noop {} {\bibfield  {journal} {\bibinfo  {journal} {Physical Review
  B}\ }\textbf {\bibinfo {volume} {62}},\ \bibinfo {pages} {1666} (\bibinfo
  {year} {2000})}\BibitemShut {NoStop}%
\bibitem [{\citenamefont {Onishi}\ and\ \citenamefont
  {Fu}(2025)}]{PhysRevResearch.7.023158}%
  \BibitemOpen
  \bibfield  {author} {\bibinfo {author} {\bibfnamefont {Y.}~\bibnamefont
  {Onishi}}\ and\ \bibinfo {author} {\bibfnamefont {L.}~\bibnamefont {Fu}},\
  }\href@noop {} {\bibfield  {journal} {\bibinfo  {journal} {Physical Review
  Research}\ }\textbf {\bibinfo {volume} {7}},\ \bibinfo {pages} {023158}
  (\bibinfo {year} {2025})}\BibitemShut {NoStop}%
\bibitem [{\citenamefont {Ahn}\ \emph {et~al.}(2022)\citenamefont {Ahn},
  \citenamefont {Guo}, \citenamefont {Nagaosa},\ and\ \citenamefont
  {Vishwanath}}]{ahn2022riemannian}%
  \BibitemOpen
  \bibfield  {author} {\bibinfo {author} {\bibfnamefont {J.}~\bibnamefont
  {Ahn}}, \bibinfo {author} {\bibfnamefont {G.-Y.}\ \bibnamefont {Guo}},
  \bibinfo {author} {\bibfnamefont {N.}~\bibnamefont {Nagaosa}},\ and\ \bibinfo
  {author} {\bibfnamefont {A.}~\bibnamefont {Vishwanath}},\ }\href@noop {}
  {\bibfield  {journal} {\bibinfo  {journal} {Nature Physics}\ }\textbf
  {\bibinfo {volume} {18}},\ \bibinfo {pages} {290} (\bibinfo {year}
  {2022})}\BibitemShut {NoStop}%
\bibitem [{\citenamefont {Verma}\ and\ \citenamefont
  {Queiroz}(2025)}]{verma2025instantaneous}%
  \BibitemOpen
  \bibfield  {author} {\bibinfo {author} {\bibfnamefont {N.}~\bibnamefont
  {Verma}}\ and\ \bibinfo {author} {\bibfnamefont {R.}~\bibnamefont
  {Queiroz}},\ }\href@noop {} {\bibfield  {journal} {\bibinfo  {journal}
  {Proceedings of the National Academy of Sciences}\ }\textbf {\bibinfo
  {volume} {122}},\ \bibinfo {pages} {e2405837122} (\bibinfo {year}
  {2025})}\BibitemShut {NoStop}%
\bibitem [{\citenamefont {Han}\ \emph {et~al.}(2024)\citenamefont {Han},
  \citenamefont {Herzog-Arbeitman}, \citenamefont {Bernevig},\ and\
  \citenamefont {Kivelson}}]{PhysRevX.14.041004}%
  \BibitemOpen
  \bibfield  {author} {\bibinfo {author} {\bibfnamefont {Z.}~\bibnamefont
  {Han}}, \bibinfo {author} {\bibfnamefont {J.}~\bibnamefont
  {Herzog-Arbeitman}}, \bibinfo {author} {\bibfnamefont {B.~A.}\ \bibnamefont
  {Bernevig}},\ and\ \bibinfo {author} {\bibfnamefont {S.~A.}\ \bibnamefont
  {Kivelson}},\ }\href@noop {} {\bibfield  {journal} {\bibinfo  {journal}
  {Physical Review X}\ }\textbf {\bibinfo {volume} {14}},\ \bibinfo {pages}
  {041004} (\bibinfo {year} {2024})}\BibitemShut {NoStop}%
\bibitem [{\citenamefont {Zhang}\ \emph {et~al.}(2026)\citenamefont {Zhang},
  \citenamefont {Wang}, \citenamefont {Balents},\ and\ \citenamefont
  {Savary}}]{zhang2026identifying}%
  \BibitemOpen
  \bibfield  {author} {\bibinfo {author} {\bibfnamefont {J.-X.}\ \bibnamefont
  {Zhang}}, \bibinfo {author} {\bibfnamefont {W.~O.}\ \bibnamefont {Wang}},
  \bibinfo {author} {\bibfnamefont {L.}~\bibnamefont {Balents}},\ and\ \bibinfo
  {author} {\bibfnamefont {L.}~\bibnamefont {Savary}},\ }\href@noop {}
  {\bibfield  {journal} {\bibinfo  {journal} {Physical Review Letters}\
  }\textbf {\bibinfo {volume} {136}},\ \bibinfo {pages} {176504} (\bibinfo
  {year} {2026})}\BibitemShut {NoStop}%
\bibitem [{\citenamefont {Rossi}(2021)}]{rossi2021quantum}%
  \BibitemOpen
  \bibfield  {author} {\bibinfo {author} {\bibfnamefont {E.}~\bibnamefont
  {Rossi}},\ }\href@noop {} {\bibfield  {journal} {\bibinfo  {journal} {Current
  Opinion in Solid State and Materials Science}\ }\textbf {\bibinfo {volume}
  {25}},\ \bibinfo {pages} {100952} (\bibinfo {year} {2021})}\BibitemShut
  {NoStop}%
\bibitem [{\citenamefont {Sukhachov}\ \emph {et~al.}(2025)\citenamefont
  {Sukhachov}, \citenamefont {Aase}, \citenamefont {M{\ae}land},\ and\
  \citenamefont {Sudb{\o}}}]{Sukhachov_2025}%
  \BibitemOpen
  \bibfield  {author} {\bibinfo {author} {\bibfnamefont {P.}~\bibnamefont
  {Sukhachov}}, \bibinfo {author} {\bibfnamefont {N.~H.}\ \bibnamefont {Aase}},
  \bibinfo {author} {\bibfnamefont {K.}~\bibnamefont {M{\ae}land}},\ and\
  \bibinfo {author} {\bibfnamefont {A.}~\bibnamefont {Sudb{\o}}},\ }\href@noop
  {} {\bibfield  {journal} {\bibinfo  {journal} {Physical Review B}\ }\textbf
  {\bibinfo {volume} {111}},\ \bibinfo {pages} {085143} (\bibinfo {year}
  {2025})}\BibitemShut {NoStop}%
\bibitem [{\citenamefont {Oh}\ \emph {et~al.}(2026)\citenamefont {Oh},
  \citenamefont {Kitamura}, \citenamefont {Daido}, \citenamefont {Rhim},\ and\
  \citenamefont {Yanase}}]{5fks-rrvg}%
  \BibitemOpen
  \bibfield  {author} {\bibinfo {author} {\bibfnamefont {C.-g.}\ \bibnamefont
  {Oh}}, \bibinfo {author} {\bibfnamefont {T.}~\bibnamefont {Kitamura}},
  \bibinfo {author} {\bibfnamefont {A.}~\bibnamefont {Daido}}, \bibinfo
  {author} {\bibfnamefont {J.-W.}\ \bibnamefont {Rhim}},\ and\ \bibinfo
  {author} {\bibfnamefont {Y.}~\bibnamefont {Yanase}},\ }\href@noop {}
  {\bibfield  {journal} {\bibinfo  {journal} {Physical Review Research}\
  }\textbf {\bibinfo {volume} {8}},\ \bibinfo {pages} {023096} (\bibinfo {year}
  {2026})}\BibitemShut {NoStop}%
\bibitem [{\citenamefont {Jiang}\ and\ \citenamefont
  {Barlas}(2023)}]{PhysRevLett.131.016002}%
  \BibitemOpen
  \bibfield  {author} {\bibinfo {author} {\bibfnamefont {G.}~\bibnamefont
  {Jiang}}\ and\ \bibinfo {author} {\bibfnamefont {Y.}~\bibnamefont {Barlas}},\
  }\href@noop {} {\bibfield  {journal} {\bibinfo  {journal} {Physical Review
  Letters}\ }\textbf {\bibinfo {volume} {131}},\ \bibinfo {pages} {016002}
  (\bibinfo {year} {2023})}\BibitemShut {NoStop}%
\bibitem [{\citenamefont {Sun}\ \emph {et~al.}(2025)\citenamefont {Sun},
  \citenamefont {Yu}, \citenamefont {Chen}, \citenamefont {Hu},\ and\
  \citenamefont {Law}}]{sun2025flat}%
  \BibitemOpen
  \bibfield  {author} {\bibinfo {author} {\bibfnamefont {Z.-T.}\ \bibnamefont
  {Sun}}, \bibinfo {author} {\bibfnamefont {R.-P.}\ \bibnamefont {Yu}},
  \bibinfo {author} {\bibfnamefont {S.~A.}\ \bibnamefont {Chen}}, \bibinfo
  {author} {\bibfnamefont {J.-X.}\ \bibnamefont {Hu}},\ and\ \bibinfo {author}
  {\bibfnamefont {K.}~\bibnamefont {Law}},\ }\href@noop {} {\bibfield
  {journal} {\bibinfo  {journal} {Quantum Frontiers}\ }\textbf {\bibinfo
  {volume} {4}},\ \bibinfo {pages} {20} (\bibinfo {year} {2025})}\BibitemShut
  {NoStop}%
\bibitem [{\citenamefont {Kitamura}\ \emph
  {et~al.}(2024{\natexlab{a}})\citenamefont {Kitamura}, \citenamefont {Daido},\
  and\ \citenamefont {Yanase}}]{PhysRevLett.132.036001}%
  \BibitemOpen
  \bibfield  {author} {\bibinfo {author} {\bibfnamefont {T.}~\bibnamefont
  {Kitamura}}, \bibinfo {author} {\bibfnamefont {A.}~\bibnamefont {Daido}},\
  and\ \bibinfo {author} {\bibfnamefont {Y.}~\bibnamefont {Yanase}},\
  }\href@noop {} {\bibfield  {journal} {\bibinfo  {journal} {Physical Review
  Letters}\ }\textbf {\bibinfo {volume} {132}},\ \bibinfo {pages} {036001}
  (\bibinfo {year} {2024}{\natexlab{a}})}\BibitemShut {NoStop}%
\bibitem [{\citenamefont {Shavit}\ and\ \citenamefont
  {Alicea}(2025)}]{shavit2025quantum}%
  \BibitemOpen
  \bibfield  {author} {\bibinfo {author} {\bibfnamefont {G.}~\bibnamefont
  {Shavit}}\ and\ \bibinfo {author} {\bibfnamefont {J.}~\bibnamefont
  {Alicea}},\ }\href@noop {} {\bibfield  {journal} {\bibinfo  {journal}
  {Physical Review Letters}\ }\textbf {\bibinfo {volume} {134}},\ \bibinfo
  {pages} {176001} (\bibinfo {year} {2025})}\BibitemShut {NoStop}%
\bibitem [{\citenamefont {Kitamura}\ \emph
  {et~al.}(2024{\natexlab{b}})\citenamefont {Kitamura}, \citenamefont {Daido},\
  and\ \citenamefont {Yanase}}]{kitamura2024spin}%
  \BibitemOpen
  \bibfield  {author} {\bibinfo {author} {\bibfnamefont {T.}~\bibnamefont
  {Kitamura}}, \bibinfo {author} {\bibfnamefont {A.}~\bibnamefont {Daido}},\
  and\ \bibinfo {author} {\bibfnamefont {Y.}~\bibnamefont {Yanase}},\
  }\href@noop {} {\bibfield  {journal} {\bibinfo  {journal} {Physical Review
  Letters}\ }\textbf {\bibinfo {volume} {132}},\ \bibinfo {pages} {036001}
  (\bibinfo {year} {2024}{\natexlab{b}})}\BibitemShut {NoStop}%
\bibitem [{\citenamefont {Hu}\ \emph {et~al.}(2026)\citenamefont {Hu},
  \citenamefont {Vafek}, \citenamefont {Haule},\ and\ \citenamefont
  {Bernevig}}]{hu2026ferromagnetism}%
  \BibitemOpen
  \bibfield  {author} {\bibinfo {author} {\bibfnamefont {H.}~\bibnamefont
  {Hu}}, \bibinfo {author} {\bibfnamefont {O.}~\bibnamefont {Vafek}}, \bibinfo
  {author} {\bibfnamefont {K.}~\bibnamefont {Haule}},\ and\ \bibinfo {author}
  {\bibfnamefont {B.~A.}\ \bibnamefont {Bernevig}},\ }\href@noop {} {\bibfield
  {journal} {\bibinfo  {journal} {Physical Review Letters}\ }\textbf {\bibinfo
  {volume} {136}},\ \bibinfo {pages} {256505} (\bibinfo {year}
  {2026})}\BibitemShut {NoStop}%
\bibitem [{\citenamefont {Hu}\ \emph {et~al.}(2025)\citenamefont {Hu},
  \citenamefont {Chen},\ and\ \citenamefont {Law}}]{hu2025anomalous}%
  \BibitemOpen
  \bibfield  {author} {\bibinfo {author} {\bibfnamefont {J.-X.}\ \bibnamefont
  {Hu}}, \bibinfo {author} {\bibfnamefont {S.~A.}\ \bibnamefont {Chen}},\ and\
  \bibinfo {author} {\bibfnamefont {K.~T.}\ \bibnamefont {Law}},\ }\href@noop
  {} {\bibfield  {journal} {\bibinfo  {journal} {Communications Physics}\
  }\textbf {\bibinfo {volume} {8}},\ \bibinfo {pages} {20} (\bibinfo {year}
  {2025})}\BibitemShut {NoStop}%
\bibitem [{\citenamefont {Guo}\ \emph {et~al.}(2025)\citenamefont {Guo},
  \citenamefont {Ma}, \citenamefont {Ying},\ and\ \citenamefont
  {Law}}]{guo2025majorana}%
  \BibitemOpen
  \bibfield  {author} {\bibinfo {author} {\bibfnamefont {X.}~\bibnamefont
  {Guo}}, \bibinfo {author} {\bibfnamefont {X.}~\bibnamefont {Ma}}, \bibinfo
  {author} {\bibfnamefont {X.}~\bibnamefont {Ying}},\ and\ \bibinfo {author}
  {\bibfnamefont {K.}~\bibnamefont {Law}},\ }\href@noop {} {\bibfield
  {journal} {\bibinfo  {journal} {Physical Review Letters}\ }\textbf {\bibinfo
  {volume} {135}},\ \bibinfo {pages} {076601} (\bibinfo {year}
  {2025})}\BibitemShut {NoStop}%
\bibitem [{\citenamefont {Li}\ \emph {et~al.}(2025)\citenamefont {Li},
  \citenamefont {Deng}, \citenamefont {Chen}, \citenamefont {Efetov},\ and\
  \citenamefont {Law}}]{PhysRevResearch.7.023273}%
  \BibitemOpen
  \bibfield  {author} {\bibinfo {author} {\bibfnamefont {Z.~C.}\ \bibnamefont
  {Li}}, \bibinfo {author} {\bibfnamefont {Y.}~\bibnamefont {Deng}}, \bibinfo
  {author} {\bibfnamefont {S.~A.}\ \bibnamefont {Chen}}, \bibinfo {author}
  {\bibfnamefont {D.~K.}\ \bibnamefont {Efetov}},\ and\ \bibinfo {author}
  {\bibfnamefont {K.}~\bibnamefont {Law}},\ }\href@noop {} {\bibfield
  {journal} {\bibinfo  {journal} {Physical Review Research}\ }\textbf {\bibinfo
  {volume} {7}},\ \bibinfo {pages} {023273} (\bibinfo {year}
  {2025})}\BibitemShut {NoStop}%
\bibitem [{\citenamefont {Chau}\ \emph {et~al.}(2026)\citenamefont {Chau},
  \citenamefont {Xiang}, \citenamefont {Chen},\ and\ \citenamefont
  {Law}}]{chau2026quantummetriclengthfundamental}%
  \BibitemOpen
  \bibfield  {author} {\bibinfo {author} {\bibfnamefont {C.~W.}\ \bibnamefont
  {Chau}}, \bibinfo {author} {\bibfnamefont {T.}~\bibnamefont {Xiang}},
  \bibinfo {author} {\bibfnamefont {S.~A.}\ \bibnamefont {Chen}},\ and\
  \bibinfo {author} {\bibfnamefont {K.~T.}\ \bibnamefont {Law}},\ }\href
  {https://arxiv.org/abs/2602.01354} {\bibinfo {title} {Quantum metric length
  as a fundamental length scale in disordered flat band materials}} (\bibinfo
  {year} {2026}),\ \Eprint {https://arxiv.org/abs/2602.01354} {arXiv:2602.01354
  [cond-mat.mes-hall]} \BibitemShut {NoStop}%
\bibitem [{\citenamefont {Ma}\ \emph {et~al.}(2025)\citenamefont {Ma},
  \citenamefont {Hu},\ and\ \citenamefont
  {Law}}]{ma2025universalboundarymodeslocalizationquantum}%
  \BibitemOpen
  \bibfield  {author} {\bibinfo {author} {\bibfnamefont {X.-L.}\ \bibnamefont
  {Ma}}, \bibinfo {author} {\bibfnamefont {J.-X.}\ \bibnamefont {Hu}},\ and\
  \bibinfo {author} {\bibfnamefont {K.~T.}\ \bibnamefont {Law}},\ }\href
  {https://arxiv.org/abs/2509.05114} {\bibinfo {title} {Universal
  boundary-modes localization from quantum metric length}} (\bibinfo {year}
  {2025}),\ \Eprint {https://arxiv.org/abs/2509.05114} {arXiv:2509.05114
  [cond-mat.mes-hall]} \BibitemShut {NoStop}%
\bibitem [{\citenamefont {Dai}\ \emph {et~al.}(2026)\citenamefont {Dai},
  \citenamefont {Zhao}, \citenamefont {Chen},\ and\ \citenamefont
  {Law}}]{dai2026quantummetriclocalizationquantum}%
  \BibitemOpen
  \bibfield  {author} {\bibinfo {author} {\bibfnamefont {W.-B.}\ \bibnamefont
  {Dai}}, \bibinfo {author} {\bibfnamefont {J.}~\bibnamefont {Zhao}}, \bibinfo
  {author} {\bibfnamefont {S.~A.}\ \bibnamefont {Chen}},\ and\ \bibinfo
  {author} {\bibfnamefont {K.~T.}\ \bibnamefont {Law}},\ }\href
  {https://arxiv.org/abs/2605.03987} {\bibinfo {title} {Quantum metric
  localization and quantum metric protection}} (\bibinfo {year} {2026}),\
  \Eprint {https://arxiv.org/abs/2605.03987} {arXiv:2605.03987
  [cond-mat.mes-hall]} \BibitemShut {NoStop}%
\bibitem [{\citenamefont {Zhao}\ \emph {et~al.}(2026)\citenamefont {Zhao},
  \citenamefont {Liu}, \citenamefont {Song},\ and\ \citenamefont
  {Law}}]{zhao2026quantummetricboundstate}%
  \BibitemOpen
  \bibfield  {author} {\bibinfo {author} {\bibfnamefont {J.}~\bibnamefont
  {Zhao}}, \bibinfo {author} {\bibfnamefont {R.}~\bibnamefont {Liu}}, \bibinfo
  {author} {\bibfnamefont {X.-Y.}\ \bibnamefont {Song}},\ and\ \bibinfo
  {author} {\bibfnamefont {K.~T.}\ \bibnamefont {Law}},\ }\href
  {https://arxiv.org/abs/2606.22479} {\bibinfo {title} {Quantum metric bound
  state of light}} (\bibinfo {year} {2026}),\ \Eprint
  {https://arxiv.org/abs/2606.22479} {arXiv:2606.22479 [cond-mat.mes-hall]}
  \BibitemShut {NoStop}%
\bibitem [{\citenamefont {Chen}\ \emph {et~al.}(2026)\citenamefont {Chen},
  \citenamefont {Yang}, \citenamefont {Cao}, \citenamefont {Lin}, \citenamefont
  {Yu},\ and\ \citenamefont {Xiao}}]{chen2026quantumgeometricquadrupolecooper}%
  \BibitemOpen
  \bibfield  {author} {\bibinfo {author} {\bibfnamefont {W.}~\bibnamefont
  {Chen}}, \bibinfo {author} {\bibfnamefont {K.}~\bibnamefont {Yang}}, \bibinfo
  {author} {\bibfnamefont {T.}~\bibnamefont {Cao}}, \bibinfo {author}
  {\bibfnamefont {S.-Z.}\ \bibnamefont {Lin}}, \bibinfo {author} {\bibfnamefont
  {J.}~\bibnamefont {Yu}},\ and\ \bibinfo {author} {\bibfnamefont
  {D.}~\bibnamefont {Xiao}},\ }\href {https://arxiv.org/abs/2605.03133}
  {\bibinfo {title} {Quantum geometric quadrupole of cooper pairs}} (\bibinfo
  {year} {2026}),\ \Eprint {https://arxiv.org/abs/2605.03133} {arXiv:2605.03133
  [cond-mat.supr-con]} \BibitemShut {NoStop}%
\bibitem [{\citenamefont {Lee}\ \emph {et~al.}(2026)\citenamefont {Lee},
  \citenamefont {Lee},\ and\ \citenamefont {Yang}}]{lee2026embedding}%
  \BibitemOpen
  \bibfield  {author} {\bibinfo {author} {\bibfnamefont {S.}~\bibnamefont
  {Lee}}, \bibinfo {author} {\bibfnamefont {S.~H.}\ \bibnamefont {Lee}},\ and\
  \bibinfo {author} {\bibfnamefont {B.-J.}\ \bibnamefont {Yang}},\ }\href@noop
  {} {\bibfield  {journal} {\bibinfo  {journal} {Physical Review Letters}\
  }\textbf {\bibinfo {volume} {137}},\ \bibinfo {pages} {016401} (\bibinfo
  {year} {2026})}\BibitemShut {NoStop}%
\bibitem [{\citenamefont {Dutreix}\ and\ \citenamefont
  {Katsnelson}(2016)}]{dutreix2016friedel}%
  \BibitemOpen
  \bibfield  {author} {\bibinfo {author} {\bibfnamefont {C.}~\bibnamefont
  {Dutreix}}\ and\ \bibinfo {author} {\bibfnamefont {M.}~\bibnamefont
  {Katsnelson}},\ }\href@noop {} {\bibfield  {journal} {\bibinfo  {journal}
  {Physical Review B}\ }\textbf {\bibinfo {volume} {93}},\ \bibinfo {pages}
  {035413} (\bibinfo {year} {2016})}\BibitemShut {NoStop}%
\bibitem [{\citenamefont {Singh}\ \emph {et~al.}(2003)\citenamefont {Singh},
  \citenamefont {Datta}, \citenamefont {Das},\ and\ \citenamefont
  {Singh}}]{singh2003ferromagnetism}%
  \BibitemOpen
  \bibfield  {author} {\bibinfo {author} {\bibfnamefont {A.}~\bibnamefont
  {Singh}}, \bibinfo {author} {\bibfnamefont {A.}~\bibnamefont {Datta}},
  \bibinfo {author} {\bibfnamefont {S.~K.}\ \bibnamefont {Das}},\ and\ \bibinfo
  {author} {\bibfnamefont {V.~A.}\ \bibnamefont {Singh}},\ }\href@noop {}
  {\bibfield  {journal} {\bibinfo  {journal} {Physical Review B}\ }\textbf
  {\bibinfo {volume} {68}},\ \bibinfo {pages} {235208} (\bibinfo {year}
  {2003})}\BibitemShut {NoStop}%
\bibitem [{\citenamefont {Chen}\ \emph {et~al.}(2024)\citenamefont {Chen},
  \citenamefont {Zhou}, \citenamefont {Zhang}, \citenamefont {Xu},\ and\
  \citenamefont {Lou}}]{chen2024impurity}%
  \BibitemOpen
  \bibfield  {author} {\bibinfo {author} {\bibfnamefont {W.}~\bibnamefont
  {Chen}}, \bibinfo {author} {\bibfnamefont {X.}~\bibnamefont {Zhou}}, \bibinfo
  {author} {\bibfnamefont {D.}~\bibnamefont {Zhang}}, \bibinfo {author}
  {\bibfnamefont {Y.-Q.}\ \bibnamefont {Xu}},\ and\ \bibinfo {author}
  {\bibfnamefont {W.-K.}\ \bibnamefont {Lou}},\ }\href@noop {} {\bibfield
  {journal} {\bibinfo  {journal} {Physical Review B}\ }\textbf {\bibinfo
  {volume} {110}},\ \bibinfo {pages} {165413} (\bibinfo {year}
  {2024})}\BibitemShut {NoStop}%
\bibitem [{\citenamefont {Sablikov}(2025)}]{sablikov2025friedel}%
  \BibitemOpen
  \bibfield  {author} {\bibinfo {author} {\bibfnamefont {V.~A.}\ \bibnamefont
  {Sablikov}},\ }\href@noop {} {\bibfield  {journal} {\bibinfo  {journal}
  {Physica E: Low-dimensional Systems and Nanostructures}\ }\textbf {\bibinfo
  {volume} {170}},\ \bibinfo {pages} {116213} (\bibinfo {year}
  {2025})}\BibitemShut {NoStop}%
\bibitem [{\citenamefont {Marzari}\ and\ \citenamefont
  {Vanderbilt}(1997)}]{marzari1997maximally}%
  \BibitemOpen
  \bibfield  {author} {\bibinfo {author} {\bibfnamefont {N.}~\bibnamefont
  {Marzari}}\ and\ \bibinfo {author} {\bibfnamefont {D.}~\bibnamefont
  {Vanderbilt}},\ }\href@noop {} {\bibfield  {journal} {\bibinfo  {journal}
  {Physical Review B}\ }\textbf {\bibinfo {volume} {56}},\ \bibinfo {pages}
  {12847} (\bibinfo {year} {1997})}\BibitemShut {NoStop}%
\bibitem [{\citenamefont {Marzari}\ \emph {et~al.}(2012)\citenamefont
  {Marzari}, \citenamefont {Mostofi}, \citenamefont {Yates}, \citenamefont
  {Souza},\ and\ \citenamefont {Vanderbilt}}]{marzari2012maximally}%
  \BibitemOpen
  \bibfield  {author} {\bibinfo {author} {\bibfnamefont {N.}~\bibnamefont
  {Marzari}}, \bibinfo {author} {\bibfnamefont {A.~A.}\ \bibnamefont
  {Mostofi}}, \bibinfo {author} {\bibfnamefont {J.~R.}\ \bibnamefont {Yates}},
  \bibinfo {author} {\bibfnamefont {I.}~\bibnamefont {Souza}},\ and\ \bibinfo
  {author} {\bibfnamefont {D.}~\bibnamefont {Vanderbilt}},\ }\href@noop {}
  {\bibfield  {journal} {\bibinfo  {journal} {Reviews of Modern Physics}\
  }\textbf {\bibinfo {volume} {84}},\ \bibinfo {pages} {1419} (\bibinfo {year}
  {2012})}\BibitemShut {NoStop}%
\bibitem [{\citenamefont {Souza}\ \emph {et~al.}(2001)\citenamefont {Souza},
  \citenamefont {Marzari},\ and\ \citenamefont
  {Vanderbilt}}]{souza2001maximally}%
  \BibitemOpen
  \bibfield  {author} {\bibinfo {author} {\bibfnamefont {I.}~\bibnamefont
  {Souza}}, \bibinfo {author} {\bibfnamefont {N.}~\bibnamefont {Marzari}},\
  and\ \bibinfo {author} {\bibfnamefont {D.}~\bibnamefont {Vanderbilt}},\
  }\href@noop {} {\bibfield  {journal} {\bibinfo  {journal} {Physical Review
  B}\ }\textbf {\bibinfo {volume} {65}},\ \bibinfo {pages} {035109} (\bibinfo
  {year} {2001})}\BibitemShut {NoStop}%
\bibitem [{\citenamefont {Liu}\ \emph {et~al.}(2025{\natexlab{b}})\citenamefont
  {Liu}, \citenamefont {Hong}, \citenamefont {Zhang}, \citenamefont {Zhu},
  \citenamefont {Dong}, \citenamefont {Watanabe}, \citenamefont {Taniguchi},
  \citenamefont {Du}, \citenamefont {Shi}, \citenamefont {Law} \emph
  {et~al.}}]{liu2025electric}%
  \BibitemOpen
  \bibfield  {author} {\bibinfo {author} {\bibfnamefont {L.}~\bibnamefont
  {Liu}}, \bibinfo {author} {\bibfnamefont {Y.}~\bibnamefont {Hong}}, \bibinfo
  {author} {\bibfnamefont {C.}~\bibnamefont {Zhang}}, \bibinfo {author}
  {\bibfnamefont {J.}~\bibnamefont {Zhu}}, \bibinfo {author} {\bibfnamefont
  {J.}~\bibnamefont {Dong}}, \bibinfo {author} {\bibfnamefont {K.}~\bibnamefont
  {Watanabe}}, \bibinfo {author} {\bibfnamefont {T.}~\bibnamefont {Taniguchi}},
  \bibinfo {author} {\bibfnamefont {L.}~\bibnamefont {Du}}, \bibinfo {author}
  {\bibfnamefont {D.}~\bibnamefont {Shi}}, \bibinfo {author} {\bibfnamefont
  {K.~T.}\ \bibnamefont {Law}}, \emph {et~al.},\ }\href@noop {} {\bibfield
  {journal} {\bibinfo  {journal} {arXiv preprint arXiv:2501.06460}\ } (\bibinfo
  {year} {2025}{\natexlab{b}})}\BibitemShut {NoStop}%
\bibitem [{\citenamefont {Geng}\ \emph {et~al.}(2026)\citenamefont {Geng},
  \citenamefont {Wang}, \citenamefont {Guo}, \citenamefont {Qiu}, \citenamefont
  {Chen}, \citenamefont {Wang}, \citenamefont {Li}, \citenamefont {Hao},
  \citenamefont {Liang}, \citenamefont {Huang} \emph
  {et~al.}}]{geng2026experimental}%
  \BibitemOpen
  \bibfield  {author} {\bibinfo {author} {\bibfnamefont {S.}~\bibnamefont
  {Geng}}, \bibinfo {author} {\bibfnamefont {X.}~\bibnamefont {Wang}}, \bibinfo
  {author} {\bibfnamefont {R.}~\bibnamefont {Guo}}, \bibinfo {author}
  {\bibfnamefont {C.}~\bibnamefont {Qiu}}, \bibinfo {author} {\bibfnamefont
  {F.}~\bibnamefont {Chen}}, \bibinfo {author} {\bibfnamefont {Q.}~\bibnamefont
  {Wang}}, \bibinfo {author} {\bibfnamefont {K.}~\bibnamefont {Li}}, \bibinfo
  {author} {\bibfnamefont {P.}~\bibnamefont {Hao}}, \bibinfo {author}
  {\bibfnamefont {H.}~\bibnamefont {Liang}}, \bibinfo {author} {\bibfnamefont
  {Y.}~\bibnamefont {Huang}}, \emph {et~al.},\ }\href@noop {} {\bibfield
  {journal} {\bibinfo  {journal} {Nature Communications}\ } (\bibinfo {year}
  {2026})}\BibitemShut {NoStop}%
\bibitem [{\citenamefont {Zhong}\ \emph {et~al.}(2026)\citenamefont {Zhong},
  \citenamefont {Geng}, \citenamefont {Ying}, \citenamefont {Li},\ and\
  \citenamefont {Zhou}}]{zhong2026yclelectridemultiorbitalcorrelated}%
  \BibitemOpen
  \bibfield  {author} {\bibinfo {author} {\bibfnamefont {J.}~\bibnamefont
  {Zhong}}, \bibinfo {author} {\bibfnamefont {S.}~\bibnamefont {Geng}},
  \bibinfo {author} {\bibfnamefont {T.-F.}\ \bibnamefont {Ying}}, \bibinfo
  {author} {\bibfnamefont {H.}~\bibnamefont {Li}},\ and\ \bibinfo {author}
  {\bibfnamefont {B.~T.}\ \bibnamefont {Zhou}},\ }\href
  {https://arxiv.org/abs/2509.05958} {\bibinfo {title} {Ycl electride as a
  multi-orbital correlated topological dice lattice system}} (\bibinfo {year}
  {2026}),\ \Eprint {https://arxiv.org/abs/2509.05958} {arXiv:2509.05958
  [cond-mat.mtrl-sci]} \BibitemShut {NoStop}%
\bibitem [{\citenamefont {Checkelsky}\ \emph {et~al.}(2024)\citenamefont
  {Checkelsky}, \citenamefont {Bernevig}, \citenamefont {Coleman},
  \citenamefont {Si},\ and\ \citenamefont {Paschen}}]{checkelsky2024flat}%
  \BibitemOpen
  \bibfield  {author} {\bibinfo {author} {\bibfnamefont {J.~G.}\ \bibnamefont
  {Checkelsky}}, \bibinfo {author} {\bibfnamefont {B.~A.}\ \bibnamefont
  {Bernevig}}, \bibinfo {author} {\bibfnamefont {P.}~\bibnamefont {Coleman}},
  \bibinfo {author} {\bibfnamefont {Q.}~\bibnamefont {Si}},\ and\ \bibinfo
  {author} {\bibfnamefont {S.}~\bibnamefont {Paschen}},\ }\href@noop {}
  {\bibfield  {journal} {\bibinfo  {journal} {Nature Reviews Materials}\
  }\textbf {\bibinfo {volume} {9}},\ \bibinfo {pages} {509} (\bibinfo {year}
  {2024})}\BibitemShut {NoStop}%
\bibitem [{\citenamefont {Bochner}(1932)}]{bochner1932vorlesungen}%
  \BibitemOpen
  \bibfield  {author} {\bibinfo {author} {\bibfnamefont {S.}~\bibnamefont
  {Bochner}},\ }\href@noop {} {\emph {\bibinfo {title} {Vorlesungen uber
  fouriersche integrale}}}\ (\bibinfo  {publisher} {Akademische
  Verlagsgesellschaft},\ \bibinfo {year} {1932})\BibitemShut {NoStop}%
\bibitem [{\citenamefont {Roy}(2014)}]{PhysRevB.90.165139}%
  \BibitemOpen
  \bibfield  {author} {\bibinfo {author} {\bibfnamefont {R.}~\bibnamefont
  {Roy}},\ }\href {https://doi.org/10.1103/PhysRevB.90.165139} {\bibfield
  {journal} {\bibinfo  {journal} {Physical Review B}\ }\textbf {\bibinfo
  {volume} {90}},\ \bibinfo {pages} {165139} (\bibinfo {year}
  {2014})}\BibitemShut {NoStop}%
\bibitem [{\citenamefont {Lieb}(1989)}]{lieb1989two}%
  \BibitemOpen
  \bibfield  {author} {\bibinfo {author} {\bibfnamefont {E.~H.}\ \bibnamefont
  {Lieb}},\ }\href@noop {} {\bibfield  {journal} {\bibinfo  {journal} {Physical
  Review Letters}\ }\textbf {\bibinfo {volume} {62}},\ \bibinfo {pages} {1201}
  (\bibinfo {year} {1989})}\BibitemShut {NoStop}%
\bibitem [{\citenamefont {Laubscher}\ \emph {et~al.}(2023)\citenamefont
  {Laubscher}, \citenamefont {Weber}, \citenamefont {H{\"u}nenberger},
  \citenamefont {Schoeller}, \citenamefont {Kennes}, \citenamefont {Loss},\
  and\ \citenamefont {Klinovaja}}]{PhysRevB.108.155429}%
  \BibitemOpen
  \bibfield  {author} {\bibinfo {author} {\bibfnamefont {K.}~\bibnamefont
  {Laubscher}}, \bibinfo {author} {\bibfnamefont {C.~S.}\ \bibnamefont
  {Weber}}, \bibinfo {author} {\bibfnamefont {M.}~\bibnamefont
  {H{\"u}nenberger}}, \bibinfo {author} {\bibfnamefont {H.}~\bibnamefont
  {Schoeller}}, \bibinfo {author} {\bibfnamefont {D.~M.}\ \bibnamefont
  {Kennes}}, \bibinfo {author} {\bibfnamefont {D.}~\bibnamefont {Loss}},\ and\
  \bibinfo {author} {\bibfnamefont {J.}~\bibnamefont {Klinovaja}},\ }\href@noop
  {} {\bibfield  {journal} {\bibinfo  {journal} {Physical Review B}\ }\textbf
  {\bibinfo {volume} {108}},\ \bibinfo {pages} {155429} (\bibinfo {year}
  {2023})}\BibitemShut {NoStop}%
\bibitem [{\citenamefont {Bouzerar}(2021)}]{PhysRevB.104.155151}%
  \BibitemOpen
  \bibfield  {author} {\bibinfo {author} {\bibfnamefont {G.}~\bibnamefont
  {Bouzerar}},\ }\href@noop {} {\bibfield  {journal} {\bibinfo  {journal}
  {Physical Review B}\ }\textbf {\bibinfo {volume} {104}},\ \bibinfo {pages}
  {155151} (\bibinfo {year} {2021})}\BibitemShut {NoStop}%
\bibitem [{\citenamefont {Luo}\ and\ \citenamefont
  {Yang}(2024)}]{luo2024influence}%
  \BibitemOpen
  \bibfield  {author} {\bibinfo {author} {\bibfnamefont {Y.-D.}\ \bibnamefont
  {Luo}}\ and\ \bibinfo {author} {\bibfnamefont {M.-F.}\ \bibnamefont {Yang}},\
  }\href@noop {} {\bibfield  {journal} {\bibinfo  {journal} {Physical Review
  B}\ }\textbf {\bibinfo {volume} {110}},\ \bibinfo {pages} {144402} (\bibinfo
  {year} {2024})}\BibitemShut {NoStop}%
\bibitem [{\citenamefont {Cao}\ \emph {et~al.}(2019)\citenamefont {Cao},
  \citenamefont {Fertig},\ and\ \citenamefont {Zhang}}]{PhysRevB.99.205430}%
  \BibitemOpen
  \bibfield  {author} {\bibinfo {author} {\bibfnamefont {J.}~\bibnamefont
  {Cao}}, \bibinfo {author} {\bibfnamefont {H.~A.}\ \bibnamefont {Fertig}},\
  and\ \bibinfo {author} {\bibfnamefont {S.}~\bibnamefont {Zhang}},\
  }\href@noop {} {\bibfield  {journal} {\bibinfo  {journal} {Physical Review
  B}\ }\textbf {\bibinfo {volume} {99}},\ \bibinfo {pages} {205430} (\bibinfo
  {year} {2019})}\BibitemShut {NoStop}%
\bibitem [{\citenamefont {Oriekhov}\ and\ \citenamefont
  {Gusynin}(2020)}]{oriekhov2020rkky}%
  \BibitemOpen
  \bibfield  {author} {\bibinfo {author} {\bibfnamefont {D.}~\bibnamefont
  {Oriekhov}}\ and\ \bibinfo {author} {\bibfnamefont {V.}~\bibnamefont
  {Gusynin}},\ }\href@noop {} {\bibfield  {journal} {\bibinfo  {journal}
  {Physical Review B}\ }\textbf {\bibinfo {volume} {101}},\ \bibinfo {pages}
  {235162} (\bibinfo {year} {2020})}\BibitemShut {NoStop}%
\bibitem [{\citenamefont {Black-Schaffer}(2010)}]{PhysRevB.81.205416}%
  \BibitemOpen
  \bibfield  {author} {\bibinfo {author} {\bibfnamefont {A.~M.}\ \bibnamefont
  {Black-Schaffer}},\ }\href@noop {} {\bibfield  {journal} {\bibinfo  {journal}
  {Physical Review B}\ }\textbf {\bibinfo {volume} {81}},\ \bibinfo {pages}
  {205416} (\bibinfo {year} {2010})}\BibitemShut {NoStop}%
\bibitem [{\citenamefont {Coleman}(2015)}]{coleman2015introduction}%
  \BibitemOpen
  \bibfield  {author} {\bibinfo {author} {\bibfnamefont {P.}~\bibnamefont
  {Coleman}},\ }\href@noop {} {\emph {\bibinfo {title} {Introduction to
  many-body physics}}}\ (\bibinfo  {publisher} {Cambridge University Press},\
  \bibinfo {year} {2015})\BibitemShut {NoStop}%
\bibitem [{\citenamefont {Deaven}\ \emph {et~al.}(1991)\citenamefont {Deaven},
  \citenamefont {Rokhsar},\ and\ \citenamefont {Johnson}}]{deaven1991simple}%
  \BibitemOpen
  \bibfield  {author} {\bibinfo {author} {\bibfnamefont {D.}~\bibnamefont
  {Deaven}}, \bibinfo {author} {\bibfnamefont {D.}~\bibnamefont {Rokhsar}},\
  and\ \bibinfo {author} {\bibfnamefont {M.}~\bibnamefont {Johnson}},\
  }\href@noop {} {\bibfield  {journal} {\bibinfo  {journal} {Physical Review
  B}\ }\textbf {\bibinfo {volume} {44}},\ \bibinfo {pages} {5977} (\bibinfo
  {year} {1991})}\BibitemShut {NoStop}%
\end{thebibliography}%

%\end{document}

	\end{document}